\documentclass[showpacs, pra,twocolumn,preprintnumbers ,amsmath, amssymb, superscriptaddress, aps]{revtex4-2}

\usepackage{color}
\usepackage{amsmath,amssymb}
\usepackage{pifont}
\usepackage{amssymb}  
\usepackage{bbold}
\usepackage{float}
\usepackage{subfloat}

\usepackage[caption=false]{subfig}
\usepackage{tikz}
\usepackage{makecell}
\usepackage{subfig}
\usepackage{pifont}   
\usepackage{graphicx} 
\usepackage{dcolumn}  
\usepackage{bm}       
\usepackage{multirow} 
\usepackage{placeins}
\usepackage[colorlinks]{hyperref}
\usepackage{mathtools}

\newcommand{\vect}[1]{\mathbf{#1}}
\newcommand{\ket}[1]{\left|{#1}\right\rangle}
\newcommand{\bra}[1]{\left\langle{#1}\right|}

\begin{document}
\title{Light driven magnetic transitions in transition metal dichalcogenide heterobilayers}

	\author{Michael Vogl}
	\email{ssss133@googlemail.com}
	\affiliation{Department of Physics, King Fahd University of Petroleum and Minerals, 31261 Dhahran, Saudi Arabia}
	\author{Swati Chaudhary}
	\affiliation{Department of Physics, The University of Texas at Austin, Austin, Texas 78712, USA}
\affiliation{Department of Physics, Northeastern University, Boston, Massachusetts 02115, USA}
\affiliation{Department of Physics, Massachusetts Institute of Technology, Cambridge, Massachusetts 02139, USA}

\author{Gregory A. Fiete}
\affiliation{Department of Physics, Northeastern University, Boston, Massachusetts 02115, USA}
\affiliation{Department of Physics, Massachusetts Institute of Technology, Cambridge, Massachusetts 02139, USA}
	\date{\today}
	\begin{abstract}
	    Motivated by the recent excitement around the physics of twisted transition metal dichalcogenide (TMD) multilayer systems, we study strongly correlated phases of TMD heterobilayers under the influence of light. We consider both waveguide light and circularly polarized light.  The former allows for longitudinally polarized light, which in the high frequency limit can be used to selectively modify interlayer hoppings in a tight-binding model. We argue based on quasi-degenerate perturbation theory that changes to the interlayer hoppings can be captured as a modulation to the strength of the moir\'e potential in a continuum model. 
	    As a consequence, waveguide light can be used to drive transitions between a myriad of different magnetic phases, including a transition from a $120^\circ$ Neel phase to a stripe ordered magnetic phase, or from a spin density wave phase to a paramagnetic phase, among others.  When the system is subjected to circularly polarized light we find that the effective mass of the active TMD layer is modified by an applied electromagnetic field. By simultaneously applying waveguide light and circularly polarized light to a system, one has a high level of control in moving through the phase diagram in-situ. Lastly,  we comment on the experimental feasibility of Floquet state preparation and argue that it is within reach of available techniques when the system is coupled to a judiciously chosen bath.
	\end{abstract}
    \maketitle

\section{Introduction}
Recent years have seen an explosion of interest in so-called moir\'e materials \cite{andrei2021marvels,he2021moire,macdonald2019bilayer,zeller2014possible,kennes2021moire}. These are materials in which two dimensional crystals are stacked in such a way that their lattices interfere to form moir\'e patterns. In the earliest works related to such a material - twisted bilayer graphene - the moir\'e patterns appear due to a relative twist between two graphene layers and have been found to lead to flat bands near a ``magic angle" \cite{bistritzer2011moire,morell2010flat,dosSantos2012}. It was speculated that this could lead to various strongly correlated phases, because flat bands imply tiny kinetic energy making any weak electron-electron interaction energy dominant \cite{Hu:prb11, Parameswaran2013816, Wu_CI:prb12, Wang:prb11, Sun:prl11, Neupert:prl11, Tang:prl11, Regnault:prx11, Kourtis:prl14}. 

In 2018 interaction driven phases in flat band systems were experimentally confirmed when superconductivity and insulating phases in twisted bilayer graphene (TBG)were reported\cite{cao2018unconventional}. Since then there have been theoretical predictions and experimental discoveries of a plethora of different strongly correlated phases\cite{cao2018unconventional,10.21468/SciPostPhys.11.4.083,xie2021fractional,cao2020tunable,wilhelm2021interplay,rademaker2018charge,yankowitz2019tuning,choi2019electronic,sherkunov2018electronic,xie2020nature,Codecidoeaaw9770,wong2019cascade,Lu2019efetov,Sharpe605,Seo_2019}. These phases can range from various magnetic states\cite{Sharpe605,Seo_2019,Lu2019efetov,wong2019cascade} to superconducting states\cite{Lu2019efetov,wong2019cascade}. This discovery naturally leads to the question of which related materials may harbor similarly exciting properties. 

One such material class that has recently risen to popularity is bilayer or few layer transition metal dichalogenides (TMDs) with a relative twist between layers\cite{Tran2019,Seyler2019,Alexeev2019,hu2021competing,PhysRevLett.121.026402,moralesduran2021nonlocal,2019wulovorn,tang2020simulation,PhysRevB.104.115154,andersen2021excitons}. Similar to TBG, theory has predicted \cite{PhysRevLett.121.026402,hu2021competing,moralesduran2021nonlocal,PhysRevB.104.115154,tang2020simulation} and experiment has confirmed\cite{wang2020correlated,andersen2021excitons} there exists a zoo of strongly correlated phases.  Analogously to TBG, these phases can be tuned via the twist angle. Compared to TBG, the twisted TMD systems offer the addition degree of freedom of material choice used to build the moir\'e heterostructure which can be used to tune different phases of matter.  Recently, it was shown theoretically that TMD heterobilayers possess very rich phase diagrams that include a multitude of different magnetically ordered phases \cite{hu2021competing}.

While TMD heterostructures  might seem more complicated than graphene based multilayer materials (each TMD layer corresponds to three atomic layers), effective descriptions for TMD heterostructures can be significantly simpler than TBG because the former allow for an effective single band description that mimics a two-dimensional non-relativistic single particle Schr\"odinger equation with a periodic potential originating in the moir\'e patterns \cite{moralesduran2021nonlocal,hu2021competing,PhysRevLett.121.026402,Angelie2021826118}. This is in stark contrast to the case of graphene based moir\'e materials such as TBG, where one needs a four component Hamiltonian similar to a massless Dirac equation to describe layer and sublattice pseudo-spin degrees of freedom. It is worth mentioning, however, that recently, the dimensionality of the twisted bilayer graphene Hamiltonian for a chiral limit has been reduced to a 2x2 Hamiltonian in an approach where the Hamiltonian is squared \cite{PhysRevB.105.115434,PhysRevB.103.245418}. Nevertheless, for TBG a mapping from position dependent interlayer hoppings to an effective potential and a single band description is problematic. In fact, such a mapping has only been made rigorously for a closely related effective one-dimensional case, where moir\'e patterns are due to a stretch of one of the layers \cite{PhysRevB.101.075413,fleischmann2019perfect,2017mvoglsemiclass}. Even in this case, the argument only applies semiclassically.  It is the simplicity of effective Hamiltonians for the TMD heterobilayers that motivates us to focus on the engineering of phase transitions for this class of materials, rather than TBG.

The TMD heterobilayer platform can be made even more manipulable by combining it with Floquet engineering which has been touted as a very powerful technique to modify the properties of quantum systems with light. In this technique, periodic drives are used to induce different material properties in a so-called prethermal phase that bears resemblence to an equilibrium phase. Recent years have witnessed incredible progress both on the theoretical as well as the experimental side of Floquet engineering. Specifically, on the theoretical side, there has been progress on estimating how long (given certain assumptions to driving strengths and frequencies) one can remain in the prethermal regime, which is often exponentially long \cite{PhysRevB.95.014112,polkovnikov2018repl}. 

There has been tremendous progress in the development of techniques that facilitate effective time-independent descriptions using so-called effective  Floquet Hamiltonians\cite{blanes2009,rahav2003,rahav2003b,bukov2015,Eckardt_2015,Feldm1984,Magnus1954,PhysRevB.95.014112,PhysRevX.4.031027,PhysRevLett.115.075301,PhysRevB.93.144307,PhysRevB.94.235419,PhysRevLett.116.125301,Vogl_2019AnalogHJ,verdeny2013,    PhysRevLett.110.200403,Vogl2020_effham,vogl2019,Rodriguez_Vega_2018,PhysRevLett.121.036402,Martiskainen2015,rigolin2008,weinberg2015,PhysRevB.96.155438,Kennes-Klinovaja-2019,PhysRevB.101.155417,novicenko2021flowequation}. Importantly, there has also been progress in our understanding of how to prepare prethermal Floquet phases\cite{Shirai_2016,10.21468/SciPostPhysCore.4.4.033,PhysRevB.104.134308}.  For instance, it is now better understood what properties a bath has to fulfill for a system to relax into a so-called Floquet-Gibbs state\cite{Shirai_2016,10.21468/SciPostPhysCore.4.4.033,PhysRevLett.123.120602}, which can tell us much about how to prepare the ground state of a Floquet Hamiltonian. In addition to the work on the theoretical foundations of Floquet engineering, there have been predictions of a multitude of exciting light induced properties and phases of matter \cite{Basov2017,Oka_2019, rudner2020_review,Giovannini_2019,McIver2020,oka2009,lindner2011,rechtsman2013,Rudner2013,PhysRevLett.105.017401,PhysRevLett.123.016806,Huang_2020,PhysRevB.99.045441,PhysRevResearch.2.013124,PhysRevB.102.094305,chaudhary2019phononinduced,Haiping2020,PhysRevLett.120.127601,PhysRevX.4.041048,PhysRevLett.118.115301,PhysRevB.98.045127,PhysRevB.93.245145,kennes2019,PhysRevLett.108.056602,PhysRevLett.124.190601,PhysRevB.94.224202,Abanin_2016,PhysRevLett.114.140401,RevModPhys.91.021001,lindner2011,dallago2015,PhysRevB.91.241404,asboth2014,calvo2011,PhysRevB.104.245135,zhenghao2011,PhysRevB.88.241112,perez2014,kundu2014,Sentef2015,Roman-Taboada2017,Dehghani2015,kitagawa2011,Usaj2014,PhysRevA.91.043625,bhattacharya2020fermionic,mentink2014,mentink2015ultrafast,mentink2017manipulating,hejazi2018,hejazi2019,Quito2021,Quito2021b,Chaudhary2019orbital,chaudhary2020controlling} (see \cite{RODRIGUEZVEGA2021168434,bao2021light} for a recent review article). Recently, Floquet engineering is gaining popularity in moir\'e materials - albeit in the non-interacting limit\cite{RODRIGUEZVEGA2021168434,ibsal2021,floquetTMDV2021,sentef2021,sentef2019,luzeng2021,PhysRevResearch.2.033494,PhysRevB.101.235411,michaelvoglInterlayer2020,PhysRevB.102.155123,PhysRevResearch.2.043275,bao2021light,Chaudhary2019orbital,chaudhary2020controlling,ge2021floquet}.

\begin{figure}[htb]
	\centering
\includegraphics[width=\linewidth]{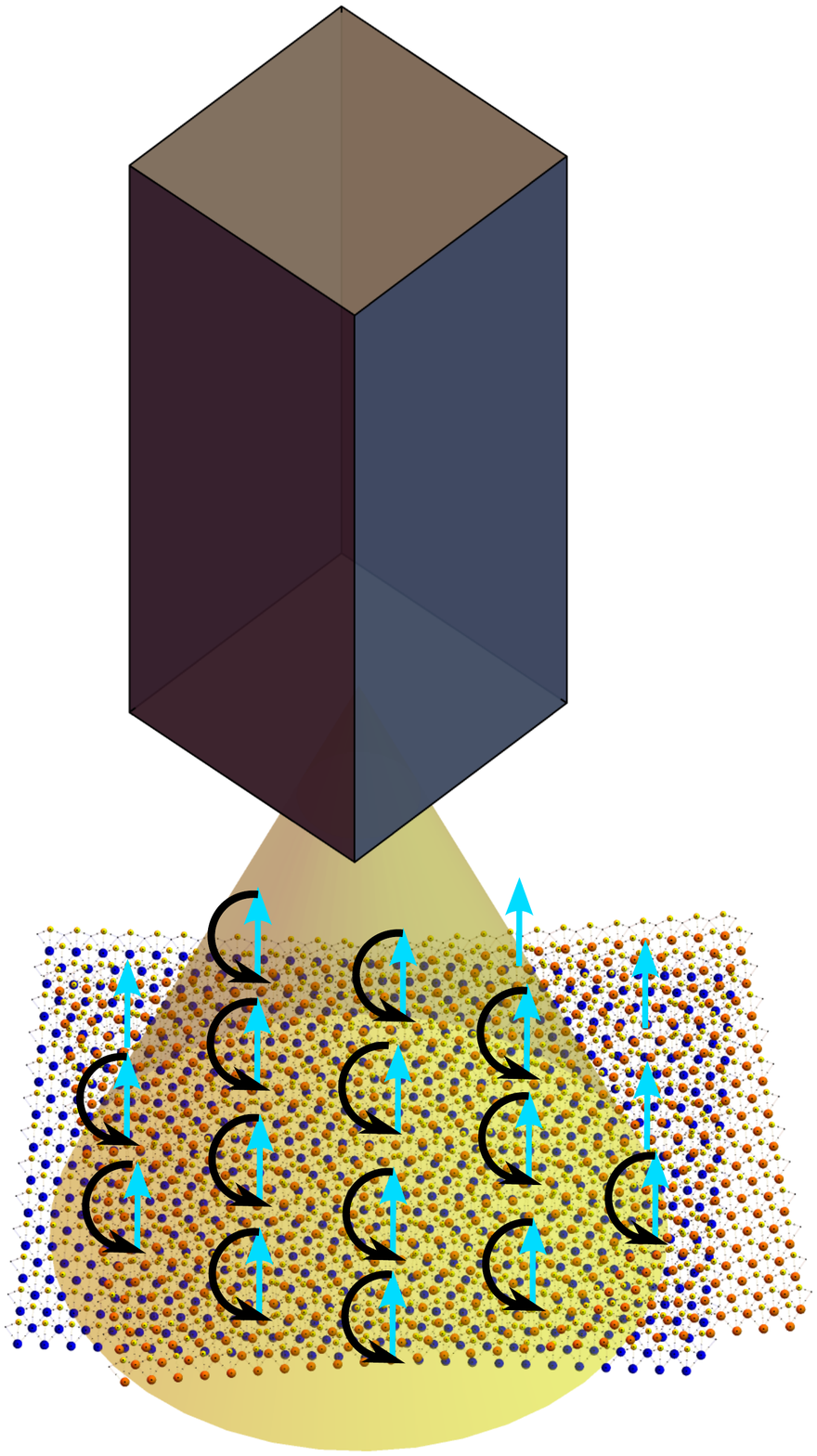}
	\caption{ Schematic representation of how the magnetic ground state of a twisted TMD heterobilayer placed at the exit of a waveguide (illustrated as a dark rectangular prism) can be affected by light emanating from the waveguide. For example, the light might cause some of the spins to reorient themselves. Blue and orange spheres correspond to transition metal atoms, yellow ones to chalcogenides. The two layers have a relative twist, as shown. Spins are represented in light blue.}\label{fig:light_magn_state}
\end{figure}

In this work, we make a first step towards Floquet engineering of different phases of matter in \emph{interacting} moir\'e materials. To maximize the accessibility of our results, we will focus on a simplified approach using a model description of a twisted TMD heterobilayer. This will allow us to focus on the universal features expected in experiment. 
A schematic depiction of the experimental scenario we envision can be seen in the Fig. \ref{fig:light_magn_state}.

Our paper is organized as follows. First, in Sec.\ref{sec:model} we give a brief review of the effective Hamiltonian used to model TMD heterobilayers. Then, in Sec.\ref{sec:engineering} we briefly describe the techniques used to derive interaction terms for an effective lattice model and discuss how a specific form of light, waveguide light possessing a longitudinal polarization, can be used to target specific parameters of the non-interacting limit.  Similarly, we show that circularly polarized light can be employed to tune the effective mass of the active TMD layer, which also appears in the non-interacting limit. A central result of our work is that by tuning the properties  and type of incident light one can reach a multitude of vastly different magnetic phases of matter in a twisted TMD without the need to change the specifics  of the sample, such as the twist angle or constituent TMDs.  One achieves dramatic control over the phase diagram. Finally, in Sec.\ref{sec:preparation} we finish with a discussion on how specifics of the environment determine how  ``Floquet ground state" phases can be prepared in practice. We conclude in Sec.\ref{sec:conclusions}.

\section{Model}
\label{sec:model}
We consider a generic non-interacting model \cite{moralesduran2021nonlocal,hu2021competing,PhysRevLett.121.026402}
\begin{equation}
    H=-\frac{\hbar^2 \vect k^2}{2m^*}+2\Delta(\vect x),
    \label{eq:non_int_ham}
\end{equation}
for ``electrons" in a TMD heterobilayer such as WSe$_2$/MoSe$_2$ or WSe$_2$/MoS$_2$. In this model, electrons are constrained to move in only one of the  transition metal layers--the so-called active layer. This means that additional layers that appear in such a TMD heterostructure are only taken into account via the effective potential $\Delta(\vect x)$ and mass $m^*$. Some interaction effects are taken into account on a density functional theory (DFT) level because these models are typically fit to DFT band structures. Furthermore, it is important to stress that the model is coarse-grained, losing some short distance features. It is most accurate at long wavelengths, specifically near the $\Gamma$ point of a single TMD layer \cite{PhysRevLett.121.026402}. 

Electronic excitations in this model have an effective mass of $m^*\sim 0.5 m_e$, where $m_e$ is the bare electron mass (details depend crucially on the type of heterobilayer, whether  WSe$_2$/MoSe$_2$ or WSe$_2$/MoS$_2$ or something else). For a twisted bilayer with small twist angle the effective periodic potential $\Delta(\vect x)$ has a six-fold rotational symmetry  due to moir\'e patterns that appear from the interference between the two lattices. It is given as \cite{moralesduran2021nonlocal,hu2021competing,PhysRevLett.121.026402}
\begin{equation}
    \Delta(\vect x)= V_m\sum_{i=1}^3 \cos(\vect b_i \vect x+\psi),
    \label{eq:potential}
\end{equation}
 where  $V_m$ is the strength of this potential and typically $V_m\sim 10meV$. Here, the $\vect b_i$ are the three shortest reciprocal lattice vectors for the moir\'e lattice that are related by a $120^\circ$ rotation and $\psi$ is a phase shift that is determined by band structure fits. 

As described in Refs. \cite{hu2021competing,PhysRevB.102.201104}, strong correlation effects for twisted TMDs can be included in a simple approximation where Wannier wavefunctions are determined using the non-interacting model Eq.~\eqref{eq:non_int_ham} only. These Wannier wavefunctions can then be used to compute effective hopping terms on the moir\'e lattice, Hubbard-$U$-like interaction terms, etc. Therefore, much of the interacting structure is crucially influenced by the properties of the non-interacting underlying model. In some cases approximations can even be found by analytical means. For instance, it has been shown that one may approximate the potential Eq.~\eqref{eq:potential} by a harmonic potential \cite{moralesduran2021nonlocal}. Harmonic oscillator wavefunctions then act as reasonably accurate approximations to the actual Wannier functions\cite{moralesduran2021nonlocal} and computations of Hubbard-$U$-like interaction terms can be done analytically, offering much insight\cite{moralesduran2021nonlocal}. 

The most important point we would like to emphasize is that the interaction parameters will depend on the parameters of the non-interacting model--particularly the moir\'e potential strength $V_m$.  This observation is especially lucid in the case of the work by Hu {\it et al.} \cite{hu2021competing}, where only one of the axes in their interacting phase diagram (see Fig.~\ref{fig:phasediagram_move}) depends on $V_m$. This observation serves as motivation for the next section.

\section{Light induced engineering of phase transitions}
\label{sec:engineering}
In this section, motivated by the work of Hu {\it et al.} \cite{hu2021competing}, we focus on how to engineer magnetic phase transitions by using two types of light.  First we consider light emanating from a waveguide  and afterwards the more standard case of circularly polarized light. 

 For light eminating from a waveguide the starting point of our discussion is a generic tight binding model. Here, the influence of light can be included through the hopping terms $t_{ij}$ according to the Peierls substitution
\begin{equation}
    t_{ij}\to e^{\int_{\vect r_i}^{\vect r_j}\vect A(t)\cdot d\vect l}t_{ij},
\end{equation}
where $\vect A$ is the vector potential. Note that if one shines \emph{normally incident longitudinally polarized light} onto a multilayered material only interlayer couplings  are affected by this substitution (for in-plane hoppings the scalar product $\vect A(t)\cdot d\vect l$ vanishes). Such an exotic form of light can be found at the exit of a waveguide \cite{michaelvoglInterlayer2020}, where it is allowed due to cavity boundary conditions. 

It was shown in Ref.\cite{michaelvoglInterlayer2020} that in the high frequency limit (the light frequency is larger than the bandwidth of the bands of interest but smaller than be bandgap) the effective time-independent interlayer couplings $t_{ij}$ given by
\begin{equation}
    t_{ij}\to J_0(a_L A)t_{ij},
    \label{eq:Peierls}
\end{equation}
where $J_0$ is the zeroth Bessel function  of the first kind that modulates the hopping amplitude. Here, $a_L$ is the interlayer distance and $A$ the strength of the incident light beam (constants like $\hbar$, etc are absorbed into $A$).

The time-averaged Peierls result, Eq.\eqref{eq:Peierls}, can be exploited to modify couplings in Eq.~\eqref{eq:non_int_ham}. First, one realizes that the non-interacting bands are relatively flat as noted in Refs. \cite{hu2021competing,PhysRevLett.121.026402,moralesduran2021nonlocal}. The bands for which the harmonic oscillator wavefunctions hold as an accurate approximation of Wannier wavefunctions are especially flat, which can also be seen quite clearly in the Fig. \ref{fig:bandstruct} below. The flat band around the Fermi level energetically separated from other bands means that  we can reasonably employ the high frequency approximation discussed above.

\begin{figure}[H]
	\centering
\includegraphics[width=0.7\linewidth]{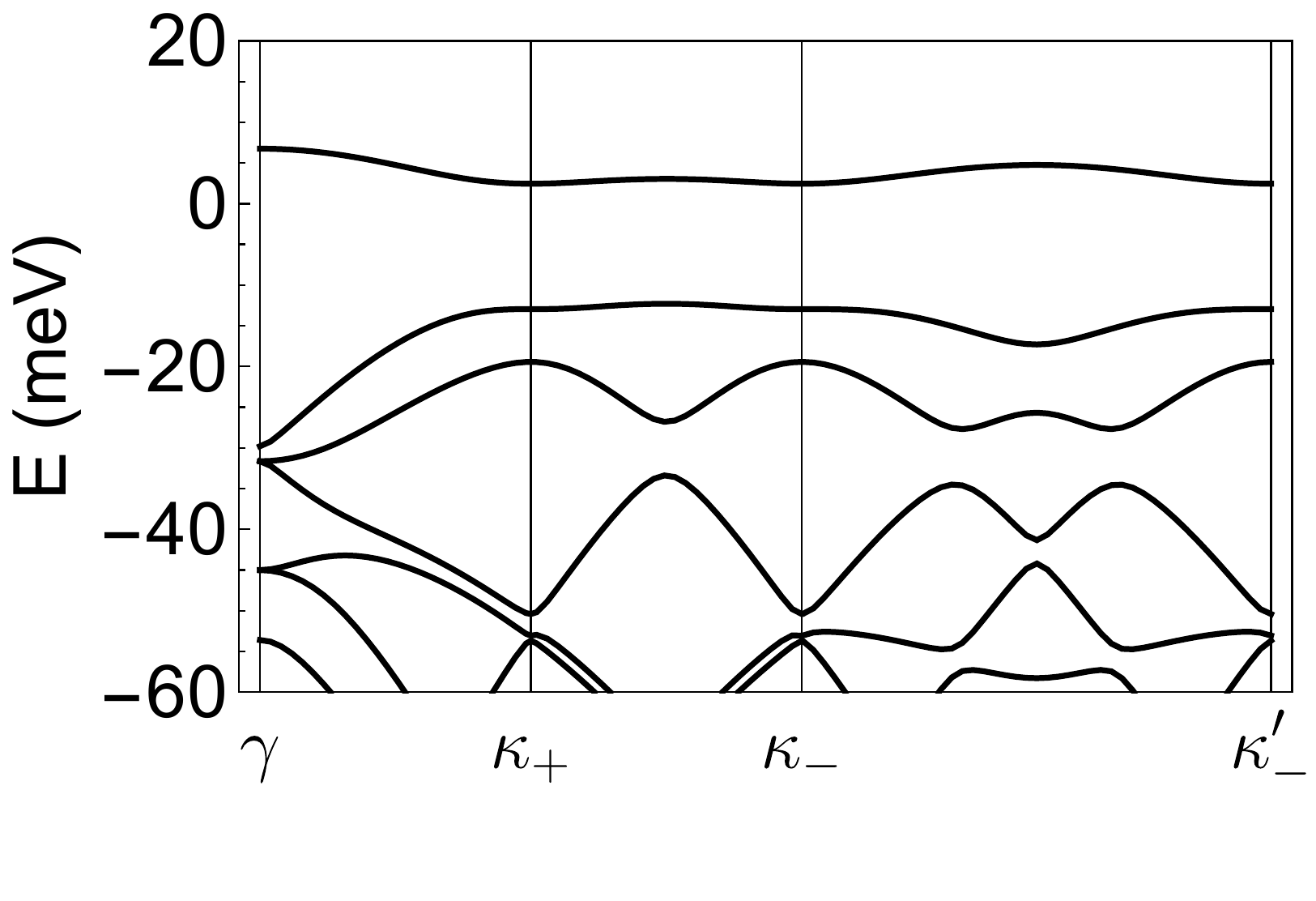} \includegraphics[width=0.7\linewidth]{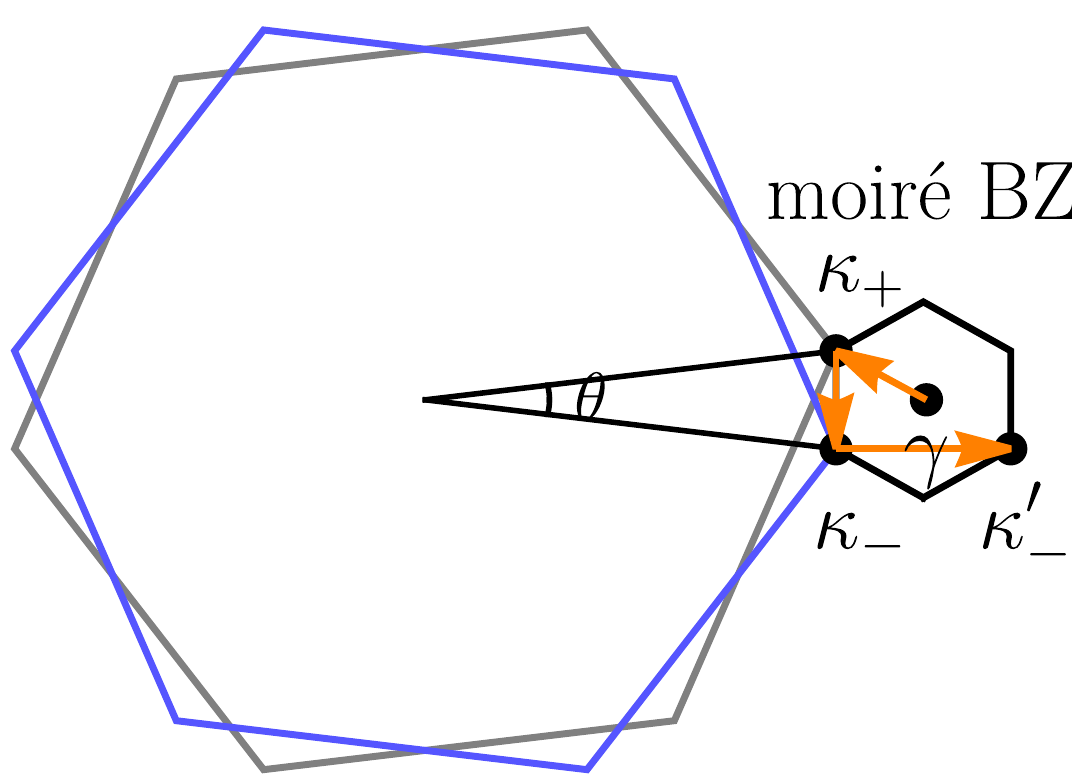}
	\caption{Top: Plot of the non-interacting band-structure for $V_m=6.6$meV, $m^*=0.35 m_e$, $a_M=12$nm and $\psi=-94^\circ$. Bottom: The moire Brillouin zone along with the path that was taken for the bandstructure plot marked in orange.}\label{fig:bandstruct}
\end{figure}

Instead of taking Eq.~\eqref{eq:non_int_ham} as a starting point for the description of a twisted TMD one could also take a tight binding model that includes both TMD layers as a starting point. This type of model could be downfolded into an effective tight binding model that only includes the active layer (AL)--the inactive layers (ILs) would be integrated out--employing a quasi-degenerate perturbation theory such as a Schrieffer-Wolff transformation or Loewdin downfolding. In this case, the dominant (2nd order in perturbation theory) contribution from interlayer tunnelings would be hopping processes like $AL\to IL\to AL$, assuming that the dominant interlayer hopping strengths are small compared to the dominant intralayer hoppings. Since hopping terms typically decrease exponentially with distance, the most dominant process is where the initial point on the active layer is the same as (right ``above") the final point on the active layer. 

Hopping processes with different initial and final points on the AL would contribute as an effective or indirect intralayer hopping, which we will neglect because direct intralayer hoppings can be expected to be much larger. This means that the dominant effect of interlayer hopping leads to on-site potential terms like $\epsilon_i\hat c_i^\dagger c_i$, where the onsite energies $\epsilon_i\propto t_{ij}^2$. A slightly more rigorous treatment of these ideas for the example of a 1D chain can be found in appendix \ref{app:Downfold_into_active_layer}. In the language of the continuum model, on-site energies can be translated as a  smoothed potential. In our case, this means that effective on-site energies that are due to second order interlayer hopping processes can be identified as the dominant origin of the moir\'e potential in Eq.~\eqref{eq:non_int_ham}. 

\begin{figure}[H]
	\centering
\includegraphics[width=\linewidth]{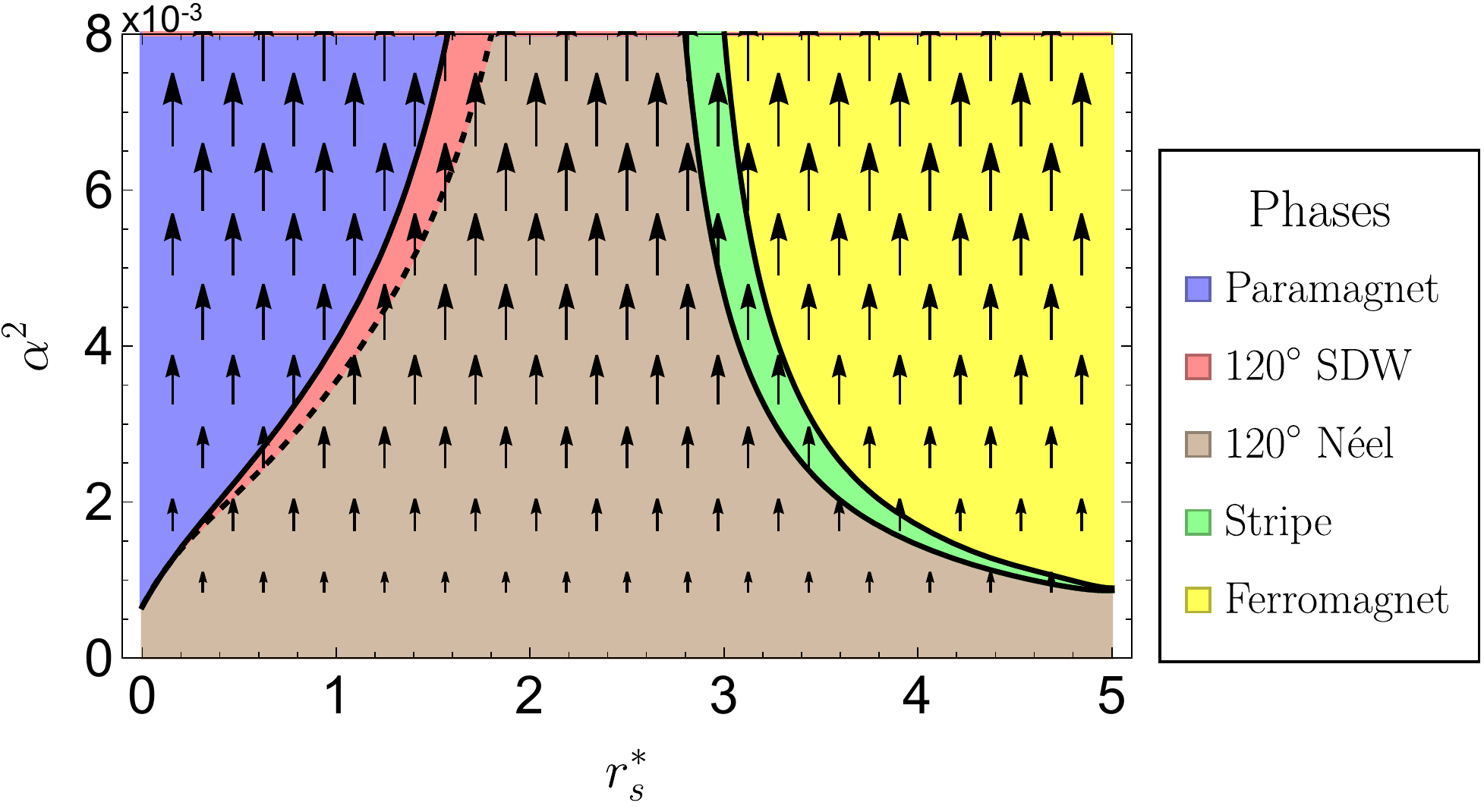} 
	\caption{Adapted and modified  redrawn version of the phase diagram from Ref.\cite{hu2021competing}.  Different magnetic phases in a twisted TMD are shown in different colors. Black arrows to indicate how waveguide light can allow one to tune vertically across phase boundaries in the phase diagram.  The parameter $\alpha^2$ appearing in Eq.\eqref{eq:rs_alpha} is increased as $V_m$ is decreased, according to Eq.\eqref{eq:replacement_moirestrength}. }\label{fig:phasediagram_move}
\end{figure}

In the high frequency limit when the material is subjected to the waveguide light
\begin{equation}
    V_m\to (J_0(a_L A))^2V_m,
    \label{eq:replacement_moirestrength}
\end{equation}
where $a_L$ is a parameter taken to be $0.85 a_{TT}$, the order of the distance between two transition metal layers $a_{TT}$. (See Appendix  \ref{app:neglected_effects}.)  Eq.~\eqref{eq:replacement_moirestrength} can now directly be used to move in the phase diagram shown in Fig.\ref{fig:phasediagram_move}. Specifically, one may introduce the following parameters \cite{hu2021competing} 
\begin{equation}
    \begin{aligned}
    &r_s^*=\frac{1}{\epsilon}\left(\frac{3}{4\pi^2}\right)^{1/4}\frac{a_M}{a_B}\frac{m^*}{m},\\
    &\alpha^2=\frac{\hbar^2}{m^*\beta V_m a_M^2},
    \end{aligned}
    \label{eq:rs_alpha}
\end{equation}
where $m$ is the bare electron mass, $m^*$ the effective mass, $a_B=\frac{\epsilon \hbar^2}{e^2m}$ 
the Bohr radius, $a_M$ the moir\'e lattice constant, $\beta=16\pi^2\cos(\psi)$ a material dependent numerical constant (material dependent through $\psi$ appearing in Eq.\eqref{eq:potential}--we assume the case $-\pi/3<\psi<\pi/3$)  and  $\epsilon$ the dielectric constant. The quantity $a_B r_s^*$ can be interpreted as the typical distance between electrons \cite{hu2021competing}. Approximating the wells of the moir\'e potential by a local harmonic potential, $\alpha$ can be related to a harmonic oscillator length scale.

A typical twisted TMD might have parameters $a_M\approx 10\,nm$, $m^*=0.3m$, $\epsilon=10$, $V_m\approx 30\,meV$, from which we get $r_s^*\approx2.97$ and $\alpha^2\approx 5.98\times 10^{-3}$, placing one near the boundary of a stripe and FM region (see. Fig.~\ref{fig:phasediagram_move}). Therefore, it may be possible to tune different phases in a twisted TMD hetero bilayer with relative ease. Indeed, in our case one is able to change $V_m$ (without additional effects in the high frequency regime) by applying waveguide light. More precisely, one can directly tune $\alpha^2$, Eq.~\eqref{eq:rs_alpha}, increasing it in size, since $V_m$ is reduced according to Eq.~\eqref{eq:replacement_moirestrength}. The effect of this can be seen below in Fig. \ref{fig:phasediagram_move}.

While the phase boundaries are the same as in Ref. \cite{hu2021competing}, our original contribution is the addition of arrows that show how the coupling parameters $\alpha^2$ and $r_s^*$ are renormalized under the drives we consider.
For instance, this allows us to show that tuning $V_m$ indirectly through the influence of light one can drive phase transitions, such as from a 120$^\circ$ Neel phase to a stripe ordered phase or to a ferromagnetic phase, for example. Note that shifting phases becomes easier as the harmonic oscillator length scale $\alpha$ is increased because for a small change to $V_m\to V_m-\delta V$ we have $\alpha^2\to(\alpha')^2\approx \alpha^2+\alpha^2\frac{\delta V}{V_m}$. We emphasize the crucial point that the drives we consider do not change the phase diagram, but allow one to experimentally tune between phases within it.  Such a situation is rare in driven systems. 

 Within our model, waveguide light only allows one to tune parameter the $\alpha$. However, one can gain greater control over the phase diagram of a twisted TMD by tuning the second parameter, $r^*_s$, which can be done with circularly polarized light modelled by a vector potential $\vect A=A(\cos(\omega t),\sin(\omega t))$. We note that $r^*_s$ can be tuned by such a choice of vector potential because it depends on the effective mass $m^*$ in the active TMD layer - a parameter that is primarily determined by values of in-plane hopping elements of a tight binding model. Circularly polarized light is a good choice if one aims to modify $m^*$ because it primarily influences the in-plane hopping elements. Typically, the most relevant out-of-plane hopping elements are almost directly perpendicular to the plane of the material and therefore they are influenced very little by circularly polarized light (recall the details of the Peierls substitution and that circularly polarized light only has in-plane vector potential components that are non-zero) - indeed hopping elements decay rapidly with distance. 

Before we discuss the effect of circularly light on the effective mass $m^*$ in Eq.\eqref{eq:non_int_ham}, we first note that the quadratic dispersion that appears in  Eq.\eqref{eq:non_int_ham} can be an approximation valid near different high-symmetry points in the hexagonal Brillouin zone. Indeed, depending on the precise choice of TMD material, low energy electrons might either be close to the $K$ points\cite{Wu2018} or the $\Gamma$ point\cite{Angelie2021826118}. In both cases, near those points the band structure can be approximated by a quadratic dispersion. 

Since $\Gamma$ and $K$ points are separated by a large distance in $k$-space, electrons at each of these points can be treated as independent excitations as noted in the appendix of \cite{PhysRevLett.122.086402}. In what follows we will therefore investigate the impact of circularly polarized light on the effective masses of the lowest energy band near the $\Gamma$ and $K$ points. In addition, we will highlight the differences at the $\Gamma$ and $K$ points. Here, it suffices to restrict our investigation to the impact on the active TMD layer.

First, we investigate the effect  of circularly polarized light on excitations near the $\Gamma$ point.  Our starting point is a third nearest neighbor three-band tight binding model \cite{Liu_2013} of a TMD that we discuss in more detail in appendix \ref{App:Downfold3Bandmodel}. From this model we find that circularly polarized light in the high frequency limit (frequencies larger than the bandwidth) modifies the effective mass $m^*$ as
\begin{equation}
    m^*\to (1+\Lambda a^2A^2)m^*,
    \label{eq:massrenorm}
\end{equation}
where we assumed a relatively weak field strength $A$, and $a$ is the in-plane lattice constant. A mass renormalization parameter $\Lambda$ that describes how strongly the effective mass $m^*$ is increased due to the electromagnetic field strength $A$ depends on the particular choice of active TMD layer. Values of $\Lambda$ for a variety of active TMD layer choices are given below in table \ref{tab:TMDmassmodParam}.
\begin{table}[H]
\centering
\begin{tabular}{|c|c|c|}\hline
     Active layer& $\Lambda$ (GGA)&  $\Lambda$ (LDA)\\\hline\hline
    MoS$_2$ & 0.998& 1.338\\\hline
     WS$_2$ & 1.021&1.183\\\hline
      MoSe$_2$ & 1.556&1.334\\\hline
      WSe$_2$ & 0.332&1.530\\\hline
      MoTe$_2$ & 8.219&1.070\\\hline
      WTe$_2$ & 1.366&1.203\\\hline
\end{tabular}
\caption{Mass renormalization parameter $\Lambda$ near the $\Gamma$ point computed based on effective third nearest neighbor tight binding with parameters that were taken from Ref.\cite{Liu_2013}. We also compare results that were obtained using different approximations to the exchange correlation potential of DFT: generalized gradient (GGA) and local density approximation (LDA). These different approximations were used in Ref.\cite{Liu_2013} to fit a third nearest neighbor tight binding model, which we used in our computations. Details of the computation to determine those parameters are found in appendix \ref{App:Downfold3Bandmodel}}
\label{tab:TMDmassmodParam}
\end{table}

We find that most computed values of the renormalization parameter $\Lambda$ are close to unity. However, $\Lambda$ seems to be more strongly dependent on a specific choice of tight binding model rather than the actual underlying material. Therefore, our results based on the tight binding parameters from Ref.\cite{Liu_2013} can be taken as an order of magnitude estimate. That is, we can only make the observation that circularly polarized light increases the effective mass near the $\Gamma$ point roughly as $m^*\to (a+a^2A^2)m^*$. Additional DFT based work might be necessary to find more reliable results for the mass renormalization parameter $\Lambda$. Such an analysis is beyond the scope of the present work.

\begin{figure}[htb]
	\centering
\includegraphics[width=\linewidth]{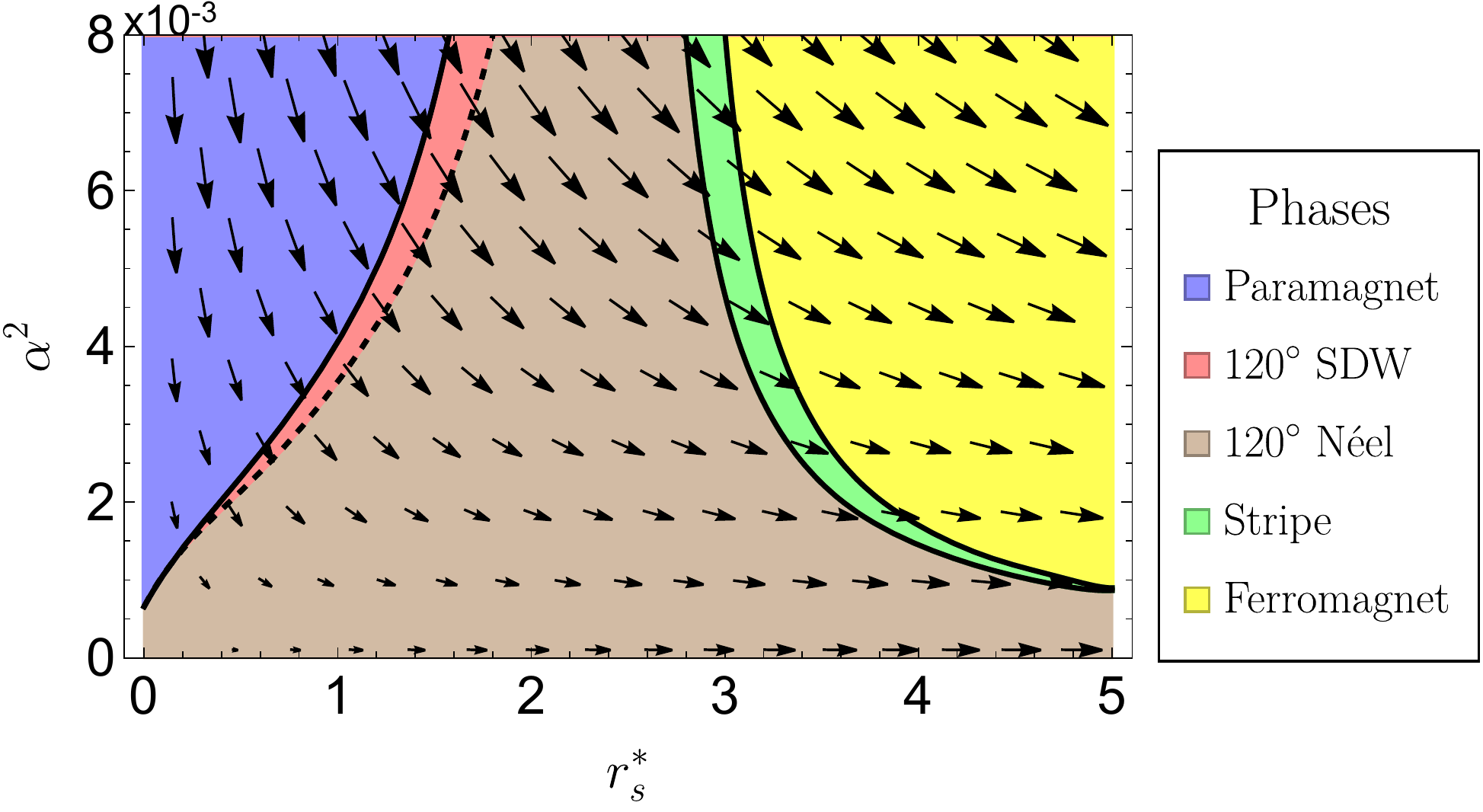} 
	\caption{Adapted and modified redrawn version of the phase diagram from Ref.\cite{hu2021competing}.  Different magnetic phases in a twisted TMD are shown in different colors. Black arrows indicate schematically how circularly polarized light can allow one to tune vertically across phase boundaries in the phase diagram.  Due to the modifications  that circularly polarized light enacts on the effective mass  $m^*$, the parameter $\alpha^2\to \alpha^2(1-a^2A^2)$ decreases depending on its position in the plot and the parameter $r^*_s\to r^*_s(1+a^2A^2)$ increases based on its position in the plot, which we indicated by arrows pointing along $(r_s^*a^2A^2,-\alpha^2 a^2A^2)$.  }\label{fig:phasediagram_move_Circ}
\end{figure}

Despite the spread of values for $\Lambda$ we find that circularly polarized light allows one to manipulate $r^*_s$, and therefore gain gain a high level of control over the phase diagram of Ref.\cite{hu2021competing}. Particularly, in Fig. \ref{fig:phasediagram_move_Circ} one can see how one may move in the phase diagram by the application of circularly polarized light. 

From the phase diagram in Fig.\ref{fig:phasediagram_move_Circ}
it becomes clear that circularly polarized light provides an additional knob that allows one to tune the magnetic phases of a twisted TMD heterobilayer independently from the tuning in the case of light emanating from a waveguide. We stress that while the phase diagram itself is unmodified from \cite{hu2021competing}, in our theoretical treatment we were able to capture renormalized couplings that allow one to tune an experimental system along the arrows we have added to the phases diagram.  The arrows, an original contribution of this work, can be interpreted as the couplings $\alpha^2$ and $r_s^*$ flowing with increasing strength of the circularly polarized light.

Next, we turn to the effect of circularly polarized light on excitations near the $K$ point. We investigate how it can also be used here to modify the effective mass $m^*$. Here, we worked with a simpler nearest neighbor three band model for a TMD that was found in Ref.\cite{Liu_2013} so that a more concise analytical expression could be obtained in Appendix \ref{App:Downfold3BandmodelK}. We were able to do so because the simpler model is a good fit to the band structure near the $K$ point - unlike the case of the $\Gamma$ point.  The model and details of our computation are discussed in the appendix \ref{App:Downfold3BandmodelK}. Relevant for the main text is the case of weak field strengths $A$ and momenta near the $K$ point. Similar to momenta near the $\Gamma$ point, we find that mass is renormalized as in Eq.\ref{eq:massrenorm}. The renormalization parameters $\Lambda$ for a variety of materials are given in Table \ref{tab:TMDmassmodParamK}.

\begin{table}[H]
\centering
\begin{tabular}{|c|c|c|}\hline
     Active layer& $\Lambda$ (GGA)&  $\Lambda$ (LDA)\\\hline\hline
    MoS$_2$ & -0.0550& -0.0507\\\hline
     WS$_2$ & -0.0065&+0.0002\\\hline
      MoSe$_2$ & -0.1674&-0.1701\\\hline
      WSe$_2$ & -0.0937&-0.0858\\\hline
      MoTe$_2$ & -0.4771&-0.5122\\\hline
      WTe$_2$ & -0.3710&-0.3597\\\hline
\end{tabular}
\caption{Mass renormalization parameter $\Lambda$ near the $K$ point computed based on nearest neighbor tight binding models that were taken from Ref.\cite{Liu_2013}. Different DFT procedures such as generalized gradient (GGA) and local density approximation (LDA) were used to fit a nearest neighbor tight binding model. Details of the computation to determine parameter $\Lambda$ are given in appendix \ref{App:Downfold3BandmodelK}. }
\label{tab:TMDmassmodParamK}
\end{table}
We find that unlike the case of the $\Gamma$ point, circularly polarized light reduces the effective mass near the $K$ point. The effect is generally not as pronounced as near the $\Gamma$ point, with a noticeably smaller $\Lambda<1$. The effect of circularly polarized light on effective masses seems to have a much stronger material dependence than in the case near the $K$ point. Lastly, results from different DFT approaches are in much better agreement with one another than in the case near the $\Gamma$ point. This might be attributed to less free parameters that appear in the nearest neighbor tight binding, which could guarantee a more consistent fit. 

Interestingly, the mass is renormalized differently near the $K$ and $\Gamma$ points. This observation could be used as an indirect probe to determine if certain material properties are dominated by the electrons near the $K$ or $\Gamma$ points. That is, if we are near a phase transition point circularly polarized light might cause a phase transition in a case dominated by the $\Gamma$ point but not the $K$ point or vice versa. This effect should be observable in magnetic response functions, most obviously the order parameter.

For completeness we note that the phase diagram in Fig.~\ref{fig:phasediagram_move} (derived in Ref.\cite{hu2021competing}) made use of self-consistent Hartree-Fock theory for a further simplified lattice model. The full interacting Hamiltonian--including the Coulomb interaction $U(\vect r)$--was projected onto the highest band in Fig. \ref{fig:bandstruct}. Therefore, the interactions were modelled by a generalized Hubbard term
\begin{equation}
    H_{int}=\frac{1}{2}\sum_{\tau,\tau^\prime,\vect R,\vect R^\prime}\tilde U(\vect R-\vect R^\prime)c_{\vect R,\tau}^\dag c_{\vect R^\prime,\tau^\prime}^\dag c_{\vect R^\prime,\tau^\prime} c_{\vect R,\tau} ,
\end{equation}
where $\tilde U(\vect R-\vect R^\prime)$ corresponds to the Coulomb interaction evaluated using Wannier functions corresponding to moire lattice sites at positions $\vect R$ and $\vect R^\prime$ \cite{PhysRevLett.121.026402}.

\section{Comments on Floquet state preparation}
\label{sec:preparation}
In the previous sections, we focused on how a periodic drive can modify exchange interactions in a TMD heterobilayer effective Hamiltonian. Our analysis was limited to studying the ground state magnetic properties of an effective time-independent Hamiltonian.  However, the question about when the dynamics of the system can be fully described by this Floquet Hamiltonian is a very challenging one in itself and cannot be fully answered by direct studies of the properties of this Hamiltonian. Furthermore, it is an open problem, in general, how specific states of a Floquet Hamiltonian can be prepared experimentally. The answers to both these questions also depend on many other additional factors which include the methods used to turn on a periodic drive and also the details of a coupling of the system to its environment. This means that there are various time scales that enter the problem. We briefly discuss some of these time-scales relevant to our proposal.

The first case we consider is one where an experimental setup allows preparation of an eigenstate of the Floquet Hamiltonian. It is well known that isolated Floquet systems due to interaction effects tend to heat to infinite temperature and thus it is challenging to keep an eigenstate of a Floquet Hamiltonian stable once it is attained. That is, the time evolution is only for a limited time governed by effective Floquet Hamiltonians - the so-called prethermal regime. After this regime a Floquet state will not be an eigenstate and therefore it will not be stable under time evolution.  Nonetheless, generically, even interacting systems can sustain an exponentially long prethermal regime \cite{PhysRevB.95.014112}  before heating dynamics take over. Sufficiently weak drives in the high frequency regime permit some rough estimates of prethermal regime times $t_{\mathrm{pre}}\propto \text{exp}(\omega/J)$, which can be made on the basis of interaction strength $J$ and drive frequency $\omega$~\cite{PhysRevB.95.014112}. 
However, not only is the length of the pre-thermal regime of experimental relevance, but also the length of time before the onset of the pre-thermal regime. The time required for the onset of the pre-thermal regime can be roughly estimated from the strength $J$ of interactions, and is roughly of order $1/J$~\cite{Bingtian2020} in this case.

The second case we consider is the problem of Floquet state preparation. While there does not seem to be a general route to prepare a Floquet state, an adiabatic switch-on protocol can be used to prepare Floquet states \cite{Weinberg_adiabatic2017} if the initial state is adiabatically connected to the target state. An alternative route to the preparation of ground states of a Floquet Hamiltonian is via coupling to a heat bath. Particularly, it has been shown~\cite{Shirai_2016}  that a periodically driven system under certain assumptions about the system-bath couplings will evolve into a so-called Floquet-Gibbs state. That is, the density matrix for relatively long times will evolve to be the density matrix for a Gibbs state $e^{-\beta H_F}$, where $H_F$ is the Floquet Hamiltonian~\cite{Shirai_2016}. This kind of state can be achieved under the condition that the heating rate of the driven system is smaller than the relaxation rate associated with the system-bath coupling.
From this perspective the problem of Floquet state preparation can be shifted to a question about the engineering of appropriate system-bath couplings. However, there are also time scales involved with how long it takes to reach a Floquet-Gibbs state. To our knowledge there do not yet exist simple estimates for this time scale.

 We will therefore focus on the time scales that we can estimate with current techniques. The prethermal regime can be achieved within a few $100\,fs$ for $J\sim10\,meV$~\cite{Bingtian2020}. High-power ultrashort lasers with pulse widths of a few 100 fs are capable of generating electric field amplitudes in the range $0.1-1\,V/nm$~\cite{shan2021giant}. This results in sufficiently large driving strengths to induce a magnetic phase transition if the system is close to a phase boundary. We conclude this section by noting that while it is challenging to apply our approach to tune magnetic phases in a twisted TMD, it is within current experimental reach.

 


\section{Conclusion}
\label{sec:conclusions}

In this work, we discussed how a periodic drive can be used to control and alter the magnetic properties of TMD heterobilayers. Many previous works~\cite{mentink2014,mentink2015ultrafast,mentink2017manipulating,hejazi2018,hejazi2019,Quito2021,Quito2021b,Chaudhary2019orbital,chaudhary2020controlling} have studied light-induced modification in exchange interactions. Most of these changes originate from the direct modulation of in-plane hopping arising from the transverse $\vect E$ field of a laser. However, in our case, we harness the additional layer degree of freedom possessed by moir\'e materials and use waveguide light to get an $\vect E$ field perpendicular to the TMD layers. In this case, we take an indirect route to modify exchange interactions. The longitudinal field of the laser renormalizes interlayer hoppings which in turn leads to a modification of the moir\'e potential. This change affects interaction parameters and in turn the magnetic properties.  Our scheme differs from previous proposals in that it is based on modifications to interaction parameters that arise from interlayer hopping renormalizations. 

 We have also studied the effect of circularly polarized light and found that it leads to a renormalization of the effective mass. This renormalization makes it possible to tune a separate parameter that determines the magnetic phase diagram. Circularly polarized light therefore serves as an additional tuning knob that allows one a high level of control over the phase diagram of a twisted TMD.
 
 We note that an interesting future direction would be to study the effect of linearly polarized transverse light on the phase diagram, which due to the breaking of rotational symmetry could lead to a wealth of new different magnetic phases.

With the rapidly expanding zoo of moir\'e materials and their magnetic properties only scarcely understood at this point, we believe these methods would find many applications in this new class of materials. Furthermore, we believe that longitudinal light can find novel applications beyond Floquet engineering. For example, it would be interesting to explore the interplay of longitudinal waveguide light  and moir\'e materials  in the context of other light-matter phenomena such as optical responses, by generalizing the work of Refs.~\cite{chaudhary2021shiftcurrent,kaplan2021,liu2020anomalous}.

\acknowledgements
M.V. gratefully acknowledge the support provided by the Deanship of Research Oversight and Coordination (DROC) at King Fahd University of Petroleum \& Minerals (KFUPM) for funding this work through start up project No.SR211001 and subsequent continued funding through exploratory research grant No. ER221002. This research was primarily supported by the National Science Foundation through the Center for Dynamics and Control of Materials: an NSF MRSEC under Cooperative Agreement No. DMR-1720595, with additional support from Grant No. DMR-2114825. G.A.F also acknowledges support from the Alexander von Humboldt Foundation.  

\bibliography{literature}

\begin{thebibliography}{161}%
\makeatletter
\providecommand \@ifxundefined [1]{%
 \@ifx{#1\undefined}
}%
\providecommand \@ifnum [1]{%
 \ifnum #1\expandafter \@firstoftwo
 \else \expandafter \@secondoftwo
 \fi
}%
\providecommand \@ifx [1]{%
 \ifx #1\expandafter \@firstoftwo
 \else \expandafter \@secondoftwo
 \fi
}%
\providecommand \natexlab [1]{#1}%
\providecommand \enquote  [1]{``#1''}%
\providecommand \bibnamefont  [1]{#1}%
\providecommand \bibfnamefont [1]{#1}%
\providecommand \citenamefont [1]{#1}%
\providecommand \href@noop [0]{\@secondoftwo}%
\providecommand \href [0]{\begingroup \@sanitize@url \@href}%
\providecommand \@href[1]{\@@startlink{#1}\@@href}%
\providecommand \@@href[1]{\endgroup#1\@@endlink}%
\providecommand \@sanitize@url [0]{\catcode `\\12\catcode `\$12\catcode
  `\&12\catcode `\#12\catcode `\^12\catcode `\_12\catcode `\%12\relax}%
\providecommand \@@startlink[1]{}%
\providecommand \@@endlink[0]{}%
\providecommand \url  [0]{\begingroup\@sanitize@url \@url }%
\providecommand \@url [1]{\endgroup\@href {#1}{\urlprefix }}%
\providecommand \urlprefix  [0]{URL }%
\providecommand \Eprint [0]{\href }%
\providecommand \doibase [0]{https://doi.org/}%
\providecommand \selectlanguage [0]{\@gobble}%
\providecommand \bibinfo  [0]{\@secondoftwo}%
\providecommand \bibfield  [0]{\@secondoftwo}%
\providecommand \translation [1]{[#1]}%
\providecommand \BibitemOpen [0]{}%
\providecommand \bibitemStop [0]{}%
\providecommand \bibitemNoStop [0]{.\EOS\space}%
\providecommand \EOS [0]{\spacefactor3000\relax}%
\providecommand \BibitemShut  [1]{\csname bibitem#1\endcsname}%
\let\auto@bib@innerbib\@empty
\bibitem [{\citenamefont {Andrei}\ \emph {et~al.}(2021)\citenamefont {Andrei},
  \citenamefont {Efetov}, \citenamefont {Jarillo-Herrero}, \citenamefont
  {MacDonald}, \citenamefont {Mak}, \citenamefont {Senthil}, \citenamefont
  {Tutuc}, \citenamefont {Yazdani},\ and\ \citenamefont
  {Young}}]{andrei2021marvels}%
  \BibitemOpen
  \bibfield  {author} {\bibinfo {author} {\bibfnamefont {E.~Y.}\ \bibnamefont
  {Andrei}}, \bibinfo {author} {\bibfnamefont {D.~K.}\ \bibnamefont {Efetov}},
  \bibinfo {author} {\bibfnamefont {P.}~\bibnamefont {Jarillo-Herrero}},
  \bibinfo {author} {\bibfnamefont {A.~H.}\ \bibnamefont {MacDonald}}, \bibinfo
  {author} {\bibfnamefont {K.~F.}\ \bibnamefont {Mak}}, \bibinfo {author}
  {\bibfnamefont {T.}~\bibnamefont {Senthil}}, \bibinfo {author} {\bibfnamefont
  {E.}~\bibnamefont {Tutuc}}, \bibinfo {author} {\bibfnamefont
  {A.}~\bibnamefont {Yazdani}},\ and\ \bibinfo {author} {\bibfnamefont {A.~F.}\
  \bibnamefont {Young}},\ }\bibfield  {title} {\bibinfo {title} {The marvels of
  moir{\'e} materials},\ }\href
  {https://www.nature.com/articles/s41578-021-00284-1} {\bibfield  {journal}
  {\bibinfo  {journal} {Nature Reviews Materials}\ }\textbf {\bibinfo {volume}
  {6}},\ \bibinfo {pages} {201} (\bibinfo {year} {2021})}\BibitemShut {NoStop}%
\bibitem [{\citenamefont {He}\ \emph {et~al.}(2021)\citenamefont {He},
  \citenamefont {Zhou}, \citenamefont {Ye}, \citenamefont {Cho}, \citenamefont
  {Jeong}, \citenamefont {Meng},\ and\ \citenamefont {Wang}}]{he2021moire}%
  \BibitemOpen
  \bibfield  {author} {\bibinfo {author} {\bibfnamefont {F.}~\bibnamefont
  {He}}, \bibinfo {author} {\bibfnamefont {Y.}~\bibnamefont {Zhou}}, \bibinfo
  {author} {\bibfnamefont {Z.}~\bibnamefont {Ye}}, \bibinfo {author}
  {\bibfnamefont {S.-H.}\ \bibnamefont {Cho}}, \bibinfo {author} {\bibfnamefont
  {J.}~\bibnamefont {Jeong}}, \bibinfo {author} {\bibfnamefont
  {X.}~\bibnamefont {Meng}},\ and\ \bibinfo {author} {\bibfnamefont
  {Y.}~\bibnamefont {Wang}},\ }\bibfield  {title} {\bibinfo {title} {Moir{\'e}
  patterns in 2d materials: A review},\ }\href@noop {} {\bibfield  {journal}
  {\bibinfo  {journal} {ACS nano}\ }\textbf {\bibinfo {volume} {15}},\ \bibinfo
  {pages} {5944} (\bibinfo {year} {2021})}\BibitemShut {NoStop}%
\bibitem [{\citenamefont {MacDonald}(2019)}]{macdonald2019bilayer}%
  \BibitemOpen
  \bibfield  {author} {\bibinfo {author} {\bibfnamefont {A.~H.}\ \bibnamefont
  {MacDonald}},\ }\bibfield  {title} {\bibinfo {title} {Bilayer graphene’s
  wicked, twisted road},\ }\href@noop {} {\bibfield  {journal} {\bibinfo
  {journal} {Physics}\ }\textbf {\bibinfo {volume} {12}},\ \bibinfo {pages}
  {12} (\bibinfo {year} {2019})}\BibitemShut {NoStop}%
\bibitem [{\citenamefont {Zeller}\ and\ \citenamefont
  {G{\"u}nther}(2014)}]{zeller2014possible}%
  \BibitemOpen
  \bibfield  {author} {\bibinfo {author} {\bibfnamefont {P.}~\bibnamefont
  {Zeller}}\ and\ \bibinfo {author} {\bibfnamefont {S.}~\bibnamefont
  {G{\"u}nther}},\ }\bibfield  {title} {\bibinfo {title} {What are the possible
  moir{\'e} patterns of graphene on hexagonally packed surfaces? universal
  solution for hexagonal coincidence lattices, derived by a geometric
  construction},\ }\href@noop {} {\bibfield  {journal} {\bibinfo  {journal}
  {New Journal of Physics}\ }\textbf {\bibinfo {volume} {16}},\ \bibinfo
  {pages} {083028} (\bibinfo {year} {2014})}\BibitemShut {NoStop}%
\bibitem [{\citenamefont {Kennes}\ \emph {et~al.}(2021)\citenamefont {Kennes},
  \citenamefont {Claassen}, \citenamefont {Xian}, \citenamefont {Georges},
  \citenamefont {Millis}, \citenamefont {Hone}, \citenamefont {Dean},
  \citenamefont {Basov}, \citenamefont {Pasupathy},\ and\ \citenamefont
  {Rubio}}]{kennes2021moire}%
  \BibitemOpen
  \bibfield  {author} {\bibinfo {author} {\bibfnamefont {D.~M.}\ \bibnamefont
  {Kennes}}, \bibinfo {author} {\bibfnamefont {M.}~\bibnamefont {Claassen}},
  \bibinfo {author} {\bibfnamefont {L.}~\bibnamefont {Xian}}, \bibinfo {author}
  {\bibfnamefont {A.}~\bibnamefont {Georges}}, \bibinfo {author} {\bibfnamefont
  {A.~J.}\ \bibnamefont {Millis}}, \bibinfo {author} {\bibfnamefont
  {J.}~\bibnamefont {Hone}}, \bibinfo {author} {\bibfnamefont {C.~R.}\
  \bibnamefont {Dean}}, \bibinfo {author} {\bibfnamefont {D.}~\bibnamefont
  {Basov}}, \bibinfo {author} {\bibfnamefont {A.~N.}\ \bibnamefont
  {Pasupathy}},\ and\ \bibinfo {author} {\bibfnamefont {A.}~\bibnamefont
  {Rubio}},\ }\bibfield  {title} {\bibinfo {title} {Moir{\'e} heterostructures
  as a condensed-matter quantum simulator},\ }\href@noop {} {\bibfield
  {journal} {\bibinfo  {journal} {Nature Physics}\ }\textbf {\bibinfo {volume}
  {17}},\ \bibinfo {pages} {155} (\bibinfo {year} {2021})}\BibitemShut
  {NoStop}%
\bibitem [{\citenamefont {Bistritzer}\ and\ \citenamefont
  {MacDonald}(2011)}]{bistritzer2011moire}%
  \BibitemOpen
  \bibfield  {author} {\bibinfo {author} {\bibfnamefont {R.}~\bibnamefont
  {Bistritzer}}\ and\ \bibinfo {author} {\bibfnamefont {A.~H.}\ \bibnamefont
  {MacDonald}},\ }\bibfield  {title} {\bibinfo {title} {Moir{\'e} bands in
  twisted double-layer graphene},\ }\href@noop {} {\bibfield  {journal}
  {\bibinfo  {journal} {Proceedings of the National Academy of Sciences}\
  }\textbf {\bibinfo {volume} {108}},\ \bibinfo {pages} {12233} (\bibinfo
  {year} {2011})}\BibitemShut {NoStop}%
\bibitem [{\citenamefont {Morell}\ \emph {et~al.}(2010)\citenamefont {Morell},
  \citenamefont {Correa}, \citenamefont {Vargas}, \citenamefont {Pacheco},\
  and\ \citenamefont {Barticevic}}]{morell2010flat}%
  \BibitemOpen
  \bibfield  {author} {\bibinfo {author} {\bibfnamefont {E.~S.}\ \bibnamefont
  {Morell}}, \bibinfo {author} {\bibfnamefont {J.}~\bibnamefont {Correa}},
  \bibinfo {author} {\bibfnamefont {P.}~\bibnamefont {Vargas}}, \bibinfo
  {author} {\bibfnamefont {M.}~\bibnamefont {Pacheco}},\ and\ \bibinfo {author}
  {\bibfnamefont {Z.}~\bibnamefont {Barticevic}},\ }\bibfield  {title}
  {\bibinfo {title} {Flat bands in slightly twisted bilayer graphene:
  Tight-binding calculations},\ }\href@noop {} {\bibfield  {journal} {\bibinfo
  {journal} {Physical Review B}\ }\textbf {\bibinfo {volume} {82}},\ \bibinfo
  {pages} {121407} (\bibinfo {year} {2010})}\BibitemShut {NoStop}%
\bibitem [{\citenamefont {Lopes~dos Santos}\ \emph {et~al.}(2012)\citenamefont
  {Lopes~dos Santos}, \citenamefont {Peres},\ and\ \citenamefont
  {Castro~Neto}}]{dosSantos2012}%
  \BibitemOpen
  \bibfield  {author} {\bibinfo {author} {\bibfnamefont {J.~M.~B.}\
  \bibnamefont {Lopes~dos Santos}}, \bibinfo {author} {\bibfnamefont
  {N.~M.~R.}\ \bibnamefont {Peres}},\ and\ \bibinfo {author} {\bibfnamefont
  {A.~H.}\ \bibnamefont {Castro~Neto}},\ }\bibfield  {title} {\bibinfo {title}
  {Continuum model of the twisted graphene bilayer},\ }\href
  {https://doi.org/10.1103/PhysRevB.86.155449} {\bibfield  {journal} {\bibinfo
  {journal} {Phys. Rev. B}\ }\textbf {\bibinfo {volume} {86}},\ \bibinfo
  {pages} {155449} (\bibinfo {year} {2012})}\BibitemShut {NoStop}%
\bibitem [{\citenamefont {Hu}\ \emph {et~al.}(2011)\citenamefont {Hu},
  \citenamefont {Kargarian},\ and\ \citenamefont {Fiete}}]{Hu:prb11}%
  \BibitemOpen
  \bibfield  {author} {\bibinfo {author} {\bibfnamefont {X.}~\bibnamefont
  {Hu}}, \bibinfo {author} {\bibfnamefont {M.}~\bibnamefont {Kargarian}},\ and\
  \bibinfo {author} {\bibfnamefont {G.~A.}\ \bibnamefont {Fiete}},\ }\bibfield
  {title} {\bibinfo {title} {Topological insulators and fractional quantum hall
  effect on the ruby lattice},\ }\href
  {https://doi.org/10.1103/PhysRevB.84.155116} {\bibfield  {journal} {\bibinfo
  {journal} {Phys. Rev. B}\ }\textbf {\bibinfo {volume} {84}},\ \bibinfo
  {pages} {155116} (\bibinfo {year} {2011})}\BibitemShut {NoStop}%
\bibitem [{\citenamefont {Parameswaran}\ \emph {et~al.}(2013)\citenamefont
  {Parameswaran}, \citenamefont {Roy},\ and\ \citenamefont
  {Sondhi}}]{Parameswaran2013816}%
  \BibitemOpen
  \bibfield  {author} {\bibinfo {author} {\bibfnamefont {S.~A.}\ \bibnamefont
  {Parameswaran}}, \bibinfo {author} {\bibfnamefont {R.}~\bibnamefont {Roy}},\
  and\ \bibinfo {author} {\bibfnamefont {S.~L.}\ \bibnamefont {Sondhi}},\
  }\bibfield  {title} {\bibinfo {title} {Fractional quantum hall physics in
  topological flat bands},\ }\href
  {https://doi.org/http://dx.doi.org/10.1016/j.crhy.2013.04.003} {\bibfield
  {journal} {\bibinfo  {journal} {Comptes Rendus Physique}\ }\textbf {\bibinfo
  {volume} {14}},\ \bibinfo {pages} {816 } (\bibinfo {year}
  {2013})}\BibitemShut {NoStop}%
\bibitem [{\citenamefont {Wu}\ \emph {et~al.}(2012)\citenamefont {Wu},
  \citenamefont {Bernevig},\ and\ \citenamefont {Regnault}}]{Wu_CI:prb12}%
  \BibitemOpen
  \bibfield  {author} {\bibinfo {author} {\bibfnamefont {Y.-L.}\ \bibnamefont
  {Wu}}, \bibinfo {author} {\bibfnamefont {B.~A.}\ \bibnamefont {Bernevig}},\
  and\ \bibinfo {author} {\bibfnamefont {N.}~\bibnamefont {Regnault}},\
  }\bibfield  {title} {\bibinfo {title} {Zoology of fractional chern
  insulators},\ }\href {https://doi.org/10.1103/PhysRevB.85.075116} {\bibfield
  {journal} {\bibinfo  {journal} {Phys. Rev. B}\ }\textbf {\bibinfo {volume}
  {85}},\ \bibinfo {pages} {075116} (\bibinfo {year} {2012})}\BibitemShut
  {NoStop}%
\bibitem [{\citenamefont {Wang}\ and\ \citenamefont {Ran}(2011)}]{Wang:prb11}%
  \BibitemOpen
  \bibfield  {author} {\bibinfo {author} {\bibfnamefont {F.}~\bibnamefont
  {Wang}}\ and\ \bibinfo {author} {\bibfnamefont {Y.}~\bibnamefont {Ran}},\
  }\bibfield  {title} {\bibinfo {title} {Nearly flat band with chern number
  $c=2$ on the dice lattice},\ }\href
  {https://doi.org/10.1103/PhysRevB.84.241103} {\bibfield  {journal} {\bibinfo
  {journal} {Phys. Rev. B}\ }\textbf {\bibinfo {volume} {84}},\ \bibinfo
  {pages} {241103} (\bibinfo {year} {2011})}\BibitemShut {NoStop}%
\bibitem [{\citenamefont {Sun}\ \emph {et~al.}(2011)\citenamefont {Sun},
  \citenamefont {Gu}, \citenamefont {Katsura},\ and\ \citenamefont
  {Das~Sarma}}]{Sun:prl11}%
  \BibitemOpen
  \bibfield  {author} {\bibinfo {author} {\bibfnamefont {K.}~\bibnamefont
  {Sun}}, \bibinfo {author} {\bibfnamefont {Z.}~\bibnamefont {Gu}}, \bibinfo
  {author} {\bibfnamefont {H.}~\bibnamefont {Katsura}},\ and\ \bibinfo {author}
  {\bibfnamefont {S.}~\bibnamefont {Das~Sarma}},\ }\bibfield  {title} {\bibinfo
  {title} {Nearly flatbands with nontrivial topology},\ }\href
  {https://doi.org/10.1103/PhysRevLett.106.236803} {\bibfield  {journal}
  {\bibinfo  {journal} {Phys. Rev. Lett.}\ }\textbf {\bibinfo {volume} {106}},\
  \bibinfo {pages} {236803} (\bibinfo {year} {2011})}\BibitemShut {NoStop}%
\bibitem [{\citenamefont {Neupert}\ \emph {et~al.}(2011)\citenamefont
  {Neupert}, \citenamefont {Santos}, \citenamefont {Chamon},\ and\
  \citenamefont {Mudry}}]{Neupert:prl11}%
  \BibitemOpen
  \bibfield  {author} {\bibinfo {author} {\bibfnamefont {T.}~\bibnamefont
  {Neupert}}, \bibinfo {author} {\bibfnamefont {L.}~\bibnamefont {Santos}},
  \bibinfo {author} {\bibfnamefont {C.}~\bibnamefont {Chamon}},\ and\ \bibinfo
  {author} {\bibfnamefont {C.}~\bibnamefont {Mudry}},\ }\bibfield  {title}
  {\bibinfo {title} {Fractional quantum hall states at zero magnetic field},\
  }\href {https://doi.org/10.1103/PhysRevLett.106.236804} {\bibfield  {journal}
  {\bibinfo  {journal} {Phys. Rev. Lett.}\ }\textbf {\bibinfo {volume} {106}},\
  \bibinfo {pages} {236804} (\bibinfo {year} {2011})}\BibitemShut {NoStop}%
\bibitem [{\citenamefont {Tang}\ \emph {et~al.}(2011)\citenamefont {Tang},
  \citenamefont {Mei},\ and\ \citenamefont {Wen}}]{Tang:prl11}%
  \BibitemOpen
  \bibfield  {author} {\bibinfo {author} {\bibfnamefont {E.}~\bibnamefont
  {Tang}}, \bibinfo {author} {\bibfnamefont {J.-W.}\ \bibnamefont {Mei}},\ and\
  \bibinfo {author} {\bibfnamefont {X.-G.}\ \bibnamefont {Wen}},\ }\bibfield
  {title} {\bibinfo {title} {High-temperature fractional quantum hall states},\
  }\href {https://doi.org/10.1103/PhysRevLett.106.236802} {\bibfield  {journal}
  {\bibinfo  {journal} {Phys. Rev. Lett.}\ }\textbf {\bibinfo {volume} {106}},\
  \bibinfo {pages} {236802} (\bibinfo {year} {2011})}\BibitemShut {NoStop}%
\bibitem [{\citenamefont {Regnault}\ and\ \citenamefont
  {Bernevig}(2011)}]{Regnault:prx11}%
  \BibitemOpen
  \bibfield  {author} {\bibinfo {author} {\bibfnamefont {N.}~\bibnamefont
  {Regnault}}\ and\ \bibinfo {author} {\bibfnamefont {B.~A.}\ \bibnamefont
  {Bernevig}},\ }\bibfield  {title} {\bibinfo {title} {Fractional chern
  insulator},\ }\href {https://doi.org/10.1103/PhysRevX.1.021014} {\bibfield
  {journal} {\bibinfo  {journal} {Phys. Rev. X}\ }\textbf {\bibinfo {volume}
  {1}},\ \bibinfo {pages} {021014} (\bibinfo {year} {2011})}\BibitemShut
  {NoStop}%
\bibitem [{\citenamefont {Kourtis}\ \emph {et~al.}(2014)\citenamefont
  {Kourtis}, \citenamefont {Neupert}, \citenamefont {Chamon},\ and\
  \citenamefont {Mudry}}]{Kourtis:prl14}%
  \BibitemOpen
  \bibfield  {author} {\bibinfo {author} {\bibfnamefont {S.}~\bibnamefont
  {Kourtis}}, \bibinfo {author} {\bibfnamefont {T.}~\bibnamefont {Neupert}},
  \bibinfo {author} {\bibfnamefont {C.}~\bibnamefont {Chamon}},\ and\ \bibinfo
  {author} {\bibfnamefont {C.}~\bibnamefont {Mudry}},\ }\bibfield  {title}
  {\bibinfo {title} {Fractional chern insulators with strong interactions that
  far exceed band gaps},\ }\href
  {https://doi.org/10.1103/PhysRevLett.112.126806} {\bibfield  {journal}
  {\bibinfo  {journal} {Phys. Rev. Lett.}\ }\textbf {\bibinfo {volume} {112}},\
  \bibinfo {pages} {126806} (\bibinfo {year} {2014})}\BibitemShut {NoStop}%
\bibitem [{\citenamefont {Cao}\ \emph {et~al.}(2018)\citenamefont {Cao},
  \citenamefont {Fatemi}, \citenamefont {Fang}, \citenamefont {Watanabe},
  \citenamefont {Taniguchi}, \citenamefont {Kaxiras},\ and\ \citenamefont
  {Jarillo-Herrero}}]{cao2018unconventional}%
  \BibitemOpen
  \bibfield  {author} {\bibinfo {author} {\bibfnamefont {Y.}~\bibnamefont
  {Cao}}, \bibinfo {author} {\bibfnamefont {V.}~\bibnamefont {Fatemi}},
  \bibinfo {author} {\bibfnamefont {S.}~\bibnamefont {Fang}}, \bibinfo {author}
  {\bibfnamefont {K.}~\bibnamefont {Watanabe}}, \bibinfo {author}
  {\bibfnamefont {T.}~\bibnamefont {Taniguchi}}, \bibinfo {author}
  {\bibfnamefont {E.}~\bibnamefont {Kaxiras}},\ and\ \bibinfo {author}
  {\bibfnamefont {P.}~\bibnamefont {Jarillo-Herrero}},\ }\bibfield  {title}
  {\bibinfo {title} {Unconventional superconductivity in magic-angle graphene
  superlattices},\ }\href@noop {} {\bibfield  {journal} {\bibinfo  {journal}
  {Nature}\ }\textbf {\bibinfo {volume} {556}},\ \bibinfo {pages} {43}
  (\bibinfo {year} {2018})}\BibitemShut {NoStop}%
\bibitem [{\citenamefont {Vahedi}\ \emph {et~al.}(2021)\citenamefont {Vahedi},
  \citenamefont {Peters}, \citenamefont {Missaoui}, \citenamefont {Honecker},\
  and\ \citenamefont {de~Laissardière}}]{10.21468/SciPostPhys.11.4.083}%
  \BibitemOpen
  \bibfield  {author} {\bibinfo {author} {\bibfnamefont {J.}~\bibnamefont
  {Vahedi}}, \bibinfo {author} {\bibfnamefont {R.}~\bibnamefont {Peters}},
  \bibinfo {author} {\bibfnamefont {A.}~\bibnamefont {Missaoui}}, \bibinfo
  {author} {\bibfnamefont {A.}~\bibnamefont {Honecker}},\ and\ \bibinfo
  {author} {\bibfnamefont {G.~T.}\ \bibnamefont {de~Laissardière}},\
  }\bibfield  {title} {\bibinfo {title} {{Magnetism of magic-angle twisted
  bilayer graphene}},\ }\href {https://doi.org/10.21468/SciPostPhys.11.4.083}
  {\bibfield  {journal} {\bibinfo  {journal} {SciPost Phys.}\ }\textbf
  {\bibinfo {volume} {11}},\ \bibinfo {pages} {83} (\bibinfo {year}
  {2021})}\BibitemShut {NoStop}%
\bibitem [{\citenamefont {Xie}\ \emph {et~al.}(2021)\citenamefont {Xie},
  \citenamefont {Pierce}, \citenamefont {Park}, \citenamefont {Parker},
  \citenamefont {Khalaf}, \citenamefont {Ledwith}, \citenamefont {Cao},
  \citenamefont {Lee}, \citenamefont {Chen}, \citenamefont {Forrester} \emph
  {et~al.}}]{xie2021fractional}%
  \BibitemOpen
  \bibfield  {author} {\bibinfo {author} {\bibfnamefont {Y.}~\bibnamefont
  {Xie}}, \bibinfo {author} {\bibfnamefont {A.~T.}\ \bibnamefont {Pierce}},
  \bibinfo {author} {\bibfnamefont {J.~M.}\ \bibnamefont {Park}}, \bibinfo
  {author} {\bibfnamefont {D.~E.}\ \bibnamefont {Parker}}, \bibinfo {author}
  {\bibfnamefont {E.}~\bibnamefont {Khalaf}}, \bibinfo {author} {\bibfnamefont
  {P.}~\bibnamefont {Ledwith}}, \bibinfo {author} {\bibfnamefont
  {Y.}~\bibnamefont {Cao}}, \bibinfo {author} {\bibfnamefont {S.~H.}\
  \bibnamefont {Lee}}, \bibinfo {author} {\bibfnamefont {S.}~\bibnamefont
  {Chen}}, \bibinfo {author} {\bibfnamefont {P.~R.}\ \bibnamefont {Forrester}},
  \emph {et~al.},\ }\bibfield  {title} {\bibinfo {title} {Fractional chern
  insulators in magic-angle twisted bilayer graphene},\ }\href@noop {}
  {\bibfield  {journal} {\bibinfo  {journal} {Nature}\ }\textbf {\bibinfo
  {volume} {600}},\ \bibinfo {pages} {439} (\bibinfo {year}
  {2021})}\BibitemShut {NoStop}%
\bibitem [{\citenamefont {Cao}\ \emph {et~al.}(2020)\citenamefont {Cao},
  \citenamefont {Rodan-Legrain}, \citenamefont {Rubies-Bigorda}, \citenamefont
  {Park}, \citenamefont {Watanabe}, \citenamefont {Taniguchi},\ and\
  \citenamefont {Jarillo-Herrero}}]{cao2020tunable}%
  \BibitemOpen
  \bibfield  {author} {\bibinfo {author} {\bibfnamefont {Y.}~\bibnamefont
  {Cao}}, \bibinfo {author} {\bibfnamefont {D.}~\bibnamefont {Rodan-Legrain}},
  \bibinfo {author} {\bibfnamefont {O.}~\bibnamefont {Rubies-Bigorda}},
  \bibinfo {author} {\bibfnamefont {J.~M.}\ \bibnamefont {Park}}, \bibinfo
  {author} {\bibfnamefont {K.}~\bibnamefont {Watanabe}}, \bibinfo {author}
  {\bibfnamefont {T.}~\bibnamefont {Taniguchi}},\ and\ \bibinfo {author}
  {\bibfnamefont {P.}~\bibnamefont {Jarillo-Herrero}},\ }\bibfield  {title}
  {\bibinfo {title} {Tunable correlated states and spin-polarized phases in
  twisted bilayer--bilayer graphene},\ }\href@noop {} {\bibfield  {journal}
  {\bibinfo  {journal} {Nature}\ }\textbf {\bibinfo {volume} {583}},\ \bibinfo
  {pages} {215} (\bibinfo {year} {2020})}\BibitemShut {NoStop}%
\bibitem [{\citenamefont {Wilhelm}\ \emph {et~al.}(2021)\citenamefont
  {Wilhelm}, \citenamefont {Lang},\ and\ \citenamefont
  {L{\"a}uchli}}]{wilhelm2021interplay}%
  \BibitemOpen
  \bibfield  {author} {\bibinfo {author} {\bibfnamefont {P.}~\bibnamefont
  {Wilhelm}}, \bibinfo {author} {\bibfnamefont {T.~C.}\ \bibnamefont {Lang}},\
  and\ \bibinfo {author} {\bibfnamefont {A.~M.}\ \bibnamefont {L{\"a}uchli}},\
  }\bibfield  {title} {\bibinfo {title} {Interplay of fractional chern
  insulator and charge density wave phases in twisted bilayer graphene},\
  }\href@noop {} {\bibfield  {journal} {\bibinfo  {journal} {Physical Review
  B}\ }\textbf {\bibinfo {volume} {103}},\ \bibinfo {pages} {125406} (\bibinfo
  {year} {2021})}\BibitemShut {NoStop}%
\bibitem [{\citenamefont {Rademaker}\ and\ \citenamefont
  {Mellado}(2018)}]{rademaker2018charge}%
  \BibitemOpen
  \bibfield  {author} {\bibinfo {author} {\bibfnamefont {L.}~\bibnamefont
  {Rademaker}}\ and\ \bibinfo {author} {\bibfnamefont {P.}~\bibnamefont
  {Mellado}},\ }\bibfield  {title} {\bibinfo {title} {Charge-transfer
  insulation in twisted bilayer graphene},\ }\href@noop {} {\bibfield
  {journal} {\bibinfo  {journal} {Physical Review B}\ }\textbf {\bibinfo
  {volume} {98}},\ \bibinfo {pages} {235158} (\bibinfo {year}
  {2018})}\BibitemShut {NoStop}%
\bibitem [{\citenamefont {Yankowitz}\ \emph {et~al.}(2019)\citenamefont
  {Yankowitz}, \citenamefont {Chen}, \citenamefont {Polshyn}, \citenamefont
  {Zhang}, \citenamefont {Watanabe}, \citenamefont {Taniguchi}, \citenamefont
  {Graf}, \citenamefont {Young},\ and\ \citenamefont
  {Dean}}]{yankowitz2019tuning}%
  \BibitemOpen
  \bibfield  {author} {\bibinfo {author} {\bibfnamefont {M.}~\bibnamefont
  {Yankowitz}}, \bibinfo {author} {\bibfnamefont {S.}~\bibnamefont {Chen}},
  \bibinfo {author} {\bibfnamefont {H.}~\bibnamefont {Polshyn}}, \bibinfo
  {author} {\bibfnamefont {Y.}~\bibnamefont {Zhang}}, \bibinfo {author}
  {\bibfnamefont {K.}~\bibnamefont {Watanabe}}, \bibinfo {author}
  {\bibfnamefont {T.}~\bibnamefont {Taniguchi}}, \bibinfo {author}
  {\bibfnamefont {D.}~\bibnamefont {Graf}}, \bibinfo {author} {\bibfnamefont
  {A.~F.}\ \bibnamefont {Young}},\ and\ \bibinfo {author} {\bibfnamefont
  {C.~R.}\ \bibnamefont {Dean}},\ }\bibfield  {title} {\bibinfo {title} {Tuning
  superconductivity in twisted bilayer graphene},\ }\href@noop {} {\bibfield
  {journal} {\bibinfo  {journal} {Science}\ }\textbf {\bibinfo {volume}
  {363}},\ \bibinfo {pages} {1059} (\bibinfo {year} {2019})}\BibitemShut
  {NoStop}%
\bibitem [{\citenamefont {Choi}\ \emph {et~al.}(2019)\citenamefont {Choi},
  \citenamefont {Kemmer}, \citenamefont {Peng}, \citenamefont {Thomson},
  \citenamefont {Arora}, \citenamefont {Polski}, \citenamefont {Zhang},
  \citenamefont {Ren}, \citenamefont {Alicea}, \citenamefont {Refael} \emph
  {et~al.}}]{choi2019electronic}%
  \BibitemOpen
  \bibfield  {author} {\bibinfo {author} {\bibfnamefont {Y.}~\bibnamefont
  {Choi}}, \bibinfo {author} {\bibfnamefont {J.}~\bibnamefont {Kemmer}},
  \bibinfo {author} {\bibfnamefont {Y.}~\bibnamefont {Peng}}, \bibinfo {author}
  {\bibfnamefont {A.}~\bibnamefont {Thomson}}, \bibinfo {author} {\bibfnamefont
  {H.}~\bibnamefont {Arora}}, \bibinfo {author} {\bibfnamefont
  {R.}~\bibnamefont {Polski}}, \bibinfo {author} {\bibfnamefont
  {Y.}~\bibnamefont {Zhang}}, \bibinfo {author} {\bibfnamefont
  {H.}~\bibnamefont {Ren}}, \bibinfo {author} {\bibfnamefont {J.}~\bibnamefont
  {Alicea}}, \bibinfo {author} {\bibfnamefont {G.}~\bibnamefont {Refael}},
  \emph {et~al.},\ }\bibfield  {title} {\bibinfo {title} {Electronic
  correlations in twisted bilayer graphene near the magic angle},\ }\href@noop
  {} {\bibfield  {journal} {\bibinfo  {journal} {Nature Physics}\ }\textbf
  {\bibinfo {volume} {15}},\ \bibinfo {pages} {1174} (\bibinfo {year}
  {2019})}\BibitemShut {NoStop}%
\bibitem [{\citenamefont {Sherkunov}\ and\ \citenamefont
  {Betouras}(2018)}]{sherkunov2018electronic}%
  \BibitemOpen
  \bibfield  {author} {\bibinfo {author} {\bibfnamefont {Y.}~\bibnamefont
  {Sherkunov}}\ and\ \bibinfo {author} {\bibfnamefont {J.~J.}\ \bibnamefont
  {Betouras}},\ }\bibfield  {title} {\bibinfo {title} {Electronic phases in
  twisted bilayer graphene at magic angles as a result of van hove
  singularities and interactions},\ }\href@noop {} {\bibfield  {journal}
  {\bibinfo  {journal} {Physical Review B}\ }\textbf {\bibinfo {volume} {98}},\
  \bibinfo {pages} {205151} (\bibinfo {year} {2018})}\BibitemShut {NoStop}%
\bibitem [{\citenamefont {Xie}\ and\ \citenamefont
  {MacDonald}(2020)}]{xie2020nature}%
  \BibitemOpen
  \bibfield  {author} {\bibinfo {author} {\bibfnamefont {M.}~\bibnamefont
  {Xie}}\ and\ \bibinfo {author} {\bibfnamefont {A.~H.}\ \bibnamefont
  {MacDonald}},\ }\bibfield  {title} {\bibinfo {title} {Nature of the
  correlated insulator states in twisted bilayer graphene},\ }\href@noop {}
  {\bibfield  {journal} {\bibinfo  {journal} {Physical review letters}\
  }\textbf {\bibinfo {volume} {124}},\ \bibinfo {pages} {097601} (\bibinfo
  {year} {2020})}\BibitemShut {NoStop}%
\bibitem [{\citenamefont {Codecido}\ \emph {et~al.}(2019)\citenamefont
  {Codecido}, \citenamefont {Wang}, \citenamefont {Koester}, \citenamefont
  {Che}, \citenamefont {Tian}, \citenamefont {Lv}, \citenamefont {Tran},
  \citenamefont {Watanabe}, \citenamefont {Taniguchi}, \citenamefont {Zhang},
  \citenamefont {Bockrath},\ and\ \citenamefont {Lau}}]{Codecidoeaaw9770}%
  \BibitemOpen
  \bibfield  {author} {\bibinfo {author} {\bibfnamefont {E.}~\bibnamefont
  {Codecido}}, \bibinfo {author} {\bibfnamefont {Q.}~\bibnamefont {Wang}},
  \bibinfo {author} {\bibfnamefont {R.}~\bibnamefont {Koester}}, \bibinfo
  {author} {\bibfnamefont {S.}~\bibnamefont {Che}}, \bibinfo {author}
  {\bibfnamefont {H.}~\bibnamefont {Tian}}, \bibinfo {author} {\bibfnamefont
  {R.}~\bibnamefont {Lv}}, \bibinfo {author} {\bibfnamefont {S.}~\bibnamefont
  {Tran}}, \bibinfo {author} {\bibfnamefont {K.}~\bibnamefont {Watanabe}},
  \bibinfo {author} {\bibfnamefont {T.}~\bibnamefont {Taniguchi}}, \bibinfo
  {author} {\bibfnamefont {F.}~\bibnamefont {Zhang}}, \bibinfo {author}
  {\bibfnamefont {M.}~\bibnamefont {Bockrath}},\ and\ \bibinfo {author}
  {\bibfnamefont {C.~N.}\ \bibnamefont {Lau}},\ }\bibfield  {title} {\bibinfo
  {title} {Correlated insulating and superconducting states in twisted bilayer
  graphene below the magic angle},\ }\bibfield  {journal} {\bibinfo  {journal}
  {Science Advances}\ }\textbf {\bibinfo {volume} {5}},\ \href
  {https://doi.org/10.1126/sciadv.aaw9770} {10.1126/sciadv.aaw9770} (\bibinfo
  {year} {2019})\BibitemShut {NoStop}%
\bibitem [{\citenamefont {Wong}\ \emph {et~al.}(2020)\citenamefont {Wong},
  \citenamefont {Nuckolls}, \citenamefont {Oh}, \citenamefont {Lian},
  \citenamefont {Xie}, \citenamefont {Jeon}, \citenamefont {Watanabe},
  \citenamefont {Taniguchi}, \citenamefont {Bernevig},\ and\ \citenamefont
  {Yazdani}}]{wong2019cascade}%
  \BibitemOpen
  \bibfield  {author} {\bibinfo {author} {\bibfnamefont {D.}~\bibnamefont
  {Wong}}, \bibinfo {author} {\bibfnamefont {K.~P.}\ \bibnamefont {Nuckolls}},
  \bibinfo {author} {\bibfnamefont {M.}~\bibnamefont {Oh}}, \bibinfo {author}
  {\bibfnamefont {B.}~\bibnamefont {Lian}}, \bibinfo {author} {\bibfnamefont
  {Y.}~\bibnamefont {Xie}}, \bibinfo {author} {\bibfnamefont {S.}~\bibnamefont
  {Jeon}}, \bibinfo {author} {\bibfnamefont {K.}~\bibnamefont {Watanabe}},
  \bibinfo {author} {\bibfnamefont {T.}~\bibnamefont {Taniguchi}}, \bibinfo
  {author} {\bibfnamefont {B.~A.}\ \bibnamefont {Bernevig}},\ and\ \bibinfo
  {author} {\bibfnamefont {A.}~\bibnamefont {Yazdani}},\ }\bibfield  {title}
  {\bibinfo {title} {Cascade of electronic transitions in magic-angle twisted
  bilayer graphene},\ }\href {https://doi.org/10.1038/s41586-020-2339-0}
  {\bibfield  {journal} {\bibinfo  {journal} {Nature}\ }\textbf {\bibinfo
  {volume} {582}},\ \bibinfo {pages} {198} (\bibinfo {year}
  {2020})}\BibitemShut {NoStop}%
\bibitem [{\citenamefont {Lu}\ \emph {et~al.}(2019)\citenamefont {Lu},
  \citenamefont {Stepanov}, \citenamefont {Yang}, \citenamefont {Xie},
  \citenamefont {Aamir}, \citenamefont {Das}, \citenamefont {Urgell},
  \citenamefont {Watanabe}, \citenamefont {Taniguchi}, \citenamefont {Zhang},
  \citenamefont {Bachtold}, \citenamefont {MacDonald},\ and\ \citenamefont
  {Efetov}}]{Lu2019efetov}%
  \BibitemOpen
  \bibfield  {author} {\bibinfo {author} {\bibfnamefont {X.}~\bibnamefont
  {Lu}}, \bibinfo {author} {\bibfnamefont {P.}~\bibnamefont {Stepanov}},
  \bibinfo {author} {\bibfnamefont {W.}~\bibnamefont {Yang}}, \bibinfo {author}
  {\bibfnamefont {M.}~\bibnamefont {Xie}}, \bibinfo {author} {\bibfnamefont
  {M.~A.}\ \bibnamefont {Aamir}}, \bibinfo {author} {\bibfnamefont
  {I.}~\bibnamefont {Das}}, \bibinfo {author} {\bibfnamefont {C.}~\bibnamefont
  {Urgell}}, \bibinfo {author} {\bibfnamefont {K.}~\bibnamefont {Watanabe}},
  \bibinfo {author} {\bibfnamefont {T.}~\bibnamefont {Taniguchi}}, \bibinfo
  {author} {\bibfnamefont {G.}~\bibnamefont {Zhang}}, \bibinfo {author}
  {\bibfnamefont {A.}~\bibnamefont {Bachtold}}, \bibinfo {author}
  {\bibfnamefont {A.~H.}\ \bibnamefont {MacDonald}},\ and\ \bibinfo {author}
  {\bibfnamefont {D.~K.}\ \bibnamefont {Efetov}},\ }\bibfield  {title}
  {\bibinfo {title} {Superconductors, orbital magnets and correlated states in
  magic-angle bilayer graphene},\ }\href
  {https://doi.org/10.1038/s41586-019-1695-0} {\bibfield  {journal} {\bibinfo
  {journal} {Nature}\ }\textbf {\bibinfo {volume} {574}},\ \bibinfo {pages}
  {653} (\bibinfo {year} {2019})}\BibitemShut {NoStop}%
\bibitem [{\citenamefont {Sharpe}\ \emph {et~al.}(2019)\citenamefont {Sharpe},
  \citenamefont {Fox}, \citenamefont {Barnard}, \citenamefont {Finney},
  \citenamefont {Watanabe}, \citenamefont {Taniguchi}, \citenamefont
  {Kastner},\ and\ \citenamefont {Goldhaber-Gordon}}]{Sharpe605}%
  \BibitemOpen
  \bibfield  {author} {\bibinfo {author} {\bibfnamefont {A.~L.}\ \bibnamefont
  {Sharpe}}, \bibinfo {author} {\bibfnamefont {E.~J.}\ \bibnamefont {Fox}},
  \bibinfo {author} {\bibfnamefont {A.~W.}\ \bibnamefont {Barnard}}, \bibinfo
  {author} {\bibfnamefont {J.}~\bibnamefont {Finney}}, \bibinfo {author}
  {\bibfnamefont {K.}~\bibnamefont {Watanabe}}, \bibinfo {author}
  {\bibfnamefont {T.}~\bibnamefont {Taniguchi}}, \bibinfo {author}
  {\bibfnamefont {M.~A.}\ \bibnamefont {Kastner}},\ and\ \bibinfo {author}
  {\bibfnamefont {D.}~\bibnamefont {Goldhaber-Gordon}},\ }\bibfield  {title}
  {\bibinfo {title} {Emergent ferromagnetism near three-quarters filling in
  twisted bilayer graphene},\ }\href {https://doi.org/10.1126/science.aaw3780}
  {\bibfield  {journal} {\bibinfo  {journal} {Science}\ }\textbf {\bibinfo
  {volume} {365}},\ \bibinfo {pages} {605} (\bibinfo {year}
  {2019})}\BibitemShut {NoStop}%
\bibitem [{\citenamefont {Seo}\ \emph {et~al.}(2019)\citenamefont {Seo},
  \citenamefont {Kotov},\ and\ \citenamefont {Uchoa}}]{Seo_2019}%
  \BibitemOpen
  \bibfield  {author} {\bibinfo {author} {\bibfnamefont {K.}~\bibnamefont
  {Seo}}, \bibinfo {author} {\bibfnamefont {V.~N.}\ \bibnamefont {Kotov}},\
  and\ \bibinfo {author} {\bibfnamefont {B.}~\bibnamefont {Uchoa}},\ }\bibfield
   {title} {\bibinfo {title} {Ferromagnetic mott state in twisted graphene
  bilayers at the magic angle},\ }\bibfield  {journal} {\bibinfo  {journal}
  {Physical Review Letters}\ }\textbf {\bibinfo {volume} {122}},\ \href
  {https://doi.org/10.1103/physrevlett.122.246402}
  {10.1103/physrevlett.122.246402} (\bibinfo {year} {2019})\BibitemShut
  {NoStop}%
\bibitem [{\citenamefont {Tran}\ \emph {et~al.}(2019)\citenamefont {Tran},
  \citenamefont {Moody}, \citenamefont {Wu}, \citenamefont {Lu}, \citenamefont
  {Choi}, \citenamefont {Kim}, \citenamefont {Rai}, \citenamefont {Sanchez},
  \citenamefont {Quan}, \citenamefont {Singh}, \citenamefont {Embley},
  \citenamefont {Zepeda}, \citenamefont {Campbell}, \citenamefont {Autry},
  \citenamefont {Taniguchi}, \citenamefont {Watanabe}, \citenamefont {Lu},
  \citenamefont {Banerjee}, \citenamefont {Silverman}, \citenamefont {Kim},
  \citenamefont {Tutuc}, \citenamefont {Yang}, \citenamefont {MacDonald},\ and\
  \citenamefont {Li}}]{Tran2019}%
  \BibitemOpen
  \bibfield  {author} {\bibinfo {author} {\bibfnamefont {K.}~\bibnamefont
  {Tran}}, \bibinfo {author} {\bibfnamefont {G.}~\bibnamefont {Moody}},
  \bibinfo {author} {\bibfnamefont {F.}~\bibnamefont {Wu}}, \bibinfo {author}
  {\bibfnamefont {X.}~\bibnamefont {Lu}}, \bibinfo {author} {\bibfnamefont
  {J.}~\bibnamefont {Choi}}, \bibinfo {author} {\bibfnamefont {K.}~\bibnamefont
  {Kim}}, \bibinfo {author} {\bibfnamefont {A.}~\bibnamefont {Rai}}, \bibinfo
  {author} {\bibfnamefont {D.~A.}\ \bibnamefont {Sanchez}}, \bibinfo {author}
  {\bibfnamefont {J.}~\bibnamefont {Quan}}, \bibinfo {author} {\bibfnamefont
  {A.}~\bibnamefont {Singh}}, \bibinfo {author} {\bibfnamefont
  {J.}~\bibnamefont {Embley}}, \bibinfo {author} {\bibfnamefont
  {A.}~\bibnamefont {Zepeda}}, \bibinfo {author} {\bibfnamefont
  {M.}~\bibnamefont {Campbell}}, \bibinfo {author} {\bibfnamefont
  {T.}~\bibnamefont {Autry}}, \bibinfo {author} {\bibfnamefont
  {T.}~\bibnamefont {Taniguchi}}, \bibinfo {author} {\bibfnamefont
  {K.}~\bibnamefont {Watanabe}}, \bibinfo {author} {\bibfnamefont
  {N.}~\bibnamefont {Lu}}, \bibinfo {author} {\bibfnamefont {S.~K.}\
  \bibnamefont {Banerjee}}, \bibinfo {author} {\bibfnamefont {K.~L.}\
  \bibnamefont {Silverman}}, \bibinfo {author} {\bibfnamefont {S.}~\bibnamefont
  {Kim}}, \bibinfo {author} {\bibfnamefont {E.}~\bibnamefont {Tutuc}}, \bibinfo
  {author} {\bibfnamefont {L.}~\bibnamefont {Yang}}, \bibinfo {author}
  {\bibfnamefont {A.~H.}\ \bibnamefont {MacDonald}},\ and\ \bibinfo {author}
  {\bibfnamefont {X.}~\bibnamefont {Li}},\ }\bibfield  {title} {\bibinfo
  {title} {Evidence for moir{\'e} excitons in van der waals heterostructures},\
  }\href {https://doi.org/10.1038/s41586-019-0975-z} {\bibfield  {journal}
  {\bibinfo  {journal} {Nature}\ }\textbf {\bibinfo {volume} {567}},\ \bibinfo
  {pages} {71} (\bibinfo {year} {2019})}\BibitemShut {NoStop}%
\bibitem [{\citenamefont {Seyler}\ \emph {et~al.}(2019)\citenamefont {Seyler},
  \citenamefont {Rivera}, \citenamefont {Yu}, \citenamefont {Wilson},
  \citenamefont {Ray}, \citenamefont {Mandrus}, \citenamefont {Yan},
  \citenamefont {Yao},\ and\ \citenamefont {Xu}}]{Seyler2019}%
  \BibitemOpen
  \bibfield  {author} {\bibinfo {author} {\bibfnamefont {K.~L.}\ \bibnamefont
  {Seyler}}, \bibinfo {author} {\bibfnamefont {P.}~\bibnamefont {Rivera}},
  \bibinfo {author} {\bibfnamefont {H.}~\bibnamefont {Yu}}, \bibinfo {author}
  {\bibfnamefont {N.~P.}\ \bibnamefont {Wilson}}, \bibinfo {author}
  {\bibfnamefont {E.~L.}\ \bibnamefont {Ray}}, \bibinfo {author} {\bibfnamefont
  {D.~G.}\ \bibnamefont {Mandrus}}, \bibinfo {author} {\bibfnamefont
  {J.}~\bibnamefont {Yan}}, \bibinfo {author} {\bibfnamefont {W.}~\bibnamefont
  {Yao}},\ and\ \bibinfo {author} {\bibfnamefont {X.}~\bibnamefont {Xu}},\
  }\bibfield  {title} {\bibinfo {title} {Signatures of moir{\'e}-trapped valley
  excitons in mose2/wse2 heterobilayers},\ }\href
  {https://doi.org/10.1038/s41586-019-0957-1} {\bibfield  {journal} {\bibinfo
  {journal} {Nature}\ }\textbf {\bibinfo {volume} {567}},\ \bibinfo {pages}
  {66} (\bibinfo {year} {2019})}\BibitemShut {NoStop}%
\bibitem [{\citenamefont {Alexeev}\ \emph {et~al.}(2019)\citenamefont
  {Alexeev}, \citenamefont {Ruiz-Tijerina}, \citenamefont {Danovich},
  \citenamefont {Hamer}, \citenamefont {Terry}, \citenamefont {Nayak},
  \citenamefont {Ahn}, \citenamefont {Pak}, \citenamefont {Lee}, \citenamefont
  {Sohn}, \citenamefont {Molas}, \citenamefont {Koperski}, \citenamefont
  {Watanabe}, \citenamefont {Taniguchi}, \citenamefont {Novoselov},
  \citenamefont {Gorbachev}, \citenamefont {Shin}, \citenamefont {Fal'ko},\
  and\ \citenamefont {Tartakovskii}}]{Alexeev2019}%
  \BibitemOpen
  \bibfield  {author} {\bibinfo {author} {\bibfnamefont {E.~M.}\ \bibnamefont
  {Alexeev}}, \bibinfo {author} {\bibfnamefont {D.~A.}\ \bibnamefont
  {Ruiz-Tijerina}}, \bibinfo {author} {\bibfnamefont {M.}~\bibnamefont
  {Danovich}}, \bibinfo {author} {\bibfnamefont {M.~J.}\ \bibnamefont {Hamer}},
  \bibinfo {author} {\bibfnamefont {D.~J.}\ \bibnamefont {Terry}}, \bibinfo
  {author} {\bibfnamefont {P.~K.}\ \bibnamefont {Nayak}}, \bibinfo {author}
  {\bibfnamefont {S.}~\bibnamefont {Ahn}}, \bibinfo {author} {\bibfnamefont
  {S.}~\bibnamefont {Pak}}, \bibinfo {author} {\bibfnamefont {J.}~\bibnamefont
  {Lee}}, \bibinfo {author} {\bibfnamefont {J.~I.}\ \bibnamefont {Sohn}},
  \bibinfo {author} {\bibfnamefont {M.~R.}\ \bibnamefont {Molas}}, \bibinfo
  {author} {\bibfnamefont {M.}~\bibnamefont {Koperski}}, \bibinfo {author}
  {\bibfnamefont {K.}~\bibnamefont {Watanabe}}, \bibinfo {author}
  {\bibfnamefont {T.}~\bibnamefont {Taniguchi}}, \bibinfo {author}
  {\bibfnamefont {K.~S.}\ \bibnamefont {Novoselov}}, \bibinfo {author}
  {\bibfnamefont {R.~V.}\ \bibnamefont {Gorbachev}}, \bibinfo {author}
  {\bibfnamefont {H.~S.}\ \bibnamefont {Shin}}, \bibinfo {author}
  {\bibfnamefont {V.~I.}\ \bibnamefont {Fal'ko}},\ and\ \bibinfo {author}
  {\bibfnamefont {A.~I.}\ \bibnamefont {Tartakovskii}},\ }\bibfield  {title}
  {\bibinfo {title} {Resonantly hybridized excitons in moir{\'e} superlattices
  in van der waals heterostructures},\ }\href
  {https://doi.org/10.1038/s41586-019-0986-9} {\bibfield  {journal} {\bibinfo
  {journal} {Nature}\ }\textbf {\bibinfo {volume} {567}},\ \bibinfo {pages}
  {81} (\bibinfo {year} {2019})}\BibitemShut {NoStop}%
\bibitem [{\citenamefont {Hu}\ and\ \citenamefont
  {MacDonald}(2021)}]{hu2021competing}%
  \BibitemOpen
  \bibfield  {author} {\bibinfo {author} {\bibfnamefont {N.~C.}\ \bibnamefont
  {Hu}}\ and\ \bibinfo {author} {\bibfnamefont {A.~H.}\ \bibnamefont
  {MacDonald}},\ }\bibfield  {title} {\bibinfo {title} {Competing magnetic
  states in transition metal dichalcogenide moir\'e materials},\ }\href
  {https://doi.org/10.1103/PhysRevB.104.214403} {\bibfield  {journal} {\bibinfo
   {journal} {Phys. Rev. B}\ }\textbf {\bibinfo {volume} {104}},\ \bibinfo
  {pages} {214403} (\bibinfo {year} {2021})}\BibitemShut {NoStop}%
\bibitem [{\citenamefont {Wu}\ \emph {et~al.}(2018{\natexlab{a}})\citenamefont
  {Wu}, \citenamefont {Lovorn}, \citenamefont {Tutuc},\ and\ \citenamefont
  {MacDonald}}]{PhysRevLett.121.026402}%
  \BibitemOpen
  \bibfield  {author} {\bibinfo {author} {\bibfnamefont {F.}~\bibnamefont
  {Wu}}, \bibinfo {author} {\bibfnamefont {T.}~\bibnamefont {Lovorn}}, \bibinfo
  {author} {\bibfnamefont {E.}~\bibnamefont {Tutuc}},\ and\ \bibinfo {author}
  {\bibfnamefont {A.~H.}\ \bibnamefont {MacDonald}},\ }\bibfield  {title}
  {\bibinfo {title} {Hubbard model physics in transition metal dichalcogenide
  moir\'e bands},\ }\href {https://doi.org/10.1103/PhysRevLett.121.026402}
  {\bibfield  {journal} {\bibinfo  {journal} {Phys. Rev. Lett.}\ }\textbf
  {\bibinfo {volume} {121}},\ \bibinfo {pages} {026402} (\bibinfo {year}
  {2018}{\natexlab{a}})}\BibitemShut {NoStop}%
\bibitem [{\citenamefont {Morales-Dur\`an}\ \emph {et~al.}(2021)\citenamefont
  {Morales-Dur\`an}, \citenamefont {Hu}, \citenamefont {Potasz},\ and\
  \citenamefont {MacDonald}}]{moralesduran2021nonlocal}%
  \BibitemOpen
  \bibfield  {author} {\bibinfo {author} {\bibfnamefont {N.}~\bibnamefont
  {Morales-Dur\`an}}, \bibinfo {author} {\bibfnamefont {N.~C.}\ \bibnamefont
  {Hu}}, \bibinfo {author} {\bibfnamefont {P.}~\bibnamefont {Potasz}},\ and\
  \bibinfo {author} {\bibfnamefont {A.~H.}\ \bibnamefont {MacDonald}},\
  }\href@noop {} {\bibinfo {title} {Non-local interactions in moir\'e hubbard
  systems}} (\bibinfo {year} {2021}),\ \Eprint
  {https://arxiv.org/abs/2108.03313} {arXiv:2108.03313 [cond-mat.str-el]}
  \BibitemShut {NoStop}%
\bibitem [{\citenamefont {Wu}\ \emph {et~al.}(2019{\natexlab{a}})\citenamefont
  {Wu}, \citenamefont {Lovorn}, \citenamefont {Tutuc}, \citenamefont {Martin},\
  and\ \citenamefont {MacDonald}}]{2019wulovorn}%
  \BibitemOpen
  \bibfield  {author} {\bibinfo {author} {\bibfnamefont {F.}~\bibnamefont
  {Wu}}, \bibinfo {author} {\bibfnamefont {T.}~\bibnamefont {Lovorn}}, \bibinfo
  {author} {\bibfnamefont {E.}~\bibnamefont {Tutuc}}, \bibinfo {author}
  {\bibfnamefont {I.}~\bibnamefont {Martin}},\ and\ \bibinfo {author}
  {\bibfnamefont {A.}~\bibnamefont {MacDonald}},\ }\bibfield  {title} {\bibinfo
  {title} {Topological insulators in twisted transition metal dichalcogenide
  homobilayers},\ }\bibfield  {journal} {\bibinfo  {journal} {Physical Review
  Letters}\ }\textbf {\bibinfo {volume} {122}},\ \href
  {https://doi.org/10.1103/physrevlett.122.086402}
  {10.1103/physrevlett.122.086402} (\bibinfo {year}
  {2019}{\natexlab{a}})\BibitemShut {NoStop}%
\bibitem [{\citenamefont {Tang}\ \emph {et~al.}(2020)\citenamefont {Tang},
  \citenamefont {Li}, \citenamefont {Li}, \citenamefont {Xu}, \citenamefont
  {Liu}, \citenamefont {Barmak}, \citenamefont {Watanabe}, \citenamefont
  {Taniguchi}, \citenamefont {MacDonald}, \citenamefont {Shan} \emph
  {et~al.}}]{tang2020simulation}%
  \BibitemOpen
  \bibfield  {author} {\bibinfo {author} {\bibfnamefont {Y.}~\bibnamefont
  {Tang}}, \bibinfo {author} {\bibfnamefont {L.}~\bibnamefont {Li}}, \bibinfo
  {author} {\bibfnamefont {T.}~\bibnamefont {Li}}, \bibinfo {author}
  {\bibfnamefont {Y.}~\bibnamefont {Xu}}, \bibinfo {author} {\bibfnamefont
  {S.}~\bibnamefont {Liu}}, \bibinfo {author} {\bibfnamefont {K.}~\bibnamefont
  {Barmak}}, \bibinfo {author} {\bibfnamefont {K.}~\bibnamefont {Watanabe}},
  \bibinfo {author} {\bibfnamefont {T.}~\bibnamefont {Taniguchi}}, \bibinfo
  {author} {\bibfnamefont {A.~H.}\ \bibnamefont {MacDonald}}, \bibinfo {author}
  {\bibfnamefont {J.}~\bibnamefont {Shan}}, \emph {et~al.},\ }\bibfield
  {title} {\bibinfo {title} {Simulation of hubbard model physics in wse 2/ws 2
  moir{\'e} superlattices},\ }\href@noop {} {\bibfield  {journal} {\bibinfo
  {journal} {Nature}\ }\textbf {\bibinfo {volume} {579}},\ \bibinfo {pages}
  {353} (\bibinfo {year} {2020})}\BibitemShut {NoStop}%
\bibitem [{\citenamefont {Zare}\ and\ \citenamefont
  {Mosadeq}(2021)}]{PhysRevB.104.115154}%
  \BibitemOpen
  \bibfield  {author} {\bibinfo {author} {\bibfnamefont {M.-H.}\ \bibnamefont
  {Zare}}\ and\ \bibinfo {author} {\bibfnamefont {H.}~\bibnamefont {Mosadeq}},\
  }\bibfield  {title} {\bibinfo {title} {Spin liquid in twisted homobilayers of
  group-vi dichalcogenides},\ }\href
  {https://doi.org/10.1103/PhysRevB.104.115154} {\bibfield  {journal} {\bibinfo
   {journal} {Phys. Rev. B}\ }\textbf {\bibinfo {volume} {104}},\ \bibinfo
  {pages} {115154} (\bibinfo {year} {2021})}\BibitemShut {NoStop}%
\bibitem [{\citenamefont {Andersen}\ \emph {et~al.}(2021)\citenamefont
  {Andersen}, \citenamefont {Scuri}, \citenamefont {Sushko}, \citenamefont
  {De~Greve}, \citenamefont {Sung}, \citenamefont {Zhou}, \citenamefont {Wild},
  \citenamefont {Gelly}, \citenamefont {Heo}, \citenamefont {B{\'e}rub{\'e}}
  \emph {et~al.}}]{andersen2021excitons}%
  \BibitemOpen
  \bibfield  {author} {\bibinfo {author} {\bibfnamefont {T.~I.}\ \bibnamefont
  {Andersen}}, \bibinfo {author} {\bibfnamefont {G.}~\bibnamefont {Scuri}},
  \bibinfo {author} {\bibfnamefont {A.}~\bibnamefont {Sushko}}, \bibinfo
  {author} {\bibfnamefont {K.}~\bibnamefont {De~Greve}}, \bibinfo {author}
  {\bibfnamefont {J.}~\bibnamefont {Sung}}, \bibinfo {author} {\bibfnamefont
  {Y.}~\bibnamefont {Zhou}}, \bibinfo {author} {\bibfnamefont {D.~S.}\
  \bibnamefont {Wild}}, \bibinfo {author} {\bibfnamefont {R.~J.}\ \bibnamefont
  {Gelly}}, \bibinfo {author} {\bibfnamefont {H.}~\bibnamefont {Heo}}, \bibinfo
  {author} {\bibfnamefont {D.}~\bibnamefont {B{\'e}rub{\'e}}}, \emph {et~al.},\
  }\bibfield  {title} {\bibinfo {title} {Excitons in a reconstructed moir{\'e}
  potential in twisted wse 2/wse 2 homobilayers},\ }\href@noop {} {\bibfield
  {journal} {\bibinfo  {journal} {Nature Materials}\ }\textbf {\bibinfo
  {volume} {20}},\ \bibinfo {pages} {480} (\bibinfo {year} {2021})}\BibitemShut
  {NoStop}%
\bibitem [{\citenamefont {Wang}\ \emph {et~al.}(2020)\citenamefont {Wang},
  \citenamefont {Shih}, \citenamefont {Ghiotto}, \citenamefont {Xian},
  \citenamefont {Rhodes}, \citenamefont {Tan}, \citenamefont {Claassen},
  \citenamefont {Kennes}, \citenamefont {Bai}, \citenamefont {Kim} \emph
  {et~al.}}]{wang2020correlated}%
  \BibitemOpen
  \bibfield  {author} {\bibinfo {author} {\bibfnamefont {L.}~\bibnamefont
  {Wang}}, \bibinfo {author} {\bibfnamefont {E.-M.}\ \bibnamefont {Shih}},
  \bibinfo {author} {\bibfnamefont {A.}~\bibnamefont {Ghiotto}}, \bibinfo
  {author} {\bibfnamefont {L.}~\bibnamefont {Xian}}, \bibinfo {author}
  {\bibfnamefont {D.~A.}\ \bibnamefont {Rhodes}}, \bibinfo {author}
  {\bibfnamefont {C.}~\bibnamefont {Tan}}, \bibinfo {author} {\bibfnamefont
  {M.}~\bibnamefont {Claassen}}, \bibinfo {author} {\bibfnamefont {D.~M.}\
  \bibnamefont {Kennes}}, \bibinfo {author} {\bibfnamefont {Y.}~\bibnamefont
  {Bai}}, \bibinfo {author} {\bibfnamefont {B.}~\bibnamefont {Kim}}, \emph
  {et~al.},\ }\bibfield  {title} {\bibinfo {title} {Correlated electronic
  phases in twisted bilayer transition metal dichalcogenides},\ }\href@noop {}
  {\bibfield  {journal} {\bibinfo  {journal} {Nature materials}\ }\textbf
  {\bibinfo {volume} {19}},\ \bibinfo {pages} {861} (\bibinfo {year}
  {2020})}\BibitemShut {NoStop}%
\bibitem [{\citenamefont {Angeli}\ and\ \citenamefont
  {MacDonald}(2021)}]{Angelie2021826118}%
  \BibitemOpen
  \bibfield  {author} {\bibinfo {author} {\bibfnamefont {M.}~\bibnamefont
  {Angeli}}\ and\ \bibinfo {author} {\bibfnamefont {A.~H.}\ \bibnamefont
  {MacDonald}},\ }\bibfield  {title} {\bibinfo {title} {$\gamma$ valley
  transition metal dichalcogenide moir{\'e} bands},\ }\href
  {https://www.pnas.org/content/118/10/e2021826118} {\bibfield  {journal}
  {\bibinfo  {journal} {Proceedings of the National Academy of Sciences}\
  }\textbf {\bibinfo {volume} {118}} (\bibinfo {year} {2021})}\BibitemShut
  {NoStop}%
\bibitem [{\citenamefont {Navarro-Labastida}\ \emph {et~al.}(2022)\citenamefont
  {Navarro-Labastida}, \citenamefont {Espinosa-Champo}, \citenamefont
  {Aguilar-Mendez},\ and\ \citenamefont {Naumis}}]{PhysRevB.105.115434}%
  \BibitemOpen
  \bibfield  {author} {\bibinfo {author} {\bibfnamefont {L.~A.}\ \bibnamefont
  {Navarro-Labastida}}, \bibinfo {author} {\bibfnamefont {A.}~\bibnamefont
  {Espinosa-Champo}}, \bibinfo {author} {\bibfnamefont {E.}~\bibnamefont
  {Aguilar-Mendez}},\ and\ \bibinfo {author} {\bibfnamefont {G.~G.}\
  \bibnamefont {Naumis}},\ }\bibfield  {title} {\bibinfo {title} {Why the first
  magic-angle is different from others in twisted graphene bilayers: Interlayer
  currents, kinetic and confinement energy, and wave-function localization},\
  }\href {https://doi.org/10.1103/PhysRevB.105.115434} {\bibfield  {journal}
  {\bibinfo  {journal} {Phys. Rev. B}\ }\textbf {\bibinfo {volume} {105}},\
  \bibinfo {pages} {115434} (\bibinfo {year} {2022})}\BibitemShut {NoStop}%
\bibitem [{\citenamefont {Naumis}\ \emph {et~al.}(2021)\citenamefont {Naumis},
  \citenamefont {Navarro-Labastida}, \citenamefont {Aguilar-M\'endez},\ and\
  \citenamefont {Espinosa-Champo}}]{PhysRevB.103.245418}%
  \BibitemOpen
  \bibfield  {author} {\bibinfo {author} {\bibfnamefont {G.~G.}\ \bibnamefont
  {Naumis}}, \bibinfo {author} {\bibfnamefont {L.~A.}\ \bibnamefont
  {Navarro-Labastida}}, \bibinfo {author} {\bibfnamefont {E.}~\bibnamefont
  {Aguilar-M\'endez}},\ and\ \bibinfo {author} {\bibfnamefont {A.}~\bibnamefont
  {Espinosa-Champo}},\ }\bibfield  {title} {\bibinfo {title} {Reduction of the
  twisted bilayer graphene chiral hamiltonian into a
  $2\ifmmode\times\else\texttimes\fi{}2$ matrix operator and physical origin of
  flat bands at magic angles},\ }\href
  {https://doi.org/10.1103/PhysRevB.103.245418} {\bibfield  {journal} {\bibinfo
   {journal} {Phys. Rev. B}\ }\textbf {\bibinfo {volume} {103}},\ \bibinfo
  {pages} {245418} (\bibinfo {year} {2021})}\BibitemShut {NoStop}%
\bibitem [{\citenamefont {Mirzakhani}\ \emph {et~al.}(2020)\citenamefont
  {Mirzakhani}, \citenamefont {Peeters},\ and\ \citenamefont
  {Zarenia}}]{PhysRevB.101.075413}%
  \BibitemOpen
  \bibfield  {author} {\bibinfo {author} {\bibfnamefont {M.}~\bibnamefont
  {Mirzakhani}}, \bibinfo {author} {\bibfnamefont {F.~M.}\ \bibnamefont
  {Peeters}},\ and\ \bibinfo {author} {\bibfnamefont {M.}~\bibnamefont
  {Zarenia}},\ }\bibfield  {title} {\bibinfo {title} {Circular quantum dots in
  twisted bilayer graphene},\ }\href
  {https://doi.org/10.1103/PhysRevB.101.075413} {\bibfield  {journal} {\bibinfo
   {journal} {Phys. Rev. B}\ }\textbf {\bibinfo {volume} {101}},\ \bibinfo
  {pages} {075413} (\bibinfo {year} {2020})}\BibitemShut {NoStop}%
\bibitem [{\citenamefont {Fleischmann}\ \emph {et~al.}(2019)\citenamefont
  {Fleischmann}, \citenamefont {Gupta}, \citenamefont {Wullschl\"ager},
  \citenamefont {Theil}, \citenamefont {Weckbecker}, \citenamefont {Meded},
  \citenamefont {Sharma}, \citenamefont {Meyer},\ and\ \citenamefont
  {Shallcross}}]{fleischmann2019perfect}%
  \BibitemOpen
  \bibfield  {author} {\bibinfo {author} {\bibfnamefont {M.}~\bibnamefont
  {Fleischmann}}, \bibinfo {author} {\bibfnamefont {R.}~\bibnamefont {Gupta}},
  \bibinfo {author} {\bibfnamefont {F.}~\bibnamefont {Wullschl\"ager}},
  \bibinfo {author} {\bibfnamefont {S.}~\bibnamefont {Theil}}, \bibinfo
  {author} {\bibfnamefont {D.}~\bibnamefont {Weckbecker}}, \bibinfo {author}
  {\bibfnamefont {V.}~\bibnamefont {Meded}}, \bibinfo {author} {\bibfnamefont
  {S.}~\bibnamefont {Sharma}}, \bibinfo {author} {\bibfnamefont
  {B.}~\bibnamefont {Meyer}},\ and\ \bibinfo {author} {\bibfnamefont
  {S.}~\bibnamefont {Shallcross}},\ }\bibfield  {title} {\bibinfo {title}
  {Perfect and controllable nesting in minimally twisted bilayer graphene},\
  }\href@noop {} {\bibfield  {journal} {\bibinfo  {journal} {Nano letters}\
  }\textbf {\bibinfo {volume} {20}},\ \bibinfo {pages} {971} (\bibinfo {year}
  {2019})}\BibitemShut {NoStop}%
\bibitem [{\citenamefont {Vogl}\ \emph {et~al.}(2017)\citenamefont {Vogl},
  \citenamefont {Pankratov},\ and\ \citenamefont
  {Shallcross}}]{2017mvoglsemiclass}%
  \BibitemOpen
  \bibfield  {author} {\bibinfo {author} {\bibfnamefont {M.}~\bibnamefont
  {Vogl}}, \bibinfo {author} {\bibfnamefont {O.}~\bibnamefont {Pankratov}},\
  and\ \bibinfo {author} {\bibfnamefont {S.}~\bibnamefont {Shallcross}},\
  }\bibfield  {title} {\bibinfo {title} {Semiclassics for matrix hamiltonians:
  The gutzwiller trace formula with applications to graphene-type systems},\
  }\bibfield  {journal} {\bibinfo  {journal} {Physical Review B}\ }\textbf
  {\bibinfo {volume} {96}},\ \href {https://doi.org/10.1103/physrevb.96.035442}
  {10.1103/physrevb.96.035442} (\bibinfo {year} {2017})\BibitemShut {NoStop}%
\bibitem [{\citenamefont {Abanin}\ \emph {et~al.}(2017)\citenamefont {Abanin},
  \citenamefont {De~Roeck}, \citenamefont {Ho},\ and\ \citenamefont
  {Huveneers}}]{PhysRevB.95.014112}%
  \BibitemOpen
  \bibfield  {author} {\bibinfo {author} {\bibfnamefont {D.~A.}\ \bibnamefont
  {Abanin}}, \bibinfo {author} {\bibfnamefont {W.}~\bibnamefont {De~Roeck}},
  \bibinfo {author} {\bibfnamefont {W.~W.}\ \bibnamefont {Ho}},\ and\ \bibinfo
  {author} {\bibfnamefont {F.~m.~c.}\ \bibnamefont {Huveneers}},\ }\bibfield
  {title} {\bibinfo {title} {Effective {Hamiltonians}, prethermalization, and
  slow energy absorption in periodically driven many-body systems},\ }\href
  {https://doi.org/10.1103/PhysRevB.95.014112} {\bibfield  {journal} {\bibinfo
  {journal} {Phys. Rev. B}\ }\textbf {\bibinfo {volume} {95}},\ \bibinfo
  {pages} {014112} (\bibinfo {year} {2017})}\BibitemShut {NoStop}%
\bibitem [{\citenamefont {Vajna}\ \emph {et~al.}(2018)\citenamefont {Vajna},
  \citenamefont {Klobas}, \citenamefont {Prosen},\ and\ \citenamefont
  {Polkovnikov}}]{polkovnikov2018repl}%
  \BibitemOpen
  \bibfield  {author} {\bibinfo {author} {\bibfnamefont {S.}~\bibnamefont
  {Vajna}}, \bibinfo {author} {\bibfnamefont {K.}~\bibnamefont {Klobas}},
  \bibinfo {author} {\bibfnamefont {T.}~\bibnamefont {Prosen}},\ and\ \bibinfo
  {author} {\bibfnamefont {A.}~\bibnamefont {Polkovnikov}},\ }\bibfield
  {title} {\bibinfo {title} {Replica resummation of the
  baker-campbell-hausdorff series},\ }\bibfield  {journal} {\bibinfo  {journal}
  {Physical Review Letters}\ }\textbf {\bibinfo {volume} {120}},\ \href
  {https://doi.org/10.1103/physrevlett.120.200607}
  {10.1103/physrevlett.120.200607} (\bibinfo {year} {2018})\BibitemShut
  {NoStop}%
\bibitem [{\citenamefont {Blanes}\ \emph {et~al.}(2009)\citenamefont {Blanes},
  \citenamefont {Casas}, \citenamefont {Oteo},\ and\ \citenamefont
  {Ros}}]{blanes2009}%
  \BibitemOpen
  \bibfield  {author} {\bibinfo {author} {\bibfnamefont {S.}~\bibnamefont
  {Blanes}}, \bibinfo {author} {\bibfnamefont {F.}~\bibnamefont {Casas}},
  \bibinfo {author} {\bibfnamefont {J.}~\bibnamefont {Oteo}},\ and\ \bibinfo
  {author} {\bibfnamefont {J.}~\bibnamefont {Ros}},\ }\bibfield  {title}
  {\bibinfo {title} {The magnus expansion and some of its applications},\
  }\href@noop {} {\bibfield  {journal} {\bibinfo  {journal} {Physics Reports}\
  }\textbf {\bibinfo {volume} {470}},\ \bibinfo {pages} {151} (\bibinfo {year}
  {2009})}\BibitemShut {NoStop}%
\bibitem [{\citenamefont {Rahav}\ \emph
  {et~al.}(2003{\natexlab{a}})\citenamefont {Rahav}, \citenamefont {Gilary},\
  and\ \citenamefont {Fishman}}]{rahav2003}%
  \BibitemOpen
  \bibfield  {author} {\bibinfo {author} {\bibfnamefont {S.}~\bibnamefont
  {Rahav}}, \bibinfo {author} {\bibfnamefont {I.}~\bibnamefont {Gilary}},\ and\
  \bibinfo {author} {\bibfnamefont {S.}~\bibnamefont {Fishman}},\ }\bibfield
  {title} {\bibinfo {title} {Effective hamiltonians for periodically driven
  systems},\ }\href@noop {} {\bibfield  {journal} {\bibinfo  {journal} {Phys.
  Rev. A}\ }\textbf {\bibinfo {volume} {68}},\ \bibinfo {pages} {013820}
  (\bibinfo {year} {2003}{\natexlab{a}})}\BibitemShut {NoStop}%
\bibitem [{\citenamefont {Rahav}\ \emph
  {et~al.}(2003{\natexlab{b}})\citenamefont {Rahav}, \citenamefont {Gilary},\
  and\ \citenamefont {Fishman}}]{rahav2003b}%
  \BibitemOpen
  \bibfield  {author} {\bibinfo {author} {\bibfnamefont {S.}~\bibnamefont
  {Rahav}}, \bibinfo {author} {\bibfnamefont {I.}~\bibnamefont {Gilary}},\ and\
  \bibinfo {author} {\bibfnamefont {S.}~\bibnamefont {Fishman}},\ }\bibfield
  {title} {\bibinfo {title} {Time independent description of rapidly
  oscillating potentials},\ }\href@noop {} {\bibfield  {journal} {\bibinfo
  {journal} {Phys. Rev. Lett.}\ }\textbf {\bibinfo {volume} {91}},\ \bibinfo
  {pages} {110404} (\bibinfo {year} {2003}{\natexlab{b}})}\BibitemShut
  {NoStop}%
\bibitem [{\citenamefont {Bukov}\ \emph {et~al.}(2015)\citenamefont {Bukov},
  \citenamefont {D'Alessio},\ and\ \citenamefont {Polkovnikov}}]{bukov2015}%
  \BibitemOpen
  \bibfield  {author} {\bibinfo {author} {\bibfnamefont {M.}~\bibnamefont
  {Bukov}}, \bibinfo {author} {\bibfnamefont {L.}~\bibnamefont {D'Alessio}},\
  and\ \bibinfo {author} {\bibfnamefont {A.}~\bibnamefont {Polkovnikov}},\
  }\bibfield  {title} {\bibinfo {title} {Universal high-frequency behavior of
  periodically driven systems: from dynamical stabilization to floquet
  engineering},\ }\href@noop {} {\bibfield  {journal} {\bibinfo  {journal}
  {Advances in Physics}\ }\textbf {\bibinfo {volume} {64}},\ \bibinfo {pages}
  {139} (\bibinfo {year} {2015})}\BibitemShut {NoStop}%
\bibitem [{\citenamefont {Eckardt}\ and\ \citenamefont
  {Anisimovas}(2015)}]{Eckardt_2015}%
  \BibitemOpen
  \bibfield  {author} {\bibinfo {author} {\bibfnamefont {A.}~\bibnamefont
  {Eckardt}}\ and\ \bibinfo {author} {\bibfnamefont {E.}~\bibnamefont
  {Anisimovas}},\ }\bibfield  {title} {\bibinfo {title} {High-frequency
  approximation for periodically driven quantum systems from a floquet-space
  perspective},\ }\href {https://doi.org/10.1088/1367-2630/17/9/093039}
  {\bibfield  {journal} {\bibinfo  {journal} {New Journal of Physics}\ }\textbf
  {\bibinfo {volume} {17}},\ \bibinfo {pages} {093039} (\bibinfo {year}
  {2015})}\BibitemShut {NoStop}%
\bibitem [{\citenamefont {Fel'dman}(1984)}]{Feldm1984}%
  \BibitemOpen
  \bibfield  {author} {\bibinfo {author} {\bibfnamefont {E.}~\bibnamefont
  {Fel'dman}},\ }\bibfield  {title} {\bibinfo {title} {On the convergence of
  the {Magnus} expansion for spin systems in periodic magnetic fields},\ }\href
  {https://doi.org/http://dx.doi.org/10.1016/0375-9601(84)90027-6} {\bibfield
  {journal} {\bibinfo  {journal} {Phys. Lett. A}\ }\textbf {\bibinfo {volume}
  {104}},\ \bibinfo {pages} {479} (\bibinfo {year} {1984})}\BibitemShut
  {NoStop}%
\bibitem [{\citenamefont {Magnus}(1954)}]{Magnus1954}%
  \BibitemOpen
  \bibfield  {author} {\bibinfo {author} {\bibfnamefont {W.}~\bibnamefont
  {Magnus}},\ }\bibfield  {title} {\bibinfo {title} {On the exponential
  solution of differential equations for a linear operator},\ }\href
  {https://doi.org/10.1002/cpa.3160070404} {\bibfield  {journal} {\bibinfo
  {journal} {Commun. Pure Appl. Math.}\ }\textbf {\bibinfo {volume} {7}},\
  \bibinfo {pages} {649} (\bibinfo {year} {1954})}\BibitemShut {NoStop}%
\bibitem [{\citenamefont {Goldman}\ and\ \citenamefont
  {Dalibard}(2014)}]{PhysRevX.4.031027}%
  \BibitemOpen
  \bibfield  {author} {\bibinfo {author} {\bibfnamefont {N.}~\bibnamefont
  {Goldman}}\ and\ \bibinfo {author} {\bibfnamefont {J.}~\bibnamefont
  {Dalibard}},\ }\bibfield  {title} {\bibinfo {title} {Periodically driven
  quantum systems: Effective hamiltonians and engineered gauge fields},\ }\href
  {https://doi.org/10.1103/PhysRevX.4.031027} {\bibfield  {journal} {\bibinfo
  {journal} {Phys. Rev. X}\ }\textbf {\bibinfo {volume} {4}},\ \bibinfo {pages}
  {031027} (\bibinfo {year} {2014})}\BibitemShut {NoStop}%
\bibitem [{\citenamefont {Itin}\ and\ \citenamefont
  {Katsnelson}(2015)}]{PhysRevLett.115.075301}%
  \BibitemOpen
  \bibfield  {author} {\bibinfo {author} {\bibfnamefont {A.~P.}\ \bibnamefont
  {Itin}}\ and\ \bibinfo {author} {\bibfnamefont {M.~I.}\ \bibnamefont
  {Katsnelson}},\ }\bibfield  {title} {\bibinfo {title} {Effective
  {Hamiltonians} for rapidly driven many-body lattice systems: Induced exchange
  interactions and density-dependent hoppings},\ }\href
  {https://doi.org/10.1103/PhysRevLett.115.075301} {\bibfield  {journal}
  {\bibinfo  {journal} {Phys. Rev. Lett.}\ }\textbf {\bibinfo {volume} {115}},\
  \bibinfo {pages} {075301} (\bibinfo {year} {2015})}\BibitemShut {NoStop}%
\bibitem [{\citenamefont {Mikami}\ \emph {et~al.}(2016)\citenamefont {Mikami},
  \citenamefont {Kitamura}, \citenamefont {Yasuda}, \citenamefont {Tsuji},
  \citenamefont {Oka},\ and\ \citenamefont {Aoki}}]{PhysRevB.93.144307}%
  \BibitemOpen
  \bibfield  {author} {\bibinfo {author} {\bibfnamefont {T.}~\bibnamefont
  {Mikami}}, \bibinfo {author} {\bibfnamefont {S.}~\bibnamefont {Kitamura}},
  \bibinfo {author} {\bibfnamefont {K.}~\bibnamefont {Yasuda}}, \bibinfo
  {author} {\bibfnamefont {N.}~\bibnamefont {Tsuji}}, \bibinfo {author}
  {\bibfnamefont {T.}~\bibnamefont {Oka}},\ and\ \bibinfo {author}
  {\bibfnamefont {H.}~\bibnamefont {Aoki}},\ }\bibfield  {title} {\bibinfo
  {title} {{Brillouin-Wigner} theory for high-frequency expansion in
  periodically driven systems: Application to {Floquet} topological
  insulators},\ }\href {https://doi.org/10.1103/PhysRevB.93.144307} {\bibfield
  {journal} {\bibinfo  {journal} {Phys. Rev. B}\ }\textbf {\bibinfo {volume}
  {93}},\ \bibinfo {pages} {144307} (\bibinfo {year} {2016})}\BibitemShut
  {NoStop}%
\bibitem [{\citenamefont {Mohan}\ \emph {et~al.}(2016)\citenamefont {Mohan},
  \citenamefont {Saxena}, \citenamefont {Kundu},\ and\ \citenamefont
  {Rao}}]{PhysRevB.94.235419}%
  \BibitemOpen
  \bibfield  {author} {\bibinfo {author} {\bibfnamefont {P.}~\bibnamefont
  {Mohan}}, \bibinfo {author} {\bibfnamefont {R.}~\bibnamefont {Saxena}},
  \bibinfo {author} {\bibfnamefont {A.}~\bibnamefont {Kundu}},\ and\ \bibinfo
  {author} {\bibfnamefont {S.}~\bibnamefont {Rao}},\ }\bibfield  {title}
  {\bibinfo {title} {{Brillouin-Wigner} theory for {Floquet} topological phase
  transitions in spin-orbit-coupled materials},\ }\href
  {https://doi.org/10.1103/PhysRevB.94.235419} {\bibfield  {journal} {\bibinfo
  {journal} {Phys. Rev. B}\ }\textbf {\bibinfo {volume} {94}},\ \bibinfo
  {pages} {235419} (\bibinfo {year} {2016})}\BibitemShut {NoStop}%
\bibitem [{\citenamefont {Bukov}\ \emph {et~al.}(2016)\citenamefont {Bukov},
  \citenamefont {Kolodrubetz},\ and\ \citenamefont
  {Polkovnikov}}]{PhysRevLett.116.125301}%
  \BibitemOpen
  \bibfield  {author} {\bibinfo {author} {\bibfnamefont {M.}~\bibnamefont
  {Bukov}}, \bibinfo {author} {\bibfnamefont {M.}~\bibnamefont {Kolodrubetz}},\
  and\ \bibinfo {author} {\bibfnamefont {A.}~\bibnamefont {Polkovnikov}},\
  }\bibfield  {title} {\bibinfo {title} {{Schrieffer-Wolff} transformation for
  periodically driven systems: Strongly correlated systems with artificial
  gauge fields},\ }\href {https://doi.org/10.1103/PhysRevLett.116.125301}
  {\bibfield  {journal} {\bibinfo  {journal} {Phys. Rev. Lett.}\ }\textbf
  {\bibinfo {volume} {116}},\ \bibinfo {pages} {125301} (\bibinfo {year}
  {2016})}\BibitemShut {NoStop}%
\bibitem [{\citenamefont {Vogl}\ \emph
  {et~al.}(2019{\natexlab{a}})\citenamefont {Vogl}, \citenamefont {Laurell},
  \citenamefont {Barr},\ and\ \citenamefont {Fiete}}]{Vogl_2019AnalogHJ}%
  \BibitemOpen
  \bibfield  {author} {\bibinfo {author} {\bibfnamefont {M.}~\bibnamefont
  {Vogl}}, \bibinfo {author} {\bibfnamefont {P.}~\bibnamefont {Laurell}},
  \bibinfo {author} {\bibfnamefont {A.~D.}\ \bibnamefont {Barr}},\ and\
  \bibinfo {author} {\bibfnamefont {G.~A.}\ \bibnamefont {Fiete}},\ }\bibfield
  {title} {\bibinfo {title} {Analog of hamilton-jacobi theory for the
  time-evolution operator},\ }\bibfield  {journal} {\bibinfo  {journal}
  {Physical Review A}\ }\textbf {\bibinfo {volume} {100}},\ \href
  {https://doi.org/10.1103/physreva.100.012132} {10.1103/physreva.100.012132}
  (\bibinfo {year} {2019}{\natexlab{a}})\BibitemShut {NoStop}%
\bibitem [{\citenamefont {Verdeny}\ \emph {et~al.}(2013)\citenamefont
  {Verdeny}, \citenamefont {Mielke},\ and\ \citenamefont
  {Mintert}}]{verdeny2013}%
  \BibitemOpen
  \bibfield  {author} {\bibinfo {author} {\bibfnamefont {A.}~\bibnamefont
  {Verdeny}}, \bibinfo {author} {\bibfnamefont {A.}~\bibnamefont {Mielke}},\
  and\ \bibinfo {author} {\bibfnamefont {F.}~\bibnamefont {Mintert}},\
  }\bibfield  {title} {\bibinfo {title} {Accurate effective hamiltonians via
  unitary flow in floquet space},\ }\href
  {https://doi.org/10.1103/PhysRevLett.111.175301} {\bibfield  {journal}
  {\bibinfo  {journal} {Phys. Rev. Lett.}\ }\textbf {\bibinfo {volume} {111}},\
  \bibinfo {pages} {175301} (\bibinfo {year} {2013})}\BibitemShut {NoStop}%
\bibitem [{\citenamefont {G\'omez-Le\'on}\ and\ \citenamefont
  {Platero}(2013)}]{PhysRevLett.110.200403}%
  \BibitemOpen
  \bibfield  {author} {\bibinfo {author} {\bibfnamefont {A.}~\bibnamefont
  {G\'omez-Le\'on}}\ and\ \bibinfo {author} {\bibfnamefont {G.}~\bibnamefont
  {Platero}},\ }\bibfield  {title} {\bibinfo {title} {Floquet-bloch theory and
  topology in periodically driven lattices},\ }\href
  {https://doi.org/10.1103/PhysRevLett.110.200403} {\bibfield  {journal}
  {\bibinfo  {journal} {Phys. Rev. Lett.}\ }\textbf {\bibinfo {volume} {110}},\
  \bibinfo {pages} {200403} (\bibinfo {year} {2013})}\BibitemShut {NoStop}%
\bibitem [{\citenamefont {Vogl}\ \emph
  {et~al.}(2020{\natexlab{a}})\citenamefont {Vogl}, \citenamefont
  {Rodriguez-Vega},\ and\ \citenamefont {Fiete}}]{Vogl2020_effham}%
  \BibitemOpen
  \bibfield  {author} {\bibinfo {author} {\bibfnamefont {M.}~\bibnamefont
  {Vogl}}, \bibinfo {author} {\bibfnamefont {M.}~\bibnamefont
  {Rodriguez-Vega}},\ and\ \bibinfo {author} {\bibfnamefont {G.~A.}\
  \bibnamefont {Fiete}},\ }\bibfield  {title} {\bibinfo {title} {Effective
  floquet hamiltonian in the low-frequency regime},\ }\href
  {https://doi.org/10.1103/PhysRevB.101.024303} {\bibfield  {journal} {\bibinfo
   {journal} {Phys. Rev. B}\ }\textbf {\bibinfo {volume} {101}},\ \bibinfo
  {pages} {024303} (\bibinfo {year} {2020}{\natexlab{a}})}\BibitemShut
  {NoStop}%
\bibitem [{\citenamefont {Vogl}\ \emph
  {et~al.}(2019{\natexlab{b}})\citenamefont {Vogl}, \citenamefont {Laurell},
  \citenamefont {Barr},\ and\ \citenamefont {Fiete}}]{vogl2019}%
  \BibitemOpen
  \bibfield  {author} {\bibinfo {author} {\bibfnamefont {M.}~\bibnamefont
  {Vogl}}, \bibinfo {author} {\bibfnamefont {P.}~\bibnamefont {Laurell}},
  \bibinfo {author} {\bibfnamefont {A.~D.}\ \bibnamefont {Barr}},\ and\
  \bibinfo {author} {\bibfnamefont {G.~A.}\ \bibnamefont {Fiete}},\ }\bibfield
  {title} {\bibinfo {title} {Flow equation approach to periodically driven
  quantum systems},\ }\href {https://doi.org/10.1103/PhysRevX.9.021037}
  {\bibfield  {journal} {\bibinfo  {journal} {Phys. Rev. X}\ }\textbf {\bibinfo
  {volume} {9}},\ \bibinfo {pages} {021037} (\bibinfo {year}
  {2019}{\natexlab{b}})}\BibitemShut {NoStop}%
\bibitem [{\citenamefont {Rodriguez-Vega}\ \emph {et~al.}(2018)\citenamefont
  {Rodriguez-Vega}, \citenamefont {Lentz},\ and\ \citenamefont
  {Seradjeh}}]{Rodriguez_Vega_2018}%
  \BibitemOpen
  \bibfield  {author} {\bibinfo {author} {\bibfnamefont {M.}~\bibnamefont
  {Rodriguez-Vega}}, \bibinfo {author} {\bibfnamefont {M.}~\bibnamefont
  {Lentz}},\ and\ \bibinfo {author} {\bibfnamefont {B.}~\bibnamefont
  {Seradjeh}},\ }\bibfield  {title} {\bibinfo {title} {Floquet perturbation
  theory: formalism and application to low-frequency limit},\ }\href
  {https://doi.org/10.1088/1367-2630/aade37} {\bibfield  {journal} {\bibinfo
  {journal} {New Journal of Physics}\ }\textbf {\bibinfo {volume} {20}},\
  \bibinfo {pages} {093022} (\bibinfo {year} {2018})}\BibitemShut {NoStop}%
\bibitem [{\citenamefont {Rodriguez-Vega}\ and\ \citenamefont
  {Seradjeh}(2018)}]{PhysRevLett.121.036402}%
  \BibitemOpen
  \bibfield  {author} {\bibinfo {author} {\bibfnamefont {M.}~\bibnamefont
  {Rodriguez-Vega}}\ and\ \bibinfo {author} {\bibfnamefont {B.}~\bibnamefont
  {Seradjeh}},\ }\bibfield  {title} {\bibinfo {title} {Universal fluctuations
  of floquet topological invariants at low frequencies},\ }\href
  {https://doi.org/10.1103/PhysRevLett.121.036402} {\bibfield  {journal}
  {\bibinfo  {journal} {Phys. Rev. Lett.}\ }\textbf {\bibinfo {volume} {121}},\
  \bibinfo {pages} {036402} (\bibinfo {year} {2018})}\BibitemShut {NoStop}%
\bibitem [{\citenamefont {Martiskainen}\ and\ \citenamefont
  {Moiseyev}(2015)}]{Martiskainen2015}%
  \BibitemOpen
  \bibfield  {author} {\bibinfo {author} {\bibfnamefont {H.}~\bibnamefont
  {Martiskainen}}\ and\ \bibinfo {author} {\bibfnamefont {N.}~\bibnamefont
  {Moiseyev}},\ }\bibfield  {title} {\bibinfo {title} {Perturbation theory for
  quasienergy floquet solutions in the low-frequency regime of the oscillating
  electric field},\ }\href {https://doi.org/10.1103/PhysRevA.91.023416}
  {\bibfield  {journal} {\bibinfo  {journal} {Phys. Rev. A}\ }\textbf {\bibinfo
  {volume} {91}},\ \bibinfo {pages} {023416} (\bibinfo {year}
  {2015})}\BibitemShut {NoStop}%
\bibitem [{\citenamefont {Rigolin}\ \emph {et~al.}(2008)\citenamefont
  {Rigolin}, \citenamefont {Ortiz},\ and\ \citenamefont {Ponce}}]{rigolin2008}%
  \BibitemOpen
  \bibfield  {author} {\bibinfo {author} {\bibfnamefont {G.}~\bibnamefont
  {Rigolin}}, \bibinfo {author} {\bibfnamefont {G.}~\bibnamefont {Ortiz}},\
  and\ \bibinfo {author} {\bibfnamefont {V.~H.}\ \bibnamefont {Ponce}},\
  }\bibfield  {title} {\bibinfo {title} {Beyond the quantum adiabatic
  approximation: Adiabatic perturbation theory},\ }\href
  {https://doi.org/10.1103/PhysRevA.78.052508} {\bibfield  {journal} {\bibinfo
  {journal} {Phys. Rev. A}\ }\textbf {\bibinfo {volume} {78}},\ \bibinfo
  {pages} {052508} (\bibinfo {year} {2008})}\BibitemShut {NoStop}%
\bibitem [{\citenamefont {Weinberg}\ \emph {et~al.}(2015)\citenamefont
  {Weinberg}, \citenamefont {\"Olschl\"ager}, \citenamefont {Str\"ater},
  \citenamefont {Prelle}, \citenamefont {Eckardt}, \citenamefont {Sengstock},\
  and\ \citenamefont {Simonet}}]{weinberg2015}%
  \BibitemOpen
  \bibfield  {author} {\bibinfo {author} {\bibfnamefont {M.}~\bibnamefont
  {Weinberg}}, \bibinfo {author} {\bibfnamefont {C.}~\bibnamefont
  {\"Olschl\"ager}}, \bibinfo {author} {\bibfnamefont {C.}~\bibnamefont
  {Str\"ater}}, \bibinfo {author} {\bibfnamefont {S.}~\bibnamefont {Prelle}},
  \bibinfo {author} {\bibfnamefont {A.}~\bibnamefont {Eckardt}}, \bibinfo
  {author} {\bibfnamefont {K.}~\bibnamefont {Sengstock}},\ and\ \bibinfo
  {author} {\bibfnamefont {J.}~\bibnamefont {Simonet}},\ }\bibfield  {title}
  {\bibinfo {title} {Multiphoton interband excitations of quantum gases in
  driven optical lattices},\ }\href
  {https://doi.org/10.1103/PhysRevA.92.043621} {\bibfield  {journal} {\bibinfo
  {journal} {Phys. Rev. A}\ }\textbf {\bibinfo {volume} {92}},\ \bibinfo
  {pages} {043621} (\bibinfo {year} {2015})}\BibitemShut {NoStop}%
\bibitem [{\citenamefont {Li}\ \emph {et~al.}(2017)\citenamefont {Li},
  \citenamefont {Lam},\ and\ \citenamefont {You}}]{PhysRevB.96.155438}%
  \BibitemOpen
  \bibfield  {author} {\bibinfo {author} {\bibfnamefont {Z.-Z.}\ \bibnamefont
  {Li}}, \bibinfo {author} {\bibfnamefont {C.-H.}\ \bibnamefont {Lam}},\ and\
  \bibinfo {author} {\bibfnamefont {J.~Q.}\ \bibnamefont {You}},\ }\bibfield
  {title} {\bibinfo {title} {Floquet engineering of long-range $p$-wave
  superconductivity: Beyond the high-frequency limit},\ }\href
  {https://doi.org/10.1103/PhysRevB.96.155438} {\bibfield  {journal} {\bibinfo
  {journal} {Phys. Rev. B}\ }\textbf {\bibinfo {volume} {96}},\ \bibinfo
  {pages} {155438} (\bibinfo {year} {2017})}\BibitemShut {NoStop}%
\bibitem [{\citenamefont {Kennes}\ \emph
  {et~al.}(2019{\natexlab{a}})\citenamefont {Kennes}, \citenamefont {M\"uller},
  \citenamefont {Pletyukhov}, \citenamefont {Weber}, \citenamefont {Bruder},
  \citenamefont {Hassler}, \citenamefont {Klinovaja}, \citenamefont {Loss},\
  and\ \citenamefont {Schoeller}}]{Kennes-Klinovaja-2019}%
  \BibitemOpen
  \bibfield  {author} {\bibinfo {author} {\bibfnamefont {D.~M.}\ \bibnamefont
  {Kennes}}, \bibinfo {author} {\bibfnamefont {N.}~\bibnamefont {M\"uller}},
  \bibinfo {author} {\bibfnamefont {M.}~\bibnamefont {Pletyukhov}}, \bibinfo
  {author} {\bibfnamefont {C.}~\bibnamefont {Weber}}, \bibinfo {author}
  {\bibfnamefont {C.}~\bibnamefont {Bruder}}, \bibinfo {author} {\bibfnamefont
  {F.}~\bibnamefont {Hassler}}, \bibinfo {author} {\bibfnamefont
  {J.}~\bibnamefont {Klinovaja}}, \bibinfo {author} {\bibfnamefont
  {D.}~\bibnamefont {Loss}},\ and\ \bibinfo {author} {\bibfnamefont
  {H.}~\bibnamefont {Schoeller}},\ }\bibfield  {title} {\bibinfo {title}
  {Chiral one-dimensional floquet topological insulators beyond the rotating
  wave approximation},\ }\href {https://doi.org/10.1103/PhysRevB.100.041103}
  {\bibfield  {journal} {\bibinfo  {journal} {Phys. Rev. B}\ }\textbf {\bibinfo
  {volume} {100}},\ \bibinfo {pages} {041103} (\bibinfo {year}
  {2019}{\natexlab{a}})}\BibitemShut {NoStop}%
\bibitem [{\citenamefont {M\"uller}\ \emph {et~al.}(2020)\citenamefont
  {M\"uller}, \citenamefont {Kennes}, \citenamefont {Klinovaja}, \citenamefont
  {Loss},\ and\ \citenamefont {Schoeller}}]{PhysRevB.101.155417}%
  \BibitemOpen
  \bibfield  {author} {\bibinfo {author} {\bibfnamefont {N.}~\bibnamefont
  {M\"uller}}, \bibinfo {author} {\bibfnamefont {D.~M.}\ \bibnamefont
  {Kennes}}, \bibinfo {author} {\bibfnamefont {J.}~\bibnamefont {Klinovaja}},
  \bibinfo {author} {\bibfnamefont {D.}~\bibnamefont {Loss}},\ and\ \bibinfo
  {author} {\bibfnamefont {H.}~\bibnamefont {Schoeller}},\ }\bibfield  {title}
  {\bibinfo {title} {Electronic transport in one-dimensional floquet
  topological insulators via topological and nontopological edge states},\
  }\href {https://doi.org/10.1103/PhysRevB.101.155417} {\bibfield  {journal}
  {\bibinfo  {journal} {Phys. Rev. B}\ }\textbf {\bibinfo {volume} {101}},\
  \bibinfo {pages} {155417} (\bibinfo {year} {2020})}\BibitemShut {NoStop}%
\bibitem [{\citenamefont {Novicenko}\ \emph {et~al.}(2021)\citenamefont
  {Novicenko}, \citenamefont {Žlabys},\ and\ \citenamefont
  {Anisimovas}}]{novicenko2021flowequation}%
  \BibitemOpen
  \bibfield  {author} {\bibinfo {author} {\bibfnamefont {V.}~\bibnamefont
  {Novicenko}}, \bibinfo {author} {\bibfnamefont {G.}~\bibnamefont {Žlabys}},\
  and\ \bibinfo {author} {\bibfnamefont {E.}~\bibnamefont {Anisimovas}},\
  }\href@noop {} {\bibinfo {title} {Flow-equation approach to quantum systems
  driven by an amplitude-modulated time-periodic force}} (\bibinfo {year}
  {2021}),\ \Eprint {https://arxiv.org/abs/2111.15368} {arXiv:2111.15368
  [quant-ph]} \BibitemShut {NoStop}%
\bibitem [{\citenamefont {Shirai}\ \emph {et~al.}(2016)\citenamefont {Shirai},
  \citenamefont {Thingna}, \citenamefont {Mori}, \citenamefont {Denisov},
  \citenamefont {Hänggi},\ and\ \citenamefont {Miyashita}}]{Shirai_2016}%
  \BibitemOpen
  \bibfield  {author} {\bibinfo {author} {\bibfnamefont {T.}~\bibnamefont
  {Shirai}}, \bibinfo {author} {\bibfnamefont {J.}~\bibnamefont {Thingna}},
  \bibinfo {author} {\bibfnamefont {T.}~\bibnamefont {Mori}}, \bibinfo {author}
  {\bibfnamefont {S.}~\bibnamefont {Denisov}}, \bibinfo {author} {\bibfnamefont
  {P.}~\bibnamefont {Hänggi}},\ and\ \bibinfo {author} {\bibfnamefont
  {S.}~\bibnamefont {Miyashita}},\ }\bibfield  {title} {\bibinfo {title}
  {Effective floquet{\textendash}gibbs states for dissipative quantum
  systems},\ }\href {https://doi.org/10.1088/1367-2630/18/5/053008} {\bibfield
  {journal} {\bibinfo  {journal} {New Journal of Physics}\ }\textbf {\bibinfo
  {volume} {18}},\ \bibinfo {pages} {053008} (\bibinfo {year}
  {2016})}\BibitemShut {NoStop}%
\bibitem [{\citenamefont {Ikeda}\ \emph {et~al.}(2021)\citenamefont {Ikeda},
  \citenamefont {Chinzei},\ and\ \citenamefont
  {Sato}}]{10.21468/SciPostPhysCore.4.4.033}%
  \BibitemOpen
  \bibfield  {author} {\bibinfo {author} {\bibfnamefont {T.~N.}\ \bibnamefont
  {Ikeda}}, \bibinfo {author} {\bibfnamefont {K.}~\bibnamefont {Chinzei}},\
  and\ \bibinfo {author} {\bibfnamefont {M.}~\bibnamefont {Sato}},\ }\bibfield
  {title} {\bibinfo {title} {{Nonequilibrium steady states in the
  Floquet-Lindblad systems: van Vleck's high-frequency expansion approach}},\
  }\href {https://doi.org/10.21468/SciPostPhysCore.4.4.033} {\bibfield
  {journal} {\bibinfo  {journal} {SciPost Phys. Core}\ }\textbf {\bibinfo
  {volume} {4}},\ \bibinfo {pages} {33} (\bibinfo {year} {2021})}\BibitemShut
  {NoStop}%
\bibitem [{\citenamefont {Ikeda}\ and\ \citenamefont
  {Polkovnikov}(2021)}]{PhysRevB.104.134308}%
  \BibitemOpen
  \bibfield  {author} {\bibinfo {author} {\bibfnamefont {T.~N.}\ \bibnamefont
  {Ikeda}}\ and\ \bibinfo {author} {\bibfnamefont {A.}~\bibnamefont
  {Polkovnikov}},\ }\bibfield  {title} {\bibinfo {title} {Fermi's golden rule
  for heating in strongly driven floquet systems},\ }\href
  {https://doi.org/10.1103/PhysRevB.104.134308} {\bibfield  {journal} {\bibinfo
   {journal} {Phys. Rev. B}\ }\textbf {\bibinfo {volume} {104}},\ \bibinfo
  {pages} {134308} (\bibinfo {year} {2021})}\BibitemShut {NoStop}%
\bibitem [{\citenamefont {Engelhardt}\ \emph {et~al.}(2019)\citenamefont
  {Engelhardt}, \citenamefont {Platero},\ and\ \citenamefont
  {Cao}}]{PhysRevLett.123.120602}%
  \BibitemOpen
  \bibfield  {author} {\bibinfo {author} {\bibfnamefont {G.}~\bibnamefont
  {Engelhardt}}, \bibinfo {author} {\bibfnamefont {G.}~\bibnamefont
  {Platero}},\ and\ \bibinfo {author} {\bibfnamefont {J.}~\bibnamefont {Cao}},\
  }\bibfield  {title} {\bibinfo {title} {Discontinuities in driven spin-boson
  systems due to coherent destruction of tunneling: Breakdown of the
  floquet-gibbs distribution},\ }\href
  {https://doi.org/10.1103/PhysRevLett.123.120602} {\bibfield  {journal}
  {\bibinfo  {journal} {Phys. Rev. Lett.}\ }\textbf {\bibinfo {volume} {123}},\
  \bibinfo {pages} {120602} (\bibinfo {year} {2019})}\BibitemShut {NoStop}%
\bibitem [{\citenamefont {Basov}\ \emph {et~al.}(2017)\citenamefont {Basov},
  \citenamefont {Averitt},\ and\ \citenamefont {Hsieh}}]{Basov2017}%
  \BibitemOpen
  \bibfield  {author} {\bibinfo {author} {\bibfnamefont {D.~N.}\ \bibnamefont
  {Basov}}, \bibinfo {author} {\bibfnamefont {R.~D.}\ \bibnamefont {Averitt}},\
  and\ \bibinfo {author} {\bibfnamefont {D.}~\bibnamefont {Hsieh}},\ }\bibfield
   {title} {\bibinfo {title} {Towards properties on demand in quantum
  materials},\ }\href {https://doi.org/10.1038/nmat5017} {\bibfield  {journal}
  {\bibinfo  {journal} {Nat. Mat.}\ }\textbf {\bibinfo {volume} {16}},\
  \bibinfo {pages} {1077} (\bibinfo {year} {2017})}\BibitemShut {NoStop}%
\bibitem [{\citenamefont {Oka}\ and\ \citenamefont
  {Kitamura}(2019)}]{Oka_2019}%
  \BibitemOpen
  \bibfield  {author} {\bibinfo {author} {\bibfnamefont {T.}~\bibnamefont
  {Oka}}\ and\ \bibinfo {author} {\bibfnamefont {S.}~\bibnamefont {Kitamura}},\
  }\bibfield  {title} {\bibinfo {title} {Floquet engineering of quantum
  materials},\ }\href
  {https://doi.org/10.1146/annurev-conmatphys-031218-013423} {\bibfield
  {journal} {\bibinfo  {journal} {Annual Review of Condensed Matter Physics}\
  }\textbf {\bibinfo {volume} {10}},\ \bibinfo {pages} {387–408} (\bibinfo
  {year} {2019})}\BibitemShut {NoStop}%
\bibitem [{\citenamefont {Rudner}\ and\ \citenamefont
  {Lindner}(2020)}]{rudner2020_review}%
  \BibitemOpen
  \bibfield  {author} {\bibinfo {author} {\bibfnamefont {M.~S.}\ \bibnamefont
  {Rudner}}\ and\ \bibinfo {author} {\bibfnamefont {N.~H.}\ \bibnamefont
  {Lindner}},\ }\bibfield  {title} {\bibinfo {title} {Band structure
  engineering and non-equilibrium dynamics in floquet topological insulators},\
  }\href {https://doi.org/10.1038/s42254-020-0170-z} {\bibfield  {journal}
  {\bibinfo  {journal} {Nature Reviews Physics}\ }\textbf {\bibinfo {volume}
  {2}},\ \bibinfo {pages} {229} (\bibinfo {year} {2020})}\BibitemShut {NoStop}%
\bibitem [{\citenamefont {Giovannini}\ and\ \citenamefont
  {Hübener}(2019)}]{Giovannini_2019}%
  \BibitemOpen
  \bibfield  {author} {\bibinfo {author} {\bibfnamefont {U.~D.}\ \bibnamefont
  {Giovannini}}\ and\ \bibinfo {author} {\bibfnamefont {H.}~\bibnamefont
  {Hübener}},\ }\bibfield  {title} {\bibinfo {title} {Floquet analysis of
  excitations in materials},\ }\href {https://doi.org/10.1088/2515-7639/ab387b}
  {\bibfield  {journal} {\bibinfo  {journal} {Journal of Physics: Materials}\
  }\textbf {\bibinfo {volume} {3}},\ \bibinfo {pages} {012001} (\bibinfo {year}
  {2019})}\BibitemShut {NoStop}%
\bibitem [{\citenamefont {McIver}\ \emph {et~al.}(2020)\citenamefont {McIver},
  \citenamefont {Schulte}, \citenamefont {Stein}, \citenamefont {Matsuyama},
  \citenamefont {Jotzu}, \citenamefont {Meier},\ and\ \citenamefont
  {Cavalleri}}]{McIver2020}%
  \BibitemOpen
  \bibfield  {author} {\bibinfo {author} {\bibfnamefont {J.~W.}\ \bibnamefont
  {McIver}}, \bibinfo {author} {\bibfnamefont {B.}~\bibnamefont {Schulte}},
  \bibinfo {author} {\bibfnamefont {F.-U.}\ \bibnamefont {Stein}}, \bibinfo
  {author} {\bibfnamefont {T.}~\bibnamefont {Matsuyama}}, \bibinfo {author}
  {\bibfnamefont {G.}~\bibnamefont {Jotzu}}, \bibinfo {author} {\bibfnamefont
  {G.}~\bibnamefont {Meier}},\ and\ \bibinfo {author} {\bibfnamefont
  {A.}~\bibnamefont {Cavalleri}},\ }\bibfield  {title} {\bibinfo {title}
  {Light-induced anomalous hall effect in graphene},\ }\href
  {https://doi.org/10.1038/s41567-019-0698-y} {\bibfield  {journal} {\bibinfo
  {journal} {Nature Physics}\ }\textbf {\bibinfo {volume} {16}},\ \bibinfo
  {pages} {38} (\bibinfo {year} {2020})}\BibitemShut {NoStop}%
\bibitem [{\citenamefont {Oka}\ and\ \citenamefont {Aoki}(2009)}]{oka2009}%
  \BibitemOpen
  \bibfield  {author} {\bibinfo {author} {\bibfnamefont {T.}~\bibnamefont
  {Oka}}\ and\ \bibinfo {author} {\bibfnamefont {H.}~\bibnamefont {Aoki}},\
  }\bibfield  {title} {\bibinfo {title} {Photovoltaic hall effect in
  graphene},\ }\href@noop {} {\bibfield  {journal} {\bibinfo  {journal} {Phys.
  Rev. B}\ }\textbf {\bibinfo {volume} {79}},\ \bibinfo {pages} {081406}
  (\bibinfo {year} {2009})}\BibitemShut {NoStop}%
\bibitem [{\citenamefont {Lindner}\ \emph {et~al.}(2011)\citenamefont
  {Lindner}, \citenamefont {Refael},\ and\ \citenamefont
  {Galitski}}]{lindner2011}%
  \BibitemOpen
  \bibfield  {author} {\bibinfo {author} {\bibfnamefont {N.~H.}\ \bibnamefont
  {Lindner}}, \bibinfo {author} {\bibfnamefont {G.}~\bibnamefont {Refael}},\
  and\ \bibinfo {author} {\bibfnamefont {V.}~\bibnamefont {Galitski}},\
  }\bibfield  {title} {\bibinfo {title} {Floquet topological insulator in
  semiconductor quantum wells},\ }\href@noop {} {\bibfield  {journal} {\bibinfo
   {journal} {Nat Phys}\ }\textbf {\bibinfo {volume} {7}},\ \bibinfo {pages}
  {490} (\bibinfo {year} {2011})}\BibitemShut {NoStop}%
\bibitem [{\citenamefont {Rechtsman}\ \emph {et~al.}(2013)\citenamefont
  {Rechtsman}, \citenamefont {Zeuner}, \citenamefont {Plotnik}, \citenamefont
  {Lumer}, \citenamefont {Podolsky}, \citenamefont {Dreisow}, \citenamefont
  {Nolte}, \citenamefont {Segev},\ and\ \citenamefont
  {Szameit}}]{rechtsman2013}%
  \BibitemOpen
  \bibfield  {author} {\bibinfo {author} {\bibfnamefont {M.~C.}\ \bibnamefont
  {Rechtsman}}, \bibinfo {author} {\bibfnamefont {J.~M.}\ \bibnamefont
  {Zeuner}}, \bibinfo {author} {\bibfnamefont {Y.}~\bibnamefont {Plotnik}},
  \bibinfo {author} {\bibfnamefont {Y.}~\bibnamefont {Lumer}}, \bibinfo
  {author} {\bibfnamefont {D.}~\bibnamefont {Podolsky}}, \bibinfo {author}
  {\bibfnamefont {F.}~\bibnamefont {Dreisow}}, \bibinfo {author} {\bibfnamefont
  {S.}~\bibnamefont {Nolte}}, \bibinfo {author} {\bibfnamefont
  {M.}~\bibnamefont {Segev}},\ and\ \bibinfo {author} {\bibfnamefont
  {A.}~\bibnamefont {Szameit}},\ }\bibfield  {title} {\bibinfo {title}
  {Photonic floquet topological insulators},\ }\href@noop {} {\bibfield
  {journal} {\bibinfo  {journal} {Nature}\ }\textbf {\bibinfo {volume} {496}},\
  \bibinfo {pages} {196} (\bibinfo {year} {2013})}\BibitemShut {NoStop}%
\bibitem [{\citenamefont {Rudner}\ \emph {et~al.}(2013)\citenamefont {Rudner},
  \citenamefont {Lindner}, \citenamefont {Berg},\ and\ \citenamefont
  {Levin}}]{Rudner2013}%
  \BibitemOpen
  \bibfield  {author} {\bibinfo {author} {\bibfnamefont {M.~S.}\ \bibnamefont
  {Rudner}}, \bibinfo {author} {\bibfnamefont {N.~H.}\ \bibnamefont {Lindner}},
  \bibinfo {author} {\bibfnamefont {E.}~\bibnamefont {Berg}},\ and\ \bibinfo
  {author} {\bibfnamefont {M.}~\bibnamefont {Levin}},\ }\bibfield  {title}
  {\bibinfo {title} {Anomalous edge states and the bulk-edge correspondence for
  periodically driven two-dimensional systems},\ }\href
  {https://doi.org/10.1103/PhysRevX.3.031005} {\bibfield  {journal} {\bibinfo
  {journal} {Phys. Rev. X}\ }\textbf {\bibinfo {volume} {3}},\ \bibinfo {pages}
  {031005} (\bibinfo {year} {2013})}\BibitemShut {NoStop}%
\bibitem [{\citenamefont {Inoue}\ and\ \citenamefont
  {Tanaka}(2010)}]{PhysRevLett.105.017401}%
  \BibitemOpen
  \bibfield  {author} {\bibinfo {author} {\bibfnamefont {J.-i.}\ \bibnamefont
  {Inoue}}\ and\ \bibinfo {author} {\bibfnamefont {A.}~\bibnamefont {Tanaka}},\
  }\bibfield  {title} {\bibinfo {title} {Photoinduced transition between
  conventional and topological insulators in two-dimensional electronic
  systems},\ }\href {https://doi.org/10.1103/PhysRevLett.105.017401} {\bibfield
   {journal} {\bibinfo  {journal} {Phys. Rev. Lett.}\ }\textbf {\bibinfo
  {volume} {105}},\ \bibinfo {pages} {017401} (\bibinfo {year}
  {2010})}\BibitemShut {NoStop}%
\bibitem [{\citenamefont {Peng}\ and\ \citenamefont
  {Refael}(2019)}]{PhysRevLett.123.016806}%
  \BibitemOpen
  \bibfield  {author} {\bibinfo {author} {\bibfnamefont {Y.}~\bibnamefont
  {Peng}}\ and\ \bibinfo {author} {\bibfnamefont {G.}~\bibnamefont {Refael}},\
  }\bibfield  {title} {\bibinfo {title} {Floquet second-order topological
  insulators from nonsymmorphic space-time symmetries},\ }\href
  {https://doi.org/10.1103/PhysRevLett.123.016806} {\bibfield  {journal}
  {\bibinfo  {journal} {Phys. Rev. Lett.}\ }\textbf {\bibinfo {volume} {123}},\
  \bibinfo {pages} {016806} (\bibinfo {year} {2019})}\BibitemShut {NoStop}%
\bibitem [{\citenamefont {Huang}\ and\ \citenamefont {Liu}(2020)}]{Huang_2020}%
  \BibitemOpen
  \bibfield  {author} {\bibinfo {author} {\bibfnamefont {B.}~\bibnamefont
  {Huang}}\ and\ \bibinfo {author} {\bibfnamefont {W.~V.}\ \bibnamefont
  {Liu}},\ }\bibfield  {title} {\bibinfo {title} {Floquet higher-order
  topological insulators with anomalous dynamical polarization},\ }\bibfield
  {journal} {\bibinfo  {journal} {Physical Review Letters}\ }\textbf {\bibinfo
  {volume} {124}},\ \href {https://doi.org/10.1103/physrevlett.124.216601}
  {10.1103/physrevlett.124.216601} (\bibinfo {year} {2020})\BibitemShut
  {NoStop}%
\bibitem [{\citenamefont {Bomantara}\ \emph {et~al.}(2019)\citenamefont
  {Bomantara}, \citenamefont {Zhou}, \citenamefont {Pan},\ and\ \citenamefont
  {Gong}}]{PhysRevB.99.045441}%
  \BibitemOpen
  \bibfield  {author} {\bibinfo {author} {\bibfnamefont {R.~W.}\ \bibnamefont
  {Bomantara}}, \bibinfo {author} {\bibfnamefont {L.}~\bibnamefont {Zhou}},
  \bibinfo {author} {\bibfnamefont {J.}~\bibnamefont {Pan}},\ and\ \bibinfo
  {author} {\bibfnamefont {J.}~\bibnamefont {Gong}},\ }\bibfield  {title}
  {\bibinfo {title} {Coupled-wire construction of static and floquet
  second-order topological insulators},\ }\href
  {https://doi.org/10.1103/PhysRevB.99.045441} {\bibfield  {journal} {\bibinfo
  {journal} {Phys. Rev. B}\ }\textbf {\bibinfo {volume} {99}},\ \bibinfo
  {pages} {045441} (\bibinfo {year} {2019})}\BibitemShut {NoStop}%
\bibitem [{\citenamefont {Peng}(2020)}]{PhysRevResearch.2.013124}%
  \BibitemOpen
  \bibfield  {author} {\bibinfo {author} {\bibfnamefont {Y.}~\bibnamefont
  {Peng}},\ }\bibfield  {title} {\bibinfo {title} {Floquet higher-order
  topological insulators and superconductors with space-time symmetries},\
  }\href {https://doi.org/10.1103/PhysRevResearch.2.013124} {\bibfield
  {journal} {\bibinfo  {journal} {Phys. Rev. Research}\ }\textbf {\bibinfo
  {volume} {2}},\ \bibinfo {pages} {013124} (\bibinfo {year}
  {2020})}\BibitemShut {NoStop}%
\bibitem [{\citenamefont {Pan}\ and\ \citenamefont
  {Zhou}(2020)}]{PhysRevB.102.094305}%
  \BibitemOpen
  \bibfield  {author} {\bibinfo {author} {\bibfnamefont {J.}~\bibnamefont
  {Pan}}\ and\ \bibinfo {author} {\bibfnamefont {L.}~\bibnamefont {Zhou}},\
  }\bibfield  {title} {\bibinfo {title} {Non-hermitian floquet second order
  topological insulators in periodically quenched lattices},\ }\href
  {https://doi.org/10.1103/PhysRevB.102.094305} {\bibfield  {journal} {\bibinfo
   {journal} {Phys. Rev. B}\ }\textbf {\bibinfo {volume} {102}},\ \bibinfo
  {pages} {094305} (\bibinfo {year} {2020})}\BibitemShut {NoStop}%
\bibitem [{\citenamefont {Chaudhary}\ \emph
  {et~al.}(2020{\natexlab{a}})\citenamefont {Chaudhary}, \citenamefont {Haim},
  \citenamefont {Peng},\ and\ \citenamefont
  {Refael}}]{chaudhary2019phononinduced}%
  \BibitemOpen
  \bibfield  {author} {\bibinfo {author} {\bibfnamefont {S.}~\bibnamefont
  {Chaudhary}}, \bibinfo {author} {\bibfnamefont {A.}~\bibnamefont {Haim}},
  \bibinfo {author} {\bibfnamefont {Y.}~\bibnamefont {Peng}},\ and\ \bibinfo
  {author} {\bibfnamefont {G.}~\bibnamefont {Refael}},\ }\bibfield  {title}
  {\bibinfo {title} {Phonon-induced floquet topological phases protected by
  space-time symmetries},\ }\href
  {https://doi.org/10.1103/PhysRevResearch.2.043431} {\bibfield  {journal}
  {\bibinfo  {journal} {Phys. Rev. Research}\ }\textbf {\bibinfo {volume}
  {2}},\ \bibinfo {pages} {043431} (\bibinfo {year}
  {2020}{\natexlab{a}})}\BibitemShut {NoStop}%
\bibitem [{\citenamefont {Hu}\ \emph {et~al.}(2020)\citenamefont {Hu},
  \citenamefont {Huang}, \citenamefont {Zhao},\ and\ \citenamefont
  {Liu}}]{Haiping2020}%
  \BibitemOpen
  \bibfield  {author} {\bibinfo {author} {\bibfnamefont {H.}~\bibnamefont
  {Hu}}, \bibinfo {author} {\bibfnamefont {B.}~\bibnamefont {Huang}}, \bibinfo
  {author} {\bibfnamefont {E.}~\bibnamefont {Zhao}},\ and\ \bibinfo {author}
  {\bibfnamefont {W.~V.}\ \bibnamefont {Liu}},\ }\bibfield  {title} {\bibinfo
  {title} {Dynamical singularities of floquet higher-order topological
  insulators},\ }\href {https://doi.org/10.1103/PhysRevLett.124.057001}
  {\bibfield  {journal} {\bibinfo  {journal} {Phys. Rev. Lett.}\ }\textbf
  {\bibinfo {volume} {124}},\ \bibinfo {pages} {057001} (\bibinfo {year}
  {2020})}\BibitemShut {NoStop}%
\bibitem [{\citenamefont {Kennes}\ \emph {et~al.}(2018)\citenamefont {Kennes},
  \citenamefont {de~la Torre}, \citenamefont {Ron}, \citenamefont {Hsieh},\
  and\ \citenamefont {Millis}}]{PhysRevLett.120.127601}%
  \BibitemOpen
  \bibfield  {author} {\bibinfo {author} {\bibfnamefont {D.~M.}\ \bibnamefont
  {Kennes}}, \bibinfo {author} {\bibfnamefont {A.}~\bibnamefont {de~la Torre}},
  \bibinfo {author} {\bibfnamefont {A.}~\bibnamefont {Ron}}, \bibinfo {author}
  {\bibfnamefont {D.}~\bibnamefont {Hsieh}},\ and\ \bibinfo {author}
  {\bibfnamefont {A.~J.}\ \bibnamefont {Millis}},\ }\bibfield  {title}
  {\bibinfo {title} {Floquet engineering in quantum chains},\ }\href
  {https://doi.org/10.1103/PhysRevLett.120.127601} {\bibfield  {journal}
  {\bibinfo  {journal} {Phys. Rev. Lett.}\ }\textbf {\bibinfo {volume} {120}},\
  \bibinfo {pages} {127601} (\bibinfo {year} {2018})}\BibitemShut {NoStop}%
\bibitem [{\citenamefont {D'Alessio}\ and\ \citenamefont
  {Rigol}(2014)}]{PhysRevX.4.041048}%
  \BibitemOpen
  \bibfield  {author} {\bibinfo {author} {\bibfnamefont {L.}~\bibnamefont
  {D'Alessio}}\ and\ \bibinfo {author} {\bibfnamefont {M.}~\bibnamefont
  {Rigol}},\ }\bibfield  {title} {\bibinfo {title} {Long-time behavior of
  isolated periodically driven interacting lattice systems},\ }\href
  {https://doi.org/10.1103/PhysRevX.4.041048} {\bibfield  {journal} {\bibinfo
  {journal} {Phys. Rev. X}\ }\textbf {\bibinfo {volume} {4}},\ \bibinfo {pages}
  {041048} (\bibinfo {year} {2014})}\BibitemShut {NoStop}%
\bibitem [{\citenamefont {Harper}\ and\ \citenamefont
  {Roy}(2017)}]{PhysRevLett.118.115301}%
  \BibitemOpen
  \bibfield  {author} {\bibinfo {author} {\bibfnamefont {F.}~\bibnamefont
  {Harper}}\ and\ \bibinfo {author} {\bibfnamefont {R.}~\bibnamefont {Roy}},\
  }\bibfield  {title} {\bibinfo {title} {Floquet topological order in
  interacting systems of bosons and fermions},\ }\href
  {https://doi.org/10.1103/PhysRevLett.118.115301} {\bibfield  {journal}
  {\bibinfo  {journal} {Phys. Rev. Lett.}\ }\textbf {\bibinfo {volume} {118}},\
  \bibinfo {pages} {115301} (\bibinfo {year} {2017})}\BibitemShut {NoStop}%
\bibitem [{\citenamefont {Reiss}\ \emph {et~al.}(2018)\citenamefont {Reiss},
  \citenamefont {Harper},\ and\ \citenamefont {Roy}}]{PhysRevB.98.045127}%
  \BibitemOpen
  \bibfield  {author} {\bibinfo {author} {\bibfnamefont {D.}~\bibnamefont
  {Reiss}}, \bibinfo {author} {\bibfnamefont {F.}~\bibnamefont {Harper}},\ and\
  \bibinfo {author} {\bibfnamefont {R.}~\bibnamefont {Roy}},\ }\bibfield
  {title} {\bibinfo {title} {Interacting floquet topological phases in three
  dimensions},\ }\href {https://doi.org/10.1103/PhysRevB.98.045127} {\bibfield
  {journal} {\bibinfo  {journal} {Phys. Rev. B}\ }\textbf {\bibinfo {volume}
  {98}},\ \bibinfo {pages} {045127} (\bibinfo {year} {2018})}\BibitemShut
  {NoStop}%
\bibitem [{\citenamefont {von Keyserlingk}\ and\ \citenamefont
  {Sondhi}(2016)}]{PhysRevB.93.245145}%
  \BibitemOpen
  \bibfield  {author} {\bibinfo {author} {\bibfnamefont {C.~W.}\ \bibnamefont
  {von Keyserlingk}}\ and\ \bibinfo {author} {\bibfnamefont {S.~L.}\
  \bibnamefont {Sondhi}},\ }\bibfield  {title} {\bibinfo {title} {Phase
  structure of one-dimensional interacting floquet systems. i. abelian
  symmetry-protected topological phases},\ }\href
  {https://doi.org/10.1103/PhysRevB.93.245145} {\bibfield  {journal} {\bibinfo
  {journal} {Phys. Rev. B}\ }\textbf {\bibinfo {volume} {93}},\ \bibinfo
  {pages} {245145} (\bibinfo {year} {2016})}\BibitemShut {NoStop}%
\bibitem [{\citenamefont {Kennes}\ \emph
  {et~al.}(2019{\natexlab{b}})\citenamefont {Kennes}, \citenamefont {Claassen},
  \citenamefont {Sentef},\ and\ \citenamefont {Karrasch}}]{kennes2019}%
  \BibitemOpen
  \bibfield  {author} {\bibinfo {author} {\bibfnamefont {D.~M.}\ \bibnamefont
  {Kennes}}, \bibinfo {author} {\bibfnamefont {M.}~\bibnamefont {Claassen}},
  \bibinfo {author} {\bibfnamefont {M.~A.}\ \bibnamefont {Sentef}},\ and\
  \bibinfo {author} {\bibfnamefont {C.}~\bibnamefont {Karrasch}},\ }\bibfield
  {title} {\bibinfo {title} {Light-induced $d$-wave superconductivity through
  floquet-engineered fermi surfaces in cuprates},\ }\href
  {https://doi.org/10.1103/PhysRevB.100.075115} {\bibfield  {journal} {\bibinfo
   {journal} {Phys. Rev. B}\ }\textbf {\bibinfo {volume} {100}},\ \bibinfo
  {pages} {075115} (\bibinfo {year} {2019}{\natexlab{b}})}\BibitemShut
  {NoStop}%
\bibitem [{\citenamefont {D\'ora}\ \emph {et~al.}(2012)\citenamefont {D\'ora},
  \citenamefont {Cayssol}, \citenamefont {Simon},\ and\ \citenamefont
  {Moessner}}]{PhysRevLett.108.056602}%
  \BibitemOpen
  \bibfield  {author} {\bibinfo {author} {\bibfnamefont {B.}~\bibnamefont
  {D\'ora}}, \bibinfo {author} {\bibfnamefont {J.}~\bibnamefont {Cayssol}},
  \bibinfo {author} {\bibfnamefont {F.}~\bibnamefont {Simon}},\ and\ \bibinfo
  {author} {\bibfnamefont {R.}~\bibnamefont {Moessner}},\ }\bibfield  {title}
  {\bibinfo {title} {Optically engineering the topological properties of a spin
  hall insulator},\ }\href {https://doi.org/10.1103/PhysRevLett.108.056602}
  {\bibfield  {journal} {\bibinfo  {journal} {Phys. Rev. Lett.}\ }\textbf
  {\bibinfo {volume} {108}},\ \bibinfo {pages} {056602} (\bibinfo {year}
  {2012})}\BibitemShut {NoStop}%
\bibitem [{\citenamefont {Decker}\ \emph {et~al.}(2020)\citenamefont {Decker},
  \citenamefont {Karrasch}, \citenamefont {Eisert},\ and\ \citenamefont
  {Kennes}}]{PhysRevLett.124.190601}%
  \BibitemOpen
  \bibfield  {author} {\bibinfo {author} {\bibfnamefont {K.~S.~C.}\
  \bibnamefont {Decker}}, \bibinfo {author} {\bibfnamefont {C.}~\bibnamefont
  {Karrasch}}, \bibinfo {author} {\bibfnamefont {J.}~\bibnamefont {Eisert}},\
  and\ \bibinfo {author} {\bibfnamefont {D.~M.}\ \bibnamefont {Kennes}},\
  }\bibfield  {title} {\bibinfo {title} {Floquet engineering topological
  many-body localized systems},\ }\href
  {https://doi.org/10.1103/PhysRevLett.124.190601} {\bibfield  {journal}
  {\bibinfo  {journal} {Phys. Rev. Lett.}\ }\textbf {\bibinfo {volume} {124}},\
  \bibinfo {pages} {190601} (\bibinfo {year} {2020})}\BibitemShut {NoStop}%
\bibitem [{\citenamefont {Zhang}\ \emph {et~al.}(2016)\citenamefont {Zhang},
  \citenamefont {Khemani},\ and\ \citenamefont {Huse}}]{PhysRevB.94.224202}%
  \BibitemOpen
  \bibfield  {author} {\bibinfo {author} {\bibfnamefont {L.}~\bibnamefont
  {Zhang}}, \bibinfo {author} {\bibfnamefont {V.}~\bibnamefont {Khemani}},\
  and\ \bibinfo {author} {\bibfnamefont {D.~A.}\ \bibnamefont {Huse}},\
  }\bibfield  {title} {\bibinfo {title} {A floquet model for the many-body
  localization transition},\ }\href
  {https://doi.org/10.1103/PhysRevB.94.224202} {\bibfield  {journal} {\bibinfo
  {journal} {Phys. Rev. B}\ }\textbf {\bibinfo {volume} {94}},\ \bibinfo
  {pages} {224202} (\bibinfo {year} {2016})}\BibitemShut {NoStop}%
\bibitem [{\citenamefont {Abanin}\ \emph {et~al.}(2016)\citenamefont {Abanin},
  \citenamefont {De~Roeck},\ and\ \citenamefont {Huveneers}}]{Abanin_2016}%
  \BibitemOpen
  \bibfield  {author} {\bibinfo {author} {\bibfnamefont {D.~A.}\ \bibnamefont
  {Abanin}}, \bibinfo {author} {\bibfnamefont {W.}~\bibnamefont {De~Roeck}},\
  and\ \bibinfo {author} {\bibfnamefont {F.}~\bibnamefont {Huveneers}},\
  }\bibfield  {title} {\bibinfo {title} {Theory of many-body localization in
  periodically driven systems},\ }\href
  {https://doi.org/10.1016/j.aop.2016.03.010} {\bibfield  {journal} {\bibinfo
  {journal} {Annals of Physics}\ }\textbf {\bibinfo {volume} {372}},\ \bibinfo
  {pages} {1–11} (\bibinfo {year} {2016})}\BibitemShut {NoStop}%
\bibitem [{\citenamefont {Ponte}\ \emph {et~al.}(2015)\citenamefont {Ponte},
  \citenamefont {Papi\ifmmode~\acute{c}\else \'{c}\fi{}}, \citenamefont
  {Huveneers},\ and\ \citenamefont {Abanin}}]{PhysRevLett.114.140401}%
  \BibitemOpen
  \bibfield  {author} {\bibinfo {author} {\bibfnamefont {P.}~\bibnamefont
  {Ponte}}, \bibinfo {author} {\bibfnamefont {Z.}~\bibnamefont
  {Papi\ifmmode~\acute{c}\else \'{c}\fi{}}}, \bibinfo {author} {\bibfnamefont
  {F.~m.~c.}\ \bibnamefont {Huveneers}},\ and\ \bibinfo {author} {\bibfnamefont
  {D.~A.}\ \bibnamefont {Abanin}},\ }\bibfield  {title} {\bibinfo {title}
  {Many-body localization in periodically driven systems},\ }\href
  {https://doi.org/10.1103/PhysRevLett.114.140401} {\bibfield  {journal}
  {\bibinfo  {journal} {Phys. Rev. Lett.}\ }\textbf {\bibinfo {volume} {114}},\
  \bibinfo {pages} {140401} (\bibinfo {year} {2015})}\BibitemShut {NoStop}%
\bibitem [{\citenamefont {Abanin}\ \emph {et~al.}(2019)\citenamefont {Abanin},
  \citenamefont {Altman}, \citenamefont {Bloch},\ and\ \citenamefont
  {Serbyn}}]{RevModPhys.91.021001}%
  \BibitemOpen
  \bibfield  {author} {\bibinfo {author} {\bibfnamefont {D.~A.}\ \bibnamefont
  {Abanin}}, \bibinfo {author} {\bibfnamefont {E.}~\bibnamefont {Altman}},
  \bibinfo {author} {\bibfnamefont {I.}~\bibnamefont {Bloch}},\ and\ \bibinfo
  {author} {\bibfnamefont {M.}~\bibnamefont {Serbyn}},\ }\bibfield  {title}
  {\bibinfo {title} {Colloquium: Many-body localization, thermalization, and
  entanglement},\ }\href {https://doi.org/10.1103/RevModPhys.91.021001}
  {\bibfield  {journal} {\bibinfo  {journal} {Rev. Mod. Phys.}\ }\textbf
  {\bibinfo {volume} {91}},\ \bibinfo {pages} {021001} (\bibinfo {year}
  {2019})}\BibitemShut {NoStop}%
\bibitem [{\citenamefont {Dal~Lago}\ \emph {et~al.}(2015)\citenamefont
  {Dal~Lago}, \citenamefont {Atala},\ and\ \citenamefont
  {Foa~Torres}}]{dallago2015}%
  \BibitemOpen
  \bibfield  {author} {\bibinfo {author} {\bibfnamefont {V.}~\bibnamefont
  {Dal~Lago}}, \bibinfo {author} {\bibfnamefont {M.}~\bibnamefont {Atala}},\
  and\ \bibinfo {author} {\bibfnamefont {L.~E.~F.}\ \bibnamefont
  {Foa~Torres}},\ }\bibfield  {title} {\bibinfo {title} {Floquet topological
  transitions in a driven one-dimensional topological insulator},\ }\href@noop
  {} {\bibfield  {journal} {\bibinfo  {journal} {Phys. Rev. A}\ }\textbf
  {\bibinfo {volume} {92}},\ \bibinfo {pages} {023624} (\bibinfo {year}
  {2015})}\BibitemShut {NoStop}%
\bibitem [{\citenamefont {Calvo}\ \emph {et~al.}(2015)\citenamefont {Calvo},
  \citenamefont {Foa~Torres}, \citenamefont {Perez-Piskunow}, \citenamefont
  {Balseiro},\ and\ \citenamefont {Usaj}}]{PhysRevB.91.241404}%
  \BibitemOpen
  \bibfield  {author} {\bibinfo {author} {\bibfnamefont {H.~L.}\ \bibnamefont
  {Calvo}}, \bibinfo {author} {\bibfnamefont {L.~E.~F.}\ \bibnamefont
  {Foa~Torres}}, \bibinfo {author} {\bibfnamefont {P.~M.}\ \bibnamefont
  {Perez-Piskunow}}, \bibinfo {author} {\bibfnamefont {C.~A.}\ \bibnamefont
  {Balseiro}},\ and\ \bibinfo {author} {\bibfnamefont {G.}~\bibnamefont
  {Usaj}},\ }\bibfield  {title} {\bibinfo {title} {Floquet interface states in
  illuminated three-dimensional topological insulators},\ }\href
  {https://doi.org/10.1103/PhysRevB.91.241404} {\bibfield  {journal} {\bibinfo
  {journal} {Phys. Rev. B}\ }\textbf {\bibinfo {volume} {91}},\ \bibinfo
  {pages} {241404} (\bibinfo {year} {2015})}\BibitemShut {NoStop}%
\bibitem [{\citenamefont {Asb\'oth}\ \emph {et~al.}(2014)\citenamefont
  {Asb\'oth}, \citenamefont {Tarasinski},\ and\ \citenamefont
  {Delplace}}]{asboth2014}%
  \BibitemOpen
  \bibfield  {author} {\bibinfo {author} {\bibfnamefont {J.~K.}\ \bibnamefont
  {Asb\'oth}}, \bibinfo {author} {\bibfnamefont {B.}~\bibnamefont
  {Tarasinski}},\ and\ \bibinfo {author} {\bibfnamefont {P.}~\bibnamefont
  {Delplace}},\ }\bibfield  {title} {\bibinfo {title} {Chiral symmetry and
  bulk-boundary correspondence in periodically driven one-dimensional
  systems},\ }\href@noop {} {\bibfield  {journal} {\bibinfo  {journal} {Phys.
  Rev. B}\ }\textbf {\bibinfo {volume} {90}},\ \bibinfo {pages} {125143}
  (\bibinfo {year} {2014})}\BibitemShut {NoStop}%
\bibitem [{\citenamefont {Calvo}\ \emph {et~al.}(2011)\citenamefont {Calvo},
  \citenamefont {Pastawski}, \citenamefont {Roche},\ and\ \citenamefont
  {Torres}}]{calvo2011}%
  \BibitemOpen
  \bibfield  {author} {\bibinfo {author} {\bibfnamefont {H.~L.}\ \bibnamefont
  {Calvo}}, \bibinfo {author} {\bibfnamefont {H.~M.}\ \bibnamefont
  {Pastawski}}, \bibinfo {author} {\bibfnamefont {S.}~\bibnamefont {Roche}},\
  and\ \bibinfo {author} {\bibfnamefont {L.~E. F.~F.}\ \bibnamefont {Torres}},\
  }\bibfield  {title} {\bibinfo {title} {Tuning laser-induced band gaps in
  graphene},\ }\href {https://doi.org/10.1063/1.3597412} {\bibfield  {journal}
  {\bibinfo  {journal} {Applied Physics Letters}\ }\textbf {\bibinfo {volume}
  {98}},\ \bibinfo {pages} {232103} (\bibinfo {year} {2011})},\ \Eprint
  {https://arxiv.org/abs/https://doi.org/10.1063/1.3597412}
  {https://doi.org/10.1063/1.3597412} \BibitemShut {NoStop}%
\bibitem [{\citenamefont {Rodriguez-Vega}\ \emph
  {et~al.}(2021{\natexlab{a}})\citenamefont {Rodriguez-Vega}, \citenamefont
  {Vogl},\ and\ \citenamefont {Fiete}}]{PhysRevB.104.245135}%
  \BibitemOpen
  \bibfield  {author} {\bibinfo {author} {\bibfnamefont {M.}~\bibnamefont
  {Rodriguez-Vega}}, \bibinfo {author} {\bibfnamefont {M.}~\bibnamefont
  {Vogl}},\ and\ \bibinfo {author} {\bibfnamefont {G.~A.}\ \bibnamefont
  {Fiete}},\ }\bibfield  {title} {\bibinfo {title} {Direct driving of
  electronic and phononic degrees of freedom in a honeycomb bilayer with
  infrared light},\ }\href {https://doi.org/10.1103/PhysRevB.104.245135}
  {\bibfield  {journal} {\bibinfo  {journal} {Phys. Rev. B}\ }\textbf {\bibinfo
  {volume} {104}},\ \bibinfo {pages} {245135} (\bibinfo {year}
  {2021}{\natexlab{a}})}\BibitemShut {NoStop}%
\bibitem [{\citenamefont {Gu}\ \emph {et~al.}(2011)\citenamefont {Gu},
  \citenamefont {Fertig}, \citenamefont {Arovas},\ and\ \citenamefont
  {Auerbach}}]{zhenghao2011}%
  \BibitemOpen
  \bibfield  {author} {\bibinfo {author} {\bibfnamefont {Z.}~\bibnamefont
  {Gu}}, \bibinfo {author} {\bibfnamefont {H.~A.}\ \bibnamefont {Fertig}},
  \bibinfo {author} {\bibfnamefont {D.~P.}\ \bibnamefont {Arovas}},\ and\
  \bibinfo {author} {\bibfnamefont {A.}~\bibnamefont {Auerbach}},\ }\bibfield
  {title} {\bibinfo {title} {Floquet spectrum and transport through an
  irradiated graphene ribbon},\ }\href@noop {} {\bibfield  {journal} {\bibinfo
  {journal} {Phys. Rev. Lett.}\ }\textbf {\bibinfo {volume} {107}},\ \bibinfo
  {pages} {216601} (\bibinfo {year} {2011})}\BibitemShut {NoStop}%
\bibitem [{\citenamefont {Syzranov}\ \emph {et~al.}(2013)\citenamefont
  {Syzranov}, \citenamefont {Rodionov}, \citenamefont {Kugel},\ and\
  \citenamefont {Nori}}]{PhysRevB.88.241112}%
  \BibitemOpen
  \bibfield  {author} {\bibinfo {author} {\bibfnamefont {S.~V.}\ \bibnamefont
  {Syzranov}}, \bibinfo {author} {\bibfnamefont {Y.~I.}\ \bibnamefont
  {Rodionov}}, \bibinfo {author} {\bibfnamefont {K.~I.}\ \bibnamefont
  {Kugel}},\ and\ \bibinfo {author} {\bibfnamefont {F.}~\bibnamefont {Nori}},\
  }\bibfield  {title} {\bibinfo {title} {Strongly anisotropic dirac
  quasiparticles in irradiated graphene},\ }\href
  {https://doi.org/10.1103/PhysRevB.88.241112} {\bibfield  {journal} {\bibinfo
  {journal} {Phys. Rev. B}\ }\textbf {\bibinfo {volume} {88}},\ \bibinfo
  {pages} {241112} (\bibinfo {year} {2013})}\BibitemShut {NoStop}%
\bibitem [{\citenamefont {Perez-Piskunow}\ \emph {et~al.}(2014)\citenamefont
  {Perez-Piskunow}, \citenamefont {Usaj}, \citenamefont {Balseiro},\ and\
  \citenamefont {Torres}}]{perez2014}%
  \BibitemOpen
  \bibfield  {author} {\bibinfo {author} {\bibfnamefont {P.~M.}\ \bibnamefont
  {Perez-Piskunow}}, \bibinfo {author} {\bibfnamefont {G.}~\bibnamefont
  {Usaj}}, \bibinfo {author} {\bibfnamefont {C.~A.}\ \bibnamefont {Balseiro}},\
  and\ \bibinfo {author} {\bibfnamefont {L.~E. F.~F.}\ \bibnamefont {Torres}},\
  }\bibfield  {title} {\bibinfo {title} {Floquet chiral edge states in
  graphene},\ }\href@noop {} {\bibfield  {journal} {\bibinfo  {journal} {Phys.
  Rev. B}\ }\textbf {\bibinfo {volume} {89}},\ \bibinfo {pages} {121401}
  (\bibinfo {year} {2014})}\BibitemShut {NoStop}%
\bibitem [{\citenamefont {Kundu}\ \emph {et~al.}(2014)\citenamefont {Kundu},
  \citenamefont {Fertig},\ and\ \citenamefont {Seradjeh}}]{kundu2014}%
  \BibitemOpen
  \bibfield  {author} {\bibinfo {author} {\bibfnamefont {A.}~\bibnamefont
  {Kundu}}, \bibinfo {author} {\bibfnamefont {H.~A.}\ \bibnamefont {Fertig}},\
  and\ \bibinfo {author} {\bibfnamefont {B.}~\bibnamefont {Seradjeh}},\
  }\bibfield  {title} {\bibinfo {title} {Effective theory of floquet
  topological transitions},\ }\href
  {https://doi.org/10.1103/PhysRevLett.113.236803} {\bibfield  {journal}
  {\bibinfo  {journal} {Phys. Rev. Lett.}\ }\textbf {\bibinfo {volume} {113}},\
  \bibinfo {pages} {236803} (\bibinfo {year} {2014})}\BibitemShut {NoStop}%
\bibitem [{\citenamefont {Sentef}\ \emph {et~al.}(2015)\citenamefont {Sentef},
  \citenamefont {Claassen}, \citenamefont {Kemper}, \citenamefont {Moritz},
  \citenamefont {Oka}, \citenamefont {Freericks},\ and\ \citenamefont
  {Devereaux}}]{Sentef2015}%
  \BibitemOpen
  \bibfield  {author} {\bibinfo {author} {\bibfnamefont {M.~A.}\ \bibnamefont
  {Sentef}}, \bibinfo {author} {\bibfnamefont {M.}~\bibnamefont {Claassen}},
  \bibinfo {author} {\bibfnamefont {A.~F.}\ \bibnamefont {Kemper}}, \bibinfo
  {author} {\bibfnamefont {B.}~\bibnamefont {Moritz}}, \bibinfo {author}
  {\bibfnamefont {T.}~\bibnamefont {Oka}}, \bibinfo {author} {\bibfnamefont
  {J.~K.}\ \bibnamefont {Freericks}},\ and\ \bibinfo {author} {\bibfnamefont
  {T.~P.}\ \bibnamefont {Devereaux}},\ }\bibfield  {title} {\bibinfo {title}
  {Theory of floquet band formation and local pseudospin textures in pump-probe
  photoemission of graphene},\ }\href {https://doi.org/10.1038/ncomms8047}
  {\bibfield  {journal} {\bibinfo  {journal} {Nature Communications}\ }\textbf
  {\bibinfo {volume} {6}},\ \bibinfo {pages} {7047} (\bibinfo {year}
  {2015})}\BibitemShut {NoStop}%
\bibitem [{\citenamefont {Roman-Taboada}\ and\ \citenamefont
  {Naumis}(2017)}]{Roman-Taboada2017}%
  \BibitemOpen
  \bibfield  {author} {\bibinfo {author} {\bibfnamefont {P.}~\bibnamefont
  {Roman-Taboada}}\ and\ \bibinfo {author} {\bibfnamefont {G.~G.}\ \bibnamefont
  {Naumis}},\ }\bibfield  {title} {\bibinfo {title} {Topological flat bands in
  time-periodically driven uniaxial strained graphene nanoribbons},\ }\href
  {https://doi.org/10.1103/PhysRevB.95.115440} {\bibfield  {journal} {\bibinfo
  {journal} {Phys. Rev. B}\ }\textbf {\bibinfo {volume} {95}},\ \bibinfo
  {pages} {115440} (\bibinfo {year} {2017})}\BibitemShut {NoStop}%
\bibitem [{\citenamefont {Dehghani}\ \emph {et~al.}(2015)\citenamefont
  {Dehghani}, \citenamefont {Oka},\ and\ \citenamefont {Mitra}}]{Dehghani2015}%
  \BibitemOpen
  \bibfield  {author} {\bibinfo {author} {\bibfnamefont {H.}~\bibnamefont
  {Dehghani}}, \bibinfo {author} {\bibfnamefont {T.}~\bibnamefont {Oka}},\ and\
  \bibinfo {author} {\bibfnamefont {A.}~\bibnamefont {Mitra}},\ }\bibfield
  {title} {\bibinfo {title} {Out-of-equilibrium electrons and the hall
  conductance of a floquet topological insulator},\ }\href
  {https://doi.org/10.1103/PhysRevB.91.155422} {\bibfield  {journal} {\bibinfo
  {journal} {Phys. Rev. B}\ }\textbf {\bibinfo {volume} {91}},\ \bibinfo
  {pages} {155422} (\bibinfo {year} {2015})}\BibitemShut {NoStop}%
\bibitem [{\citenamefont {Kitagawa}\ \emph {et~al.}(2011)\citenamefont
  {Kitagawa}, \citenamefont {Oka}, \citenamefont {Brataas}, \citenamefont
  {Fu},\ and\ \citenamefont {Demler}}]{kitagawa2011}%
  \BibitemOpen
  \bibfield  {author} {\bibinfo {author} {\bibfnamefont {T.}~\bibnamefont
  {Kitagawa}}, \bibinfo {author} {\bibfnamefont {T.}~\bibnamefont {Oka}},
  \bibinfo {author} {\bibfnamefont {A.}~\bibnamefont {Brataas}}, \bibinfo
  {author} {\bibfnamefont {L.}~\bibnamefont {Fu}},\ and\ \bibinfo {author}
  {\bibfnamefont {E.}~\bibnamefont {Demler}},\ }\bibfield  {title} {\bibinfo
  {title} {Transport properties of nonequilibrium systems under the application
  of light: Photoinduced quantum hall insulators without landau levels},\
  }\href@noop {} {\bibfield  {journal} {\bibinfo  {journal} {Phys. Rev. B}\
  }\textbf {\bibinfo {volume} {84}},\ \bibinfo {pages} {235108} (\bibinfo
  {year} {2011})}\BibitemShut {NoStop}%
\bibitem [{\citenamefont {Usaj}\ \emph {et~al.}(2014)\citenamefont {Usaj},
  \citenamefont {Perez-Piskunow}, \citenamefont {Foa~Torres},\ and\
  \citenamefont {Balseiro}}]{Usaj2014}%
  \BibitemOpen
  \bibfield  {author} {\bibinfo {author} {\bibfnamefont {G.}~\bibnamefont
  {Usaj}}, \bibinfo {author} {\bibfnamefont {P.~M.}\ \bibnamefont
  {Perez-Piskunow}}, \bibinfo {author} {\bibfnamefont {L.~E.~F.}\ \bibnamefont
  {Foa~Torres}},\ and\ \bibinfo {author} {\bibfnamefont {C.~A.}\ \bibnamefont
  {Balseiro}},\ }\bibfield  {title} {\bibinfo {title} {Irradiated graphene as a
  tunable floquet topological insulator},\ }\href
  {https://doi.org/10.1103/PhysRevB.90.115423} {\bibfield  {journal} {\bibinfo
  {journal} {Phys. Rev. B}\ }\textbf {\bibinfo {volume} {90}},\ \bibinfo
  {pages} {115423} (\bibinfo {year} {2014})}\BibitemShut {NoStop}%
\bibitem [{\citenamefont {Perez-Piskunow}\ \emph {et~al.}(2015)\citenamefont
  {Perez-Piskunow}, \citenamefont {Foa~Torres},\ and\ \citenamefont
  {Usaj}}]{PhysRevA.91.043625}%
  \BibitemOpen
  \bibfield  {author} {\bibinfo {author} {\bibfnamefont {P.~M.}\ \bibnamefont
  {Perez-Piskunow}}, \bibinfo {author} {\bibfnamefont {L.~E.~F.}\ \bibnamefont
  {Foa~Torres}},\ and\ \bibinfo {author} {\bibfnamefont {G.}~\bibnamefont
  {Usaj}},\ }\bibfield  {title} {\bibinfo {title} {Hierarchy of floquet gaps
  and edge states for driven honeycomb lattices},\ }\href
  {https://doi.org/10.1103/PhysRevA.91.043625} {\bibfield  {journal} {\bibinfo
  {journal} {Phys. Rev. A}\ }\textbf {\bibinfo {volume} {91}},\ \bibinfo
  {pages} {043625} (\bibinfo {year} {2015})}\BibitemShut {NoStop}%
\bibitem [{\citenamefont {Bhattacharya}\ \emph {et~al.}(2022)\citenamefont
  {Bhattacharya}, \citenamefont {Chaudhary}, \citenamefont {Grass},
  \citenamefont {Johnson}, \citenamefont {Wall},\ and\ \citenamefont
  {Lewenstein}}]{bhattacharya2020fermionic}%
  \BibitemOpen
  \bibfield  {author} {\bibinfo {author} {\bibfnamefont {U.}~\bibnamefont
  {Bhattacharya}}, \bibinfo {author} {\bibfnamefont {S.}~\bibnamefont
  {Chaudhary}}, \bibinfo {author} {\bibfnamefont {T.}~\bibnamefont {Grass}},
  \bibinfo {author} {\bibfnamefont {A.~S.}\ \bibnamefont {Johnson}}, \bibinfo
  {author} {\bibfnamefont {S.}~\bibnamefont {Wall}},\ and\ \bibinfo {author}
  {\bibfnamefont {M.}~\bibnamefont {Lewenstein}},\ }\bibfield  {title}
  {\bibinfo {title} {Fermionic chern insulator from twisted light with linear
  polarization},\ }\href {https://doi.org/10.1103/PhysRevB.105.L081406}
  {\bibfield  {journal} {\bibinfo  {journal} {Phys. Rev. B}\ }\textbf {\bibinfo
  {volume} {105}},\ \bibinfo {pages} {L081406} (\bibinfo {year}
  {2022})}\BibitemShut {NoStop}%
\bibitem [{\citenamefont {Mentink}\ and\ \citenamefont
  {Eckstein}(2014)}]{mentink2014}%
  \BibitemOpen
  \bibfield  {author} {\bibinfo {author} {\bibfnamefont {J.~H.}\ \bibnamefont
  {Mentink}}\ and\ \bibinfo {author} {\bibfnamefont {M.}~\bibnamefont
  {Eckstein}},\ }\bibfield  {title} {\bibinfo {title} {Ultrafast quenching of
  the exchange interaction in a mott insulator},\ }\href
  {https://doi.org/10.1103/PhysRevLett.113.057201} {\bibfield  {journal}
  {\bibinfo  {journal} {Phys. Rev. Lett.}\ }\textbf {\bibinfo {volume} {113}},\
  \bibinfo {pages} {057201} (\bibinfo {year} {2014})}\BibitemShut {NoStop}%
\bibitem [{\citenamefont {Mentink}\ \emph {et~al.}(2015)\citenamefont
  {Mentink}, \citenamefont {Balzer},\ and\ \citenamefont
  {Eckstein}}]{mentink2015ultrafast}%
  \BibitemOpen
  \bibfield  {author} {\bibinfo {author} {\bibfnamefont {J.}~\bibnamefont
  {Mentink}}, \bibinfo {author} {\bibfnamefont {K.}~\bibnamefont {Balzer}},\
  and\ \bibinfo {author} {\bibfnamefont {M.}~\bibnamefont {Eckstein}},\
  }\bibfield  {title} {\bibinfo {title} {Ultrafast and reversible control of
  the exchange interaction in mott insulators},\ }\href
  {https://www.nature.com/articles/ncomms7708} {\bibfield  {journal} {\bibinfo
  {journal} {Nature communications}\ }\textbf {\bibinfo {volume} {6}},\
  \bibinfo {pages} {1} (\bibinfo {year} {2015})}\BibitemShut {NoStop}%
\bibitem [{\citenamefont {Mentink}(2017)}]{mentink2017manipulating}%
  \BibitemOpen
  \bibfield  {author} {\bibinfo {author} {\bibfnamefont {J.}~\bibnamefont
  {Mentink}},\ }\bibfield  {title} {\bibinfo {title} {Manipulating magnetism by
  ultrafast control of the exchange interaction},\ }\href
  {https://iopscience.iop.org/article/10.1088/1361-648X/aa8abf/meta} {\bibfield
   {journal} {\bibinfo  {journal} {Journal of Physics: Condensed Matter}\
  }\textbf {\bibinfo {volume} {29}},\ \bibinfo {pages} {453001} (\bibinfo
  {year} {2017})}\BibitemShut {NoStop}%
\bibitem [{\citenamefont {Liu}\ \emph {et~al.}(2018)\citenamefont {Liu},
  \citenamefont {Hejazi},\ and\ \citenamefont {Balents}}]{hejazi2018}%
  \BibitemOpen
  \bibfield  {author} {\bibinfo {author} {\bibfnamefont {J.}~\bibnamefont
  {Liu}}, \bibinfo {author} {\bibfnamefont {K.}~\bibnamefont {Hejazi}},\ and\
  \bibinfo {author} {\bibfnamefont {L.}~\bibnamefont {Balents}},\ }\bibfield
  {title} {\bibinfo {title} {Floquet engineering of multiorbital mott
  insulators: Applications to orthorhombic titanates},\ }\href
  {https://doi.org/10.1103/PhysRevLett.121.107201} {\bibfield  {journal}
  {\bibinfo  {journal} {Phys. Rev. Lett.}\ }\textbf {\bibinfo {volume} {121}},\
  \bibinfo {pages} {107201} (\bibinfo {year} {2018})}\BibitemShut {NoStop}%
\bibitem [{\citenamefont {Hejazi}\ \emph {et~al.}(2019)\citenamefont {Hejazi},
  \citenamefont {Liu},\ and\ \citenamefont {Balents}}]{hejazi2019}%
  \BibitemOpen
  \bibfield  {author} {\bibinfo {author} {\bibfnamefont {K.}~\bibnamefont
  {Hejazi}}, \bibinfo {author} {\bibfnamefont {J.}~\bibnamefont {Liu}},\ and\
  \bibinfo {author} {\bibfnamefont {L.}~\bibnamefont {Balents}},\ }\bibfield
  {title} {\bibinfo {title} {Floquet spin and spin-orbital hamiltonians and
  doublon-holon generations in periodically driven mott insulators},\ }\href
  {https://doi.org/10.1103/PhysRevB.99.205111} {\bibfield  {journal} {\bibinfo
  {journal} {Phys. Rev. B}\ }\textbf {\bibinfo {volume} {99}},\ \bibinfo
  {pages} {205111} (\bibinfo {year} {2019})}\BibitemShut {NoStop}%
\bibitem [{\citenamefont {Quito}\ and\ \citenamefont
  {Flint}(2021{\natexlab{a}})}]{Quito2021}%
  \BibitemOpen
  \bibfield  {author} {\bibinfo {author} {\bibfnamefont {V.~L.}\ \bibnamefont
  {Quito}}\ and\ \bibinfo {author} {\bibfnamefont {R.}~\bibnamefont {Flint}},\
  }\bibfield  {title} {\bibinfo {title} {Floquet engineering correlated
  materials with unpolarized light},\ }\href
  {https://doi.org/10.1103/PhysRevLett.126.177201} {\bibfield  {journal}
  {\bibinfo  {journal} {Phys. Rev. Lett.}\ }\textbf {\bibinfo {volume} {126}},\
  \bibinfo {pages} {177201} (\bibinfo {year} {2021}{\natexlab{a}})}\BibitemShut
  {NoStop}%
\bibitem [{\citenamefont {Quito}\ and\ \citenamefont
  {Flint}(2021{\natexlab{b}})}]{Quito2021b}%
  \BibitemOpen
  \bibfield  {author} {\bibinfo {author} {\bibfnamefont {V.~L.}\ \bibnamefont
  {Quito}}\ and\ \bibinfo {author} {\bibfnamefont {R.}~\bibnamefont {Flint}},\
  }\bibfield  {title} {\bibinfo {title} {Polarization as a tuning parameter for
  floquet engineering: Magnetism in the honeycomb, square, and triangular mott
  insulators},\ }\href {https://doi.org/10.1103/PhysRevB.103.134435} {\bibfield
   {journal} {\bibinfo  {journal} {Phys. Rev. B}\ }\textbf {\bibinfo {volume}
  {103}},\ \bibinfo {pages} {134435} (\bibinfo {year}
  {2021}{\natexlab{b}})}\BibitemShut {NoStop}%
\bibitem [{\citenamefont {Chaudhary}\ \emph {et~al.}(2019)\citenamefont
  {Chaudhary}, \citenamefont {Hsieh},\ and\ \citenamefont
  {Refael}}]{Chaudhary2019orbital}%
  \BibitemOpen
  \bibfield  {author} {\bibinfo {author} {\bibfnamefont {S.}~\bibnamefont
  {Chaudhary}}, \bibinfo {author} {\bibfnamefont {D.}~\bibnamefont {Hsieh}},\
  and\ \bibinfo {author} {\bibfnamefont {G.}~\bibnamefont {Refael}},\
  }\bibfield  {title} {\bibinfo {title} {Orbital floquet engineering of
  exchange interactions in magnetic materials},\ }\href
  {https://doi.org/10.1103/PhysRevB.100.220403} {\bibfield  {journal} {\bibinfo
   {journal} {Phys. Rev. B}\ }\textbf {\bibinfo {volume} {100}},\ \bibinfo
  {pages} {220403} (\bibinfo {year} {2019})}\BibitemShut {NoStop}%
\bibitem [{\citenamefont {Chaudhary}\ \emph
  {et~al.}(2020{\natexlab{b}})\citenamefont {Chaudhary}, \citenamefont {Ron},
  \citenamefont {Hsieh},\ and\ \citenamefont
  {Refael}}]{chaudhary2020controlling}%
  \BibitemOpen
  \bibfield  {author} {\bibinfo {author} {\bibfnamefont {S.}~\bibnamefont
  {Chaudhary}}, \bibinfo {author} {\bibfnamefont {A.}~\bibnamefont {Ron}},
  \bibinfo {author} {\bibfnamefont {D.}~\bibnamefont {Hsieh}},\ and\ \bibinfo
  {author} {\bibfnamefont {G.}~\bibnamefont {Refael}},\ }\bibfield  {title}
  {\bibinfo {title} {Controlling ligand-mediated exchange interactions in
  periodically driven magnetic materials},\ }\href
  {https://arxiv.org/abs/2009.00813} {\bibfield  {journal} {\bibinfo  {journal}
  {arXiv preprint arXiv:2009.00813}\ } (\bibinfo {year}
  {2020}{\natexlab{b}})}\BibitemShut {NoStop}%
\bibitem [{\citenamefont {Rodriguez-Vega}\ \emph
  {et~al.}(2021{\natexlab{b}})\citenamefont {Rodriguez-Vega}, \citenamefont
  {Vogl},\ and\ \citenamefont {Fiete}}]{RODRIGUEZVEGA2021168434}%
  \BibitemOpen
  \bibfield  {author} {\bibinfo {author} {\bibfnamefont {M.}~\bibnamefont
  {Rodriguez-Vega}}, \bibinfo {author} {\bibfnamefont {M.}~\bibnamefont
  {Vogl}},\ and\ \bibinfo {author} {\bibfnamefont {G.~A.}\ \bibnamefont
  {Fiete}},\ }\bibfield  {title} {\bibinfo {title} {Low-frequency and
  moiré–floquet engineering: A review},\ }\href
  {https://doi.org/https://doi.org/10.1016/j.aop.2021.168434} {\bibfield
  {journal} {\bibinfo  {journal} {Annals of Physics}\ }\textbf {\bibinfo
  {volume} {435}},\ \bibinfo {pages} {168434} (\bibinfo {year}
  {2021}{\natexlab{b}})},\ \bibinfo {note} {special issue on Philip W.
  Anderson}\BibitemShut {NoStop}%
\bibitem [{\citenamefont {Bao}\ \emph {et~al.}(2021)\citenamefont {Bao},
  \citenamefont {Tang}, \citenamefont {Sun},\ and\ \citenamefont
  {Zhou}}]{bao2021light}%
  \BibitemOpen
  \bibfield  {author} {\bibinfo {author} {\bibfnamefont {C.}~\bibnamefont
  {Bao}}, \bibinfo {author} {\bibfnamefont {P.}~\bibnamefont {Tang}}, \bibinfo
  {author} {\bibfnamefont {D.}~\bibnamefont {Sun}},\ and\ \bibinfo {author}
  {\bibfnamefont {S.}~\bibnamefont {Zhou}},\ }\bibfield  {title} {\bibinfo
  {title} {Light-induced emergent phenomena in 2d materials and topological
  materials},\ }\href@noop {} {\bibfield  {journal} {\bibinfo  {journal}
  {Nature Reviews Physics}\ ,\ \bibinfo {pages} {1}} (\bibinfo {year}
  {2021})}\BibitemShut {NoStop}%
\bibitem [{\citenamefont {Assi}\ \emph {et~al.}(2021)\citenamefont {Assi},
  \citenamefont {LeBlanc}, \citenamefont {Rodriguez-Vega}, \citenamefont
  {Bahlouli},\ and\ \citenamefont {Vogl}}]{ibsal2021}%
  \BibitemOpen
  \bibfield  {author} {\bibinfo {author} {\bibfnamefont {I.~A.}\ \bibnamefont
  {Assi}}, \bibinfo {author} {\bibfnamefont {J.~P.~F.}\ \bibnamefont
  {LeBlanc}}, \bibinfo {author} {\bibfnamefont {M.}~\bibnamefont
  {Rodriguez-Vega}}, \bibinfo {author} {\bibfnamefont {H.}~\bibnamefont
  {Bahlouli}},\ and\ \bibinfo {author} {\bibfnamefont {M.}~\bibnamefont
  {Vogl}},\ }\bibfield  {title} {\bibinfo {title} {Floquet engineering and
  nonequilibrium topological maps in twisted trilayer graphene},\ }\bibfield
  {journal} {\bibinfo  {journal} {Physical Review B}\ }\textbf {\bibinfo
  {volume} {104}},\ \href {https://doi.org/10.1103/physrevb.104.195429}
  {10.1103/physrevb.104.195429} (\bibinfo {year} {2021})\BibitemShut {NoStop}%
\bibitem [{\citenamefont {Vogl}\ \emph {et~al.}(2021)\citenamefont {Vogl},
  \citenamefont {Rodriguez-Vega}, \citenamefont {Flebus}, \citenamefont
  {MacDonald},\ and\ \citenamefont {Fiete}}]{floquetTMDV2021}%
  \BibitemOpen
  \bibfield  {author} {\bibinfo {author} {\bibfnamefont {M.}~\bibnamefont
  {Vogl}}, \bibinfo {author} {\bibfnamefont {M.}~\bibnamefont
  {Rodriguez-Vega}}, \bibinfo {author} {\bibfnamefont {B.}~\bibnamefont
  {Flebus}}, \bibinfo {author} {\bibfnamefont {A.~H.}\ \bibnamefont
  {MacDonald}},\ and\ \bibinfo {author} {\bibfnamefont {G.~A.}\ \bibnamefont
  {Fiete}},\ }\bibfield  {title} {\bibinfo {title} {Floquet engineering of
  topological transitions in a twisted transition metal dichalcogenide
  homobilayer},\ }\bibfield  {journal} {\bibinfo  {journal} {Physical Review
  B}\ }\textbf {\bibinfo {volume} {103}},\ \href
  {https://doi.org/10.1103/physrevb.103.014310} {10.1103/physrevb.103.014310}
  (\bibinfo {year} {2021})\BibitemShut {NoStop}%
\bibitem [{\citenamefont {Topp}\ \emph {et~al.}(2021)\citenamefont {Topp},
  \citenamefont {Eckhardt}, \citenamefont {Kennes}, \citenamefont {Sentef},\
  and\ \citenamefont {Törmä}}]{sentef2021}%
  \BibitemOpen
  \bibfield  {author} {\bibinfo {author} {\bibfnamefont {G.~E.}\ \bibnamefont
  {Topp}}, \bibinfo {author} {\bibfnamefont {C.~J.}\ \bibnamefont {Eckhardt}},
  \bibinfo {author} {\bibfnamefont {D.~M.}\ \bibnamefont {Kennes}}, \bibinfo
  {author} {\bibfnamefont {M.~A.}\ \bibnamefont {Sentef}},\ and\ \bibinfo
  {author} {\bibfnamefont {P.}~\bibnamefont {Törmä}},\ }\bibfield  {title}
  {\bibinfo {title} {Light-matter coupling and quantum geometry in moiré
  materials},\ }\bibfield  {journal} {\bibinfo  {journal} {Physical Review B}\
  }\textbf {\bibinfo {volume} {104}},\ \href
  {https://doi.org/10.1103/physrevb.104.064306} {10.1103/physrevb.104.064306}
  (\bibinfo {year} {2021})\BibitemShut {NoStop}%
\bibitem [{\citenamefont {Topp}\ \emph {et~al.}(2019)\citenamefont {Topp},
  \citenamefont {Jotzu}, \citenamefont {McIver}, \citenamefont {Xian},
  \citenamefont {Rubio},\ and\ \citenamefont {Sentef}}]{sentef2019}%
  \BibitemOpen
  \bibfield  {author} {\bibinfo {author} {\bibfnamefont {G.~E.}\ \bibnamefont
  {Topp}}, \bibinfo {author} {\bibfnamefont {G.}~\bibnamefont {Jotzu}},
  \bibinfo {author} {\bibfnamefont {J.~W.}\ \bibnamefont {McIver}}, \bibinfo
  {author} {\bibfnamefont {L.}~\bibnamefont {Xian}}, \bibinfo {author}
  {\bibfnamefont {A.}~\bibnamefont {Rubio}},\ and\ \bibinfo {author}
  {\bibfnamefont {M.~A.}\ \bibnamefont {Sentef}},\ }\bibfield  {title}
  {\bibinfo {title} {Topological floquet engineering of twisted bilayer
  graphene},\ }\bibfield  {journal} {\bibinfo  {journal} {Physical Review
  Research}\ }\textbf {\bibinfo {volume} {1}},\ \href
  {https://doi.org/10.1103/physrevresearch.1.023031}
  {10.1103/physrevresearch.1.023031} (\bibinfo {year} {2019})\BibitemShut
  {NoStop}%
\bibitem [{\citenamefont {Lu}\ \emph {et~al.}(2021)\citenamefont {Lu},
  \citenamefont {Zeng}, \citenamefont {Liu}, \citenamefont {Gao},\ and\
  \citenamefont {Xie}}]{luzeng2021}%
  \BibitemOpen
  \bibfield  {author} {\bibinfo {author} {\bibfnamefont {M.}~\bibnamefont
  {Lu}}, \bibinfo {author} {\bibfnamefont {J.}~\bibnamefont {Zeng}}, \bibinfo
  {author} {\bibfnamefont {H.}~\bibnamefont {Liu}}, \bibinfo {author}
  {\bibfnamefont {J.-H.}\ \bibnamefont {Gao}},\ and\ \bibinfo {author}
  {\bibfnamefont {X.~C.}\ \bibnamefont {Xie}},\ }\bibfield  {title} {\bibinfo
  {title} {Valley-selective floquet chern flat bands in twisted multilayer
  graphene},\ }\bibfield  {journal} {\bibinfo  {journal} {Physical Review B}\
  }\textbf {\bibinfo {volume} {103}},\ \href
  {https://doi.org/10.1103/physrevb.103.195146} {10.1103/physrevb.103.195146}
  (\bibinfo {year} {2021})\BibitemShut {NoStop}%
\bibitem [{\citenamefont {Rodriguez-Vega}\ \emph {et~al.}(2020)\citenamefont
  {Rodriguez-Vega}, \citenamefont {Vogl},\ and\ \citenamefont
  {Fiete}}]{PhysRevResearch.2.033494}%
  \BibitemOpen
  \bibfield  {author} {\bibinfo {author} {\bibfnamefont {M.}~\bibnamefont
  {Rodriguez-Vega}}, \bibinfo {author} {\bibfnamefont {M.}~\bibnamefont
  {Vogl}},\ and\ \bibinfo {author} {\bibfnamefont {G.~A.}\ \bibnamefont
  {Fiete}},\ }\bibfield  {title} {\bibinfo {title} {Floquet engineering of
  twisted double bilayer graphene},\ }\href
  {https://doi.org/10.1103/PhysRevResearch.2.033494} {\bibfield  {journal}
  {\bibinfo  {journal} {Phys. Rev. Research}\ }\textbf {\bibinfo {volume}
  {2}},\ \bibinfo {pages} {033494} (\bibinfo {year} {2020})}\BibitemShut
  {NoStop}%
\bibitem [{\citenamefont {Vogl}\ \emph
  {et~al.}(2020{\natexlab{b}})\citenamefont {Vogl}, \citenamefont
  {Rodriguez-Vega},\ and\ \citenamefont {Fiete}}]{PhysRevB.101.235411}%
  \BibitemOpen
  \bibfield  {author} {\bibinfo {author} {\bibfnamefont {M.}~\bibnamefont
  {Vogl}}, \bibinfo {author} {\bibfnamefont {M.}~\bibnamefont
  {Rodriguez-Vega}},\ and\ \bibinfo {author} {\bibfnamefont {G.~A.}\
  \bibnamefont {Fiete}},\ }\bibfield  {title} {\bibinfo {title} {Effective
  floquet hamiltonians for periodically driven twisted bilayer graphene},\
  }\href {https://doi.org/10.1103/PhysRevB.101.235411} {\bibfield  {journal}
  {\bibinfo  {journal} {Phys. Rev. B}\ }\textbf {\bibinfo {volume} {101}},\
  \bibinfo {pages} {235411} (\bibinfo {year} {2020}{\natexlab{b}})}\BibitemShut
  {NoStop}%
\bibitem [{\citenamefont {Vogl}\ \emph
  {et~al.}(2020{\natexlab{c}})\citenamefont {Vogl}, \citenamefont
  {Rodriguez-Vega},\ and\ \citenamefont {Fiete}}]{michaelvoglInterlayer2020}%
  \BibitemOpen
  \bibfield  {author} {\bibinfo {author} {\bibfnamefont {M.}~\bibnamefont
  {Vogl}}, \bibinfo {author} {\bibfnamefont {M.}~\bibnamefont
  {Rodriguez-Vega}},\ and\ \bibinfo {author} {\bibfnamefont {G.~A.}\
  \bibnamefont {Fiete}},\ }\bibfield  {title} {\bibinfo {title} {Floquet
  engineering of interlayer couplings: Tuning the magic angle of twisted
  bilayer graphene at the exit of a waveguide},\ }\href
  {https://doi.org/10.1103/PhysRevB.101.241408} {\bibfield  {journal} {\bibinfo
   {journal} {Phys. Rev. B}\ }\textbf {\bibinfo {volume} {101}},\ \bibinfo
  {pages} {241408} (\bibinfo {year} {2020}{\natexlab{c}})}\BibitemShut
  {NoStop}%
\bibitem [{\citenamefont {Katz}\ \emph {et~al.}(2020)\citenamefont {Katz},
  \citenamefont {Refael},\ and\ \citenamefont {Lindner}}]{PhysRevB.102.155123}%
  \BibitemOpen
  \bibfield  {author} {\bibinfo {author} {\bibfnamefont {O.}~\bibnamefont
  {Katz}}, \bibinfo {author} {\bibfnamefont {G.}~\bibnamefont {Refael}},\ and\
  \bibinfo {author} {\bibfnamefont {N.~H.}\ \bibnamefont {Lindner}},\
  }\bibfield  {title} {\bibinfo {title} {Optically induced flat bands in
  twisted bilayer graphene},\ }\href
  {https://doi.org/10.1103/PhysRevB.102.155123} {\bibfield  {journal} {\bibinfo
   {journal} {Phys. Rev. B}\ }\textbf {\bibinfo {volume} {102}},\ \bibinfo
  {pages} {155123} (\bibinfo {year} {2020})}\BibitemShut {NoStop}%
\bibitem [{\citenamefont {Li}\ \emph {et~al.}(2020)\citenamefont {Li},
  \citenamefont {Fertig},\ and\ \citenamefont
  {Seradjeh}}]{PhysRevResearch.2.043275}%
  \BibitemOpen
  \bibfield  {author} {\bibinfo {author} {\bibfnamefont {Y.}~\bibnamefont
  {Li}}, \bibinfo {author} {\bibfnamefont {H.~A.}\ \bibnamefont {Fertig}},\
  and\ \bibinfo {author} {\bibfnamefont {B.}~\bibnamefont {Seradjeh}},\
  }\bibfield  {title} {\bibinfo {title} {Floquet-engineered topological flat
  bands in irradiated twisted bilayer graphene},\ }\href
  {https://doi.org/10.1103/PhysRevResearch.2.043275} {\bibfield  {journal}
  {\bibinfo  {journal} {Phys. Rev. Research}\ }\textbf {\bibinfo {volume}
  {2}},\ \bibinfo {pages} {043275} (\bibinfo {year} {2020})}\BibitemShut
  {NoStop}%
\bibitem [{\citenamefont {Ge}\ and\ \citenamefont
  {Kolodrubetz}(2021)}]{ge2021floquet}%
  \BibitemOpen
  \bibfield  {author} {\bibinfo {author} {\bibfnamefont {R.-C.}\ \bibnamefont
  {Ge}}\ and\ \bibinfo {author} {\bibfnamefont {M.}~\bibnamefont
  {Kolodrubetz}},\ }\bibfield  {title} {\bibinfo {title} {Floquet engineering
  of lattice structure and dimensionality in twisted moir$\backslash$'e
  heterobilayers},\ }\href {https://arxiv.org/abs/2103.09874} {\bibfield
  {journal} {\bibinfo  {journal} {arXiv preprint arXiv:2103.09874}\ } (\bibinfo
  {year} {2021})}\BibitemShut {NoStop}%
\bibitem [{\citenamefont {Pan}\ \emph {et~al.}(2020)\citenamefont {Pan},
  \citenamefont {Wu},\ and\ \citenamefont {Das~Sarma}}]{PhysRevB.102.201104}%
  \BibitemOpen
  \bibfield  {author} {\bibinfo {author} {\bibfnamefont {H.}~\bibnamefont
  {Pan}}, \bibinfo {author} {\bibfnamefont {F.}~\bibnamefont {Wu}},\ and\
  \bibinfo {author} {\bibfnamefont {S.}~\bibnamefont {Das~Sarma}},\ }\bibfield
  {title} {\bibinfo {title} {Quantum phase diagram of a moir\'e-hubbard
  model},\ }\href {https://doi.org/10.1103/PhysRevB.102.201104} {\bibfield
  {journal} {\bibinfo  {journal} {Phys. Rev. B}\ }\textbf {\bibinfo {volume}
  {102}},\ \bibinfo {pages} {201104} (\bibinfo {year} {2020})}\BibitemShut
  {NoStop}%
\bibitem [{\citenamefont {Wu}\ \emph {et~al.}(2018{\natexlab{b}})\citenamefont
  {Wu}, \citenamefont {Lovorn}, \citenamefont {Tutuc},\ and\ \citenamefont
  {MacDonald}}]{Wu2018}%
  \BibitemOpen
  \bibfield  {author} {\bibinfo {author} {\bibfnamefont {F.}~\bibnamefont
  {Wu}}, \bibinfo {author} {\bibfnamefont {T.}~\bibnamefont {Lovorn}}, \bibinfo
  {author} {\bibfnamefont {E.}~\bibnamefont {Tutuc}},\ and\ \bibinfo {author}
  {\bibfnamefont {A.~H.}\ \bibnamefont {MacDonald}},\ }\bibfield  {title}
  {\bibinfo {title} {Hubbard model physics in transition metal dichalcogenide
  moir\'e bands},\ }\href {https://doi.org/10.1103/PhysRevLett.121.026402}
  {\bibfield  {journal} {\bibinfo  {journal} {Phys. Rev. Lett.}\ }\textbf
  {\bibinfo {volume} {121}},\ \bibinfo {pages} {026402} (\bibinfo {year}
  {2018}{\natexlab{b}})}\BibitemShut {NoStop}%
\bibitem [{\citenamefont {Wu}\ \emph {et~al.}(2019{\natexlab{b}})\citenamefont
  {Wu}, \citenamefont {Lovorn}, \citenamefont {Tutuc}, \citenamefont {Martin},\
  and\ \citenamefont {MacDonald}}]{PhysRevLett.122.086402}%
  \BibitemOpen
  \bibfield  {author} {\bibinfo {author} {\bibfnamefont {F.}~\bibnamefont
  {Wu}}, \bibinfo {author} {\bibfnamefont {T.}~\bibnamefont {Lovorn}}, \bibinfo
  {author} {\bibfnamefont {E.}~\bibnamefont {Tutuc}}, \bibinfo {author}
  {\bibfnamefont {I.}~\bibnamefont {Martin}},\ and\ \bibinfo {author}
  {\bibfnamefont {A.~H.}\ \bibnamefont {MacDonald}},\ }\bibfield  {title}
  {\bibinfo {title} {Topological insulators in twisted transition metal
  dichalcogenide homobilayers},\ }\href
  {https://doi.org/10.1103/PhysRevLett.122.086402} {\bibfield  {journal}
  {\bibinfo  {journal} {Phys. Rev. Lett.}\ }\textbf {\bibinfo {volume} {122}},\
  \bibinfo {pages} {086402} (\bibinfo {year} {2019}{\natexlab{b}})}\BibitemShut
  {NoStop}%
\bibitem [{\citenamefont {Liu}\ \emph {et~al.}(2013)\citenamefont {Liu},
  \citenamefont {Shan}, \citenamefont {Yao}, \citenamefont {Yao},\ and\
  \citenamefont {Xiao}}]{Liu_2013}%
  \BibitemOpen
  \bibfield  {author} {\bibinfo {author} {\bibfnamefont {G.-B.}\ \bibnamefont
  {Liu}}, \bibinfo {author} {\bibfnamefont {W.-Y.}\ \bibnamefont {Shan}},
  \bibinfo {author} {\bibfnamefont {Y.}~\bibnamefont {Yao}}, \bibinfo {author}
  {\bibfnamefont {W.}~\bibnamefont {Yao}},\ and\ \bibinfo {author}
  {\bibfnamefont {D.}~\bibnamefont {Xiao}},\ }\bibfield  {title} {\bibinfo
  {title} {Three-band tight-binding model for monolayers of group-{VIB}
  transition metal dichalcogenides},\ }\bibfield  {journal} {\bibinfo
  {journal} {Physical Review B}\ }\textbf {\bibinfo {volume} {88}},\ \href
  {https://doi.org/10.1103/physrevb.88.085433} {10.1103/physrevb.88.085433}
  (\bibinfo {year} {2013})\BibitemShut {NoStop}%
\bibitem [{\citenamefont {Ye}\ \emph {et~al.}(2020)\citenamefont {Ye},
  \citenamefont {Machado}, \citenamefont {White}, \citenamefont {Mong},\ and\
  \citenamefont {Yao}}]{Bingtian2020}%
  \BibitemOpen
  \bibfield  {author} {\bibinfo {author} {\bibfnamefont {B.}~\bibnamefont
  {Ye}}, \bibinfo {author} {\bibfnamefont {F.}~\bibnamefont {Machado}},
  \bibinfo {author} {\bibfnamefont {C.~D.}\ \bibnamefont {White}}, \bibinfo
  {author} {\bibfnamefont {R.~S.~K.}\ \bibnamefont {Mong}},\ and\ \bibinfo
  {author} {\bibfnamefont {N.~Y.}\ \bibnamefont {Yao}},\ }\bibfield  {title}
  {\bibinfo {title} {Emergent hydrodynamics in nonequilibrium quantum
  systems},\ }\href {https://doi.org/10.1103/PhysRevLett.125.030601} {\bibfield
   {journal} {\bibinfo  {journal} {Phys. Rev. Lett.}\ }\textbf {\bibinfo
  {volume} {125}},\ \bibinfo {pages} {030601} (\bibinfo {year}
  {2020})}\BibitemShut {NoStop}%
\bibitem [{\citenamefont {Weinberg}\ \emph {et~al.}(2017)\citenamefont
  {Weinberg}, \citenamefont {Bukov}, \citenamefont {D’Alessio}, \citenamefont
  {Polkovnikov}, \citenamefont {Vajna},\ and\ \citenamefont
  {Kolodrubetz}}]{Weinberg_adiabatic2017}%
  \BibitemOpen
  \bibfield  {author} {\bibinfo {author} {\bibfnamefont {P.}~\bibnamefont
  {Weinberg}}, \bibinfo {author} {\bibfnamefont {M.}~\bibnamefont {Bukov}},
  \bibinfo {author} {\bibfnamefont {L.}~\bibnamefont {D’Alessio}}, \bibinfo
  {author} {\bibfnamefont {A.}~\bibnamefont {Polkovnikov}}, \bibinfo {author}
  {\bibfnamefont {S.}~\bibnamefont {Vajna}},\ and\ \bibinfo {author}
  {\bibfnamefont {M.}~\bibnamefont {Kolodrubetz}},\ }\bibfield  {title}
  {\bibinfo {title} {Adiabatic perturbation theory and geometry of
  periodically-driven systems},\ }\href
  {https://doi.org/10.1016/j.physrep.2017.05.003} {\bibfield  {journal}
  {\bibinfo  {journal} {Physics Reports}\ }\textbf {\bibinfo {volume} {688}},\
  \bibinfo {pages} {1–35} (\bibinfo {year} {2017})}\BibitemShut {NoStop}%
\bibitem [{\citenamefont {Shan}\ \emph {et~al.}(2021)\citenamefont {Shan},
  \citenamefont {Ye}, \citenamefont {Chu}, \citenamefont {Lee}, \citenamefont
  {Park}, \citenamefont {Balents},\ and\ \citenamefont
  {Hsieh}}]{shan2021giant}%
  \BibitemOpen
  \bibfield  {author} {\bibinfo {author} {\bibfnamefont {J.-Y.}\ \bibnamefont
  {Shan}}, \bibinfo {author} {\bibfnamefont {M.}~\bibnamefont {Ye}}, \bibinfo
  {author} {\bibfnamefont {H.}~\bibnamefont {Chu}}, \bibinfo {author}
  {\bibfnamefont {S.}~\bibnamefont {Lee}}, \bibinfo {author} {\bibfnamefont
  {J.-G.}\ \bibnamefont {Park}}, \bibinfo {author} {\bibfnamefont
  {L.}~\bibnamefont {Balents}},\ and\ \bibinfo {author} {\bibfnamefont
  {D.}~\bibnamefont {Hsieh}},\ }\bibfield  {title} {\bibinfo {title} {Giant
  modulation of optical nonlinearity by floquet engineering},\ }\href
  {https://www.nature.com/articles/s41586-021-04051-8} {\bibfield  {journal}
  {\bibinfo  {journal} {Nature}\ }\textbf {\bibinfo {volume} {600}},\ \bibinfo
  {pages} {235} (\bibinfo {year} {2021})}\BibitemShut {NoStop}%
\bibitem [{\citenamefont {Chaudhary}\ \emph {et~al.}(2022)\citenamefont
  {Chaudhary}, \citenamefont {Lewandowski},\ and\ \citenamefont
  {Refael}}]{chaudhary2021shiftcurrent}%
  \BibitemOpen
  \bibfield  {author} {\bibinfo {author} {\bibfnamefont {S.}~\bibnamefont
  {Chaudhary}}, \bibinfo {author} {\bibfnamefont {C.}~\bibnamefont
  {Lewandowski}},\ and\ \bibinfo {author} {\bibfnamefont {G.}~\bibnamefont
  {Refael}},\ }\bibfield  {title} {\bibinfo {title} {Shift-current response as
  a probe of quantum geometry and electron-electron interactions in twisted
  bilayer graphene},\ }\href {https://doi.org/10.1103/PhysRevResearch.4.013164}
  {\bibfield  {journal} {\bibinfo  {journal} {Phys. Rev. Research}\ }\textbf
  {\bibinfo {volume} {4}},\ \bibinfo {pages} {013164} (\bibinfo {year}
  {2022})}\BibitemShut {NoStop}%
\bibitem [{\citenamefont {Kaplan}\ \emph {et~al.}(2021)\citenamefont {Kaplan},
  \citenamefont {Holder},\ and\ \citenamefont {Yan}}]{kaplan2021}%
  \BibitemOpen
  \bibfield  {author} {\bibinfo {author} {\bibfnamefont {D.}~\bibnamefont
  {Kaplan}}, \bibinfo {author} {\bibfnamefont {T.}~\bibnamefont {Holder}},\
  and\ \bibinfo {author} {\bibfnamefont {B.}~\bibnamefont {Yan}},\ }\bibfield
  {title} {\bibinfo {title} {Momentum shift current at terahertz frequencies in
  twisted bilayer graphene},\ }\href {https://arxiv.org/abs/2101.07539}
  {\bibfield  {journal} {\bibinfo  {journal} {arXiv:2101.07539}\ } (\bibinfo
  {year} {2021})}\BibitemShut {NoStop}%
\bibitem [{\citenamefont {Liu}\ and\ \citenamefont
  {Dai}(2020)}]{liu2020anomalous}%
  \BibitemOpen
  \bibfield  {author} {\bibinfo {author} {\bibfnamefont {J.}~\bibnamefont
  {Liu}}\ and\ \bibinfo {author} {\bibfnamefont {X.}~\bibnamefont {Dai}},\
  }\bibfield  {title} {\bibinfo {title} {Anomalous hall effect, magneto-optical
  properties, and nonlinear optical properties of twisted graphene systems},\
  }\href {https://www.nature.com/articles/s41524-020-0299-4} {\bibfield
  {journal} {\bibinfo  {journal} {npj Computational Materials}\ }\textbf
  {\bibinfo {volume} {6}},\ \bibinfo {pages} {1} (\bibinfo {year}
  {2020})}\BibitemShut {NoStop}%
\bibitem [{\citenamefont {Xu}\ \emph {et~al.}(2017)\citenamefont {Xu},
  \citenamefont {Zhang}, \citenamefont {Zhang},\ and\ \citenamefont
  {Lee}}]{https://doi.org/10.1002/aenm.201700571}%
  \BibitemOpen
  \bibfield  {author} {\bibinfo {author} {\bibfnamefont {J.}~\bibnamefont
  {Xu}}, \bibinfo {author} {\bibfnamefont {J.}~\bibnamefont {Zhang}}, \bibinfo
  {author} {\bibfnamefont {W.}~\bibnamefont {Zhang}},\ and\ \bibinfo {author}
  {\bibfnamefont {C.-S.}\ \bibnamefont {Lee}},\ }\bibfield  {title} {\bibinfo
  {title} {Interlayer nanoarchitectonics of two-dimensional transition-metal
  dichalcogenides nanosheets for energy storage and conversion applications},\
  }\href {https://onlinelibrary.wiley.com/doi/abs/10.1002/aenm.201700571}
  {\bibfield  {journal} {\bibinfo  {journal} {Advanced Energy Materials}\
  }\textbf {\bibinfo {volume} {7}},\ \bibinfo {pages} {1700571} (\bibinfo
  {year} {2017})}\BibitemShut {NoStop}%
\bibitem [{\citenamefont {Li}\ \emph {et~al.}(2016)\citenamefont {Li},
  \citenamefont {Shi}, \citenamefont {Liu}, \citenamefont {Yu}, \citenamefont
  {Xi},\ and\ \citenamefont {Wang}}]{li2016experimental}%
  \BibitemOpen
  \bibfield  {author} {\bibinfo {author} {\bibfnamefont {M.}~\bibnamefont
  {Li}}, \bibinfo {author} {\bibfnamefont {J.}~\bibnamefont {Shi}}, \bibinfo
  {author} {\bibfnamefont {L.}~\bibnamefont {Liu}}, \bibinfo {author}
  {\bibfnamefont {P.}~\bibnamefont {Yu}}, \bibinfo {author} {\bibfnamefont
  {N.}~\bibnamefont {Xi}},\ and\ \bibinfo {author} {\bibfnamefont
  {Y.}~\bibnamefont {Wang}},\ }\bibfield  {title} {\bibinfo {title}
  {Experimental study and modeling of atomic-scale friction in zigzag and
  armchair lattice orientations of mos2},\ }\href@noop {} {\bibfield  {journal}
  {\bibinfo  {journal} {Science and Technology of advanced MaTerialS}\ }\textbf
  {\bibinfo {volume} {17}},\ \bibinfo {pages} {189} (\bibinfo {year}
  {2016})}\BibitemShut {NoStop}%
\bibitem [{\citenamefont {Schutte}\ \emph {et~al.}(1987)\citenamefont
  {Schutte}, \citenamefont {{De Boer}},\ and\ \citenamefont
  {Jellinek}}]{SCHUTTE1987207}%
  \BibitemOpen
  \bibfield  {author} {\bibinfo {author} {\bibfnamefont {W.}~\bibnamefont
  {Schutte}}, \bibinfo {author} {\bibfnamefont {J.}~\bibnamefont {{De Boer}}},\
  and\ \bibinfo {author} {\bibfnamefont {F.}~\bibnamefont {Jellinek}},\
  }\bibfield  {title} {\bibinfo {title} {Crystal structures of tungsten
  disulfide and diselenide},\ }\href
  {https://doi.org/https://doi.org/10.1016/0022-4596(87)90057-0} {\bibfield
  {journal} {\bibinfo  {journal} {Journal of Solid State Chemistry}\ }\textbf
  {\bibinfo {volume} {70}},\ \bibinfo {pages} {207} (\bibinfo {year}
  {1987})}\BibitemShut {NoStop}%
\end{thebibliography}%

\onecolumngrid
\appendix

\section{Mechanism for on-site energy from interlayer hopping}
\label{app:Downfold_into_active_layer}
In this appendix we will discuss in more detail the mechanism for indirect on-site hopping that eventually leads to the moir\'e potential in Eq. \eqref{eq:non_int_ham}.

For simplicity of the discussion, let us consider a one dimensional system of two initially disjoint chains denoted by $A$ and $B$:
\begin{equation}
    H_0=H_A+H_B=\sum_i(\varepsilon_Aa_i^\dagger a_i)+(t_Aa_{i+1}^\dagger a_i+h.c)+\sum_i(\varepsilon_Bb_i^\dagger b_i)+(t_Bb_{i+1}^\dagger b_i+h.c),
\end{equation}
with $\varepsilon_B\gg \varepsilon_A$ and $\varepsilon_{B}-\varepsilon_A\gg t_A, t_B$. That is the system has a relatively large energy gap, which makes it possible to find a simpler model by separating the problem into high- and low-energy sectors. We will be interested in the low-energy sector of this hamiltonian, which is spanned by:
\begin{equation}
    \ket{k_a}=a_k^\dagger\ket{0}=\sum_j e^{ikj}a^\dagger_j\ket{0},
\end{equation}
with energy $E_{k}^a=\varepsilon_A+2t_A\cos k$. The corresponding high-energy sector is spanned by
\begin{equation}
    \ket{k_b}=b_k^\dagger\ket{0}=\sum_j e^{ikj}b^\dagger_j\ket{0},
\end{equation}
with energy $E_{k}^b=\varepsilon_B+2t_B\cos k$. 

Now, We will consider the impact that a coupling between the two chains will have. Particularly, let us consider a perturbation arising from a site-dependent interchain hopping. Let's assume that the spatial dependence of this interchain hopping has a period $n>1$. We can model this term as:
\begin{equation}
V=\sum_i t_{AB}^i (a_i^\dagger b_i+h.c)=\sum_i v^i (a_i^\dagger b_i+h.c),
\end{equation}
with $v^{i+n}=v^i$. This perturbation is clearly off-diagonal between $\mathcal{P}_A$ and $\mathcal{P}_B$  where $\mathcal{P}_A$ and $\mathcal{P}_B$ are projection operators for chain $A$ and $B$, respectively.  Furthermore, the periodic nature of this perturbation allows us to write:
\begin{equation}
    v^m=\sum_q e^{i\frac{2\pi m}{n}q}v_q.
    \label{moirehopping}
\end{equation}
Now, we are interested in finding out how this perturbation affects the effective low-energy hamiltonian. Here, we can employ a Schrieffer-Wolff transformations to derive the effective low-energy hamiltonian. The second-order correction is given by:
\begin{equation}
    H_\text{eff,2}=\frac{1}{2}P_A[S_1,V]P_A,
\end{equation}
where
\begin{equation}
    S_1=\sum_{k',k}\left(\frac{\bra{k'_a}V\ket{k_b}}{E_{k'}^a-E_k^b}\ket{k'_a}\bra{k_b}+\frac{\bra{k_b}V\ket{k'_a}}{E_{k}^b-E_{k'}^a}\ket{k_b}\bra{k'_a}\right).
\end{equation}
Now, 
\begin{equation}
    \bra{k'_a}V\ket{k_b}=V_{k'k}^{ab}=\sum_{j}e^{i(k-k')j}v^j=\sum_{j,q}e^{i(k-k'+\frac{2\pi}{n}q)j}v_q=\sum_q\delta\left(k'-(k+\frac{2\pi}{n}q)\right)v_q,
\end{equation}
and equivalently
\begin{equation}
    \bra{k_b}V\ket{k'_a}=V_{kk'}^{ba}=\sum_{j}e^{i(k'-k)j}v^j=\sum_{j,q}e^{i(k'-k+\frac{2\pi}{n}q)j}v_q=\sum_q\delta\left(k-(k'+\frac{2\pi}{n}q)\right)v_q,
\end{equation}
\begin{equation}
    H_\text{eff,2}^{aa}=\frac{1}{2}\sum_{k'',k',k}\left(\frac{V_{k'k}^{ab}}{E_{k'}^a-E_k^b}V_{kk''}^{ba}-V_{k'k}^{ab}\frac{V_{kk''}^{ba}}{E_k^b-E_{k''}^a}\right)\ket{k'_a}\bra{k''_a},
\end{equation}
\begin{equation}
\begin{split}
    H_\text{eff,2}^{aa}=\frac{1}{2}\sum_{k'',k',k,q,q'}\left(\frac{1}{E_{k'}^a-E_k^b}\delta\left(k'-(k+\frac{2\pi}{n}q)\right)\delta\left(k-(k''+\frac{2\pi}{n}q')\right)v_qv_{q'}\right)\ket{k'_a}\bra{k''_a}\\-\frac{1}{2}\sum_{k'',k',k,q,q'}\left(\frac{1}{E_k^b-E_{k''}^a}\delta\left(k'-(k+\frac{2\pi}{n}q)\right)\delta\left(k-(k''+\frac{2\pi}{n}q')\right)v_qv_{q'}\right)\ket{k'_a}\bra{k''_a},
\end{split}
\end{equation}
\begin{equation}
\begin{split}
    H_\text{eff,2}^{aa}=\frac{1}{2}\sum_{k,q,q'}\left(\frac{1}{E_{k+\frac{2\pi}{n}q}^a-E_k^b}v_qv_{q'}\right)\ket{\left(k+\frac{2\pi}{n}q\right)_a}\bra{\left(k-\frac{2\pi}{n}q'\right)_a}\\+\frac{1}{2}\sum_{k,q,q'}\left(\frac{1}{E_{k-\frac{2\pi}{n}q'}^a-E_k^b}v_qv_{q'}\right)\ket{\left(k+\frac{2\pi}{n}q\right)_a}\bra{\left(k-\frac{2\pi}{n}q'\right)_a},
\end{split}
\end{equation}

\begin{equation}
    \ket{\left(k+\frac{2\pi}{n}q\right)_a}\bra{\left(k-\frac{2\pi}{n}q'\right)_a}=\sum_{l,m}e^{ik(l-m)}e^{i\frac{2\pi}{n}(ql+q'm)}\ket{l_a}\bra{m_a}.
\end{equation}
For the case, where $\varepsilon_B\gg\varepsilon_A, t_A,t_B$, we can simply assume $E_{k'}^a-E_k^b\approx -\varepsilon_B$.
\begin{equation}
\begin{split}
    H_\text{eff,2}^{aa}=-\frac{1}{\varepsilon_B}\sum_{k,q,q',l,m}v_qv_{q'}e^{ik(l-m)}e^{i\frac{2\pi}{n}(ql+q'm)}\ket{l_a}\bra{m_a},
\end{split}
\end{equation}
\begin{equation}
\begin{split}
    H_\text{eff,2}^{aa}=-\frac{1}{\varepsilon_B}\sum_{q,q',l,m}v_qv_{q'}\delta_{lm}e^{i\frac{2\pi}{n}(ql+q'm)}\ket{l_a}\bra{m_a},\\
    \implies H_\text{eff,2}^{aa}=-\frac{1}{\varepsilon_B}\sum_{q,q',l}v_qv_{q'}e^{i\frac{2\pi}{n}(q+q')l}\ket{l_a}\bra{l_a}.
    \end{split}
\end{equation}

Now, using Eq.~\ref{moirehopping}, we can show
\begin{equation}
\begin{split}
     H_\text{eff,2}^{aa}=-\frac{1}{\varepsilon_B}\sum_{l}\underbrace{\left(\sum_q{e^{i\frac{2\pi}{n}ql}}v_q\right)}_{v^l}\underbrace{\left(\sum_{q'}{e^{i\frac{2\pi}{n}q'l}}v_q'\right)}_{v^l}\ket{l_a}\bra{l_a}=-\sum_{l}\frac{(v^l)^2}{\varepsilon_B}\ket{l_a}\bra{l_a},
\end{split}
\end{equation}
which is an on-site energy shift proportional to the square of inter-layer at the given site. 
In second-quantization picture, we can write:
\begin{equation}
    H_\text{eff,2}=-\sum_i \frac{(t_{AB}^i)^2}{\varepsilon_B}a^\dagger_i a_i.
\end{equation}
We notice the on-site potential is modified only at sites where interchain hopping is non-zero. These results can be simply generalized for  a two-dimensional set up where we can just replace index $i$ with a pair of indices $i,j$ denoting the lattice site as long as we have a tight-binding hamiltonian with nearest-neighbor interaction only. 
\color{black}


\section{Neglected effects of indirect hopping on-site potential}
\label{app:neglected_effects}
In our discussion we have focused on just two layers of the twisted TMD. However, it is important to note that a TMD bilayer does not consist of two atomic layers but actually six. In the example of $WSe_2/MoTe_2$ one has a layer structure like $Se-W-Se-Te-Mo-Te$. For concreteness we will keep referring to the material combination $WSe_2/MoTe_2$, although for a general discussion one may take $W/Mo$ as being any two different transition metals $M$ and $Se/Te$ as any chalcogenide $X$. If we now assume that $W$ is the active layer \cite{moralesduran2021nonlocal} we can then see that both hoppings $W\leftrightarrow Te$ and $W\leftrightarrow Mo$ contribute to the interlayer hopping that can lead to moir\'e potential $\Delta(\vect x)$. It is therefore not completely correct to just replace $V_m\to J_0(a_LA)^2$ because there is not just one interlayer lattice constant $a_L$. However, for the purposes of a clear discussion and to find a simplified and easy to interpret model we will make the approximation $a_L=c a_{W-Mo}$, which essentially boils down to an order of magnitude estimate. We can estimate that $c\sim 0.85$. 

The reason for this approximation is as follows: First we see that the spacing between the centers of two TMD layers is approximately $6\AA$\cite{https://doi.org/10.1002/aenm.201700571}. Next, we note that the spacing between two the layers of $M$ and $X$ in a TMD $MX_2$ is approximately $1.5\AA$\cite{li2016experimental,SCHUTTE1987207}. Therefore, if the hopping $W\leftrightarrow Mo$ and $W\leftrightarrow Te$ for the closer of the $Te$ is dominant then both interlayer distances are somewhere between $0.7 a_{W-Mo}$ and $1 a_{W-Mo}$. Therefore, a compromise is $a_L= 0.85 a_{W-Mo}$. This approximation will lead to small errors of $\lesssim 5\%$  if we compare $J_0(a_LA)^2$ to $J_0(_{W-Mo}A)^2$ or $J_0(_{W-Te}A)^2$  if driving strengths in the range $a_LA<0.5$ are considered. 

\section{Downfolding the three band model of a TMD subjected to circularly polarized light in the high frequency limit near the \texorpdfstring{$\Gamma$}{G} point}
\label{App:Downfold3Bandmodel}
In this appendix, we study the impact of circularly polarized light on a twisted TMD hetero bilayer. We are interested in the effective mass near the $\Gamma$ point. Since circularly polarized light mostly modifies in-plane hopping elements we first consider the case of an isolated TMD layer, which in our case is assumed to describe the active TMD layer in our twisted TMD heterobilayer. The starting point of our discussion is the third nearest neighbor three band tight binding model of a TMD that was derived in Ref.\cite{Liu_2013}. The Hamiltonian is 
\begin{equation}
\begin{aligned}
&H^{\text{TNN}}(\vect k)=\begin{pmatrix}V_{0} & V_{1} & V_{2}\\
V_{1}^{*} & V_{11} & V_{12}\\
V_{2}^{*} & V_{12}^{*} & V_{22}
\end{pmatrix}
\end{aligned}.
\end{equation}
This Hamiltonian is written in the basis of d-orbitals $\{d_{z^2},d_{xy},d_{x^2-y^2}\}$.
The different contributions to the Hamiltonian are given as
\begin{equation}
\begin{aligned}
V_{0}&=\varepsilon_{1}+2t_{0}(2\cos X_k\cos Y_k+\cos2X_k)+2r_{0}(2\cos3X_k\cos Y_k+\cos2Y_k)+2u_{0}(2\cos2X_k\cos2Y_k+\cos4X_k)\\
\end{aligned}
\end{equation}

\begin{equation}
\begin{aligned}
V_{1}=&-2\sqrt{3}t_{2}\sin X_k\sin Y_k+2(r_{1}+r_{2})\sin3X_k\sin Y_k-2\sqrt{3}u_{2}\sin2X_k\sin2Y_k\\
&+i\Big[2t_{1}\sin X_k(2\cos X_k+\cos Y_k)+2(r_{1}-r_{2})\sin3X_k\cos Y_k+2u_{1}\sin2X_k(2\cos2X_k+\cos2Y_k)\Big]
\end{aligned}
\end{equation}

\begin{equation}
\begin{aligned}
V_{2}&=2t_{2}(\cos2X_k-\cos X_k\cos Y_k)-\frac{2}{\sqrt{3}}(r_{1}+r_{2})(\cos3X_k\cos Y_k-\cos2Y_k)+2u_{2}(\cos4X_k-\cos2X_k\cos2Y_k)\\
&+i\Big[2\sqrt{3}t_{1}\cos X_k\sin Y_k+\frac{2}{\sqrt{3}}\sin Y_k(r_{1}-r_{2})(\cos3X_k+2\cos Y_k)+2\sqrt{3}u_{1}\cos2X_k\sin2Y_k\Big]
\end{aligned}
\end{equation}

\begin{equation}
\begin{aligned}
V_{11}=&\varepsilon_{2}+(t_{11}+3t_{22})\cos X_k\cos Y_k+2t_{11}\cos2X_k+4r_{11}\cos3X_k\cos Y_k+2(r_{11}+\sqrt{3}r_{12})\cos2Y_k\\
&+(u_{11}+3u_{22})\cos2X_k\cos2Y_k+2u_{11}\cos4X_k,
\end{aligned}
\end{equation}
\begin{equation}
\begin{aligned}
V_{12}=&\sqrt{3}(t_{22}-t_{11})\sin X_k\sin Y_k+4r_{12}\sin3X_k\sin Y_k+\sqrt{3}(u_{22}-u_{11})\sin2X_k\sin2Y_k\\
&+i\Big[4t_{12}\sin X_k(\cos X_k-\cos Y_k)+4u_{12}\sin2X_k(\cos2X_k-\cos2Y_k)\Big]
\end{aligned}
\end{equation}
 with the shorthand
\begin{equation}
	X_k=\frac{1}{2}k_xa;\quad Y_k=\frac{\sqrt{3}}{2}k_ya.
\end{equation}
The different couplings $\varepsilon_i$, $t_i$, $r_i$ and $u_i$ for 
a variety of different TMDs can be found in Ref.\cite{Liu_2013}.

Starting from this Hamiltonian circularly polarized light is introduced in the standard way via a vector potential $\vect A(t)=A(\cos(\omega t),\sin(\omega t))$ and employing the minimal substitution procedure $\vect k\to \vect k+\vect A$. This allows us to define
\begin{equation}
    H(\vect k,t)=H^{\text{TNN}}(\vect k+\vect A(t)).
\end{equation}

In the high frequency limit to leading order we may include the effect of circularly polarized light by employing an average Hamiltonian procedure 
\begin{equation}
    H_{\mathrm{av}}(\vect k)=\frac{1}{T}\int_0^T dt H(\vect k,t),
\end{equation}
which corresponds to the leading order approximation of various high frequency expansions\cite{RODRIGUEZVEGA2021168434,bukov2015}.

The resulting Hamiltonian $H_{\mathrm{av}}(\vect k)$ includes various Bessel function factors $J_0(aA)$, $J_0(\sqrt{3}aA)$ and $J_0(2aA)$ but offers few insights and is unwieldy. Therefore, we do not explicitly include this expression here. However, it is important to note that it is diagonal at the $\Gamma$ point, $\vect k=\Gamma$. Therefore, one may apply second order perturbation theory in $\vect k$ to determine the effective mass $m^*$ of the lowest band from a comparison to the energy expression $E=c+\frac{\hbar}{2m^*}k^2$ (small non-rotationally symmetric parts were dropped since they do not appreciably modify the band structure). A plot of such an approximation is shown in  Fig. \ref{fig:quadraticapprox} below.

\begin{figure}[H]
	\centering
\includegraphics[width=0.7\linewidth]{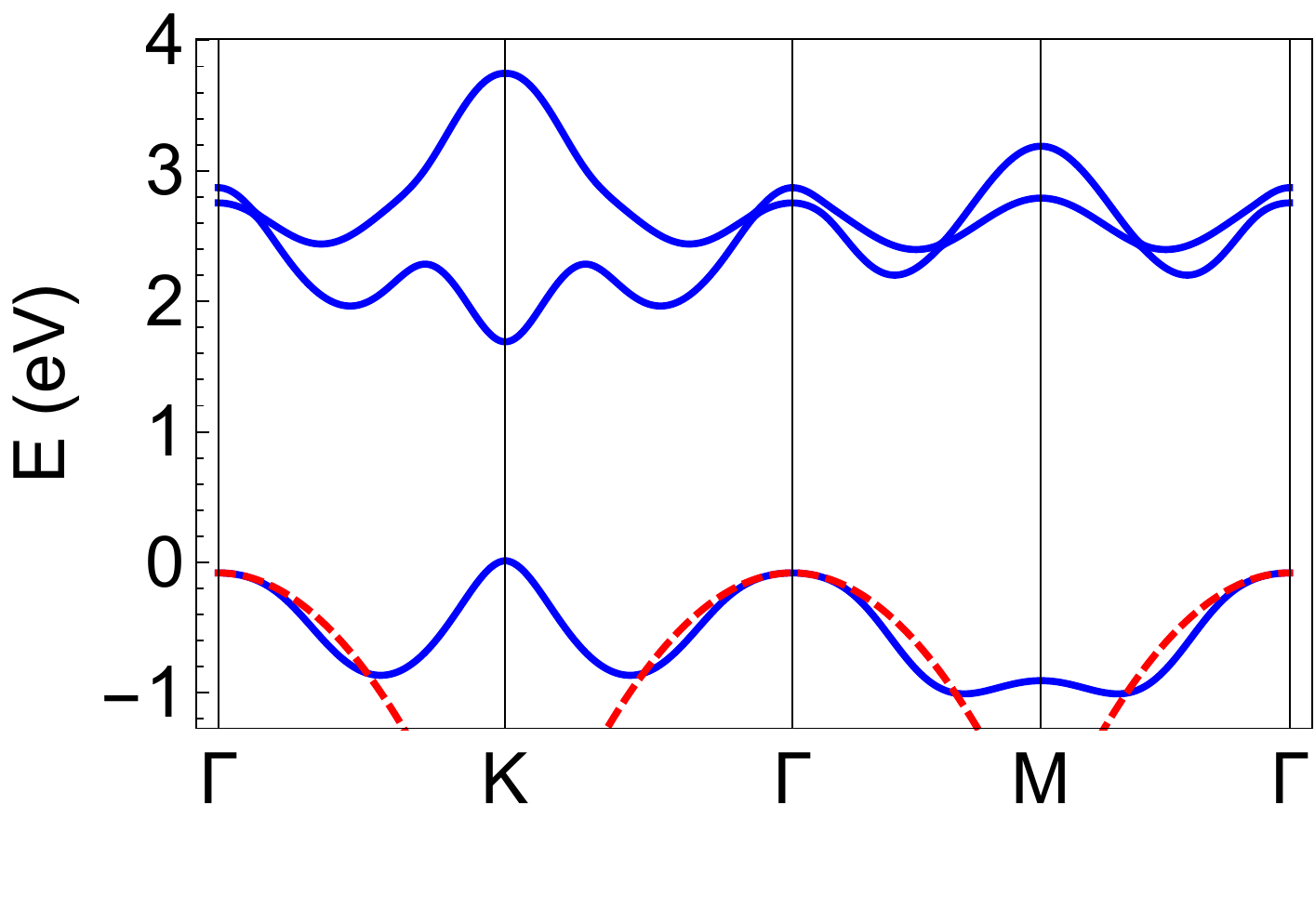} 
	\caption{Plot of the band structure of WS$_2$ under the influence of circularly polarized light of strength $A=0.1$ using GGA parameters from \cite{Liu_2013}. Dashed in red we see the quadratic.approximation to lowest band of the Hamiltonian that is valid near the $\Gamma$ point }
	\label{fig:quadraticapprox}
\end{figure}

The same procedure can of course also be applied in the case of no circularly polarized light, $A=0$. Doing so one may find how light modifies the effective mass $m^*$. In particular one finds the expression
\begin{equation}
    \frac{m^*(A)}{m^*(A=0)}=1+\Lambda a^2A^2,
\end{equation}
where we have employed an expansion for small field strengths $A$.
An explicit expression for $\Lambda$ is given below:

\begin{equation}
\begin{aligned}
    \Lambda=&\frac{1}{4}\left(\frac{9 ( r_1 -r_2 +t_1 +2 u_1 )^2 \left(6 r_0 -6 r_{11} -2 \sqrt{3} r_{12} +2 t_0 -t_{11} -t_{22} +8 u_0 -4 u_{11} -4 u_{22} \right)}{\left(-6 r_0 +6 r_{11} +2 \sqrt{3} r_{12} -6 t_0 +3 t_{11} +3 t_{22} -6 u_0 +3 u_{11} +3 u_{22} -\varepsilon_1 +\varepsilon_2 \right)^2}\right.\\
    &\left.+\frac{27 ( r_1 -r_2 +t_1 +2 u_1 )^2 \left(18 r_0 -18 r_{11} -4 \sqrt{3} r_{12} +6 t_0 -3 t_{11} -3 t_{22} +24 u_0 -12 u_{11} -12 u_{22} \right)}{\left(-18 r_0 +18 r_{11} +4 \sqrt{3} r_{12} -18 t_0 +9 t_{11} +9 t_{22} -18 u_0 +9 u_{11} +9 u_{22} -3 \varepsilon_1 +3 \varepsilon_2 \right)^2}\right.\\
    &\left.-\frac{6 (3  r_1 -3 r_2 +t_1 +8 u_1 ) ( r_1 -r_2 +t_1 +2 u_1 )}{6 r_0 -6 r_{11} -2 \sqrt{3} r_{12} +6 t_0 -3 t_{11} -3 t_{22} +6 u_0 -3 u_{11} -3 u_{22} +\varepsilon_1 -\varepsilon_2 }\right.\\
    &\left.-\frac{18 (3  r_1 -3 r_2 +t_1 +8 u_1 ) ( r_1 -r_2 +t_1 +2 u_1 )}{18 r_0 -18 r_{11} -4 \sqrt{3} r_{12} +18 t_0 -9 t_{11} -9 t_{22} +18 u_0 -9 u_{11} -9 u_{22} +3 \varepsilon_1 -3 \varepsilon_2 }+9 r_0 +t_0 +16 u_0 \right)\\
    &\times \left(\frac{3 ( r_1 -r_2 +t_1 +2 u_1 )^2}{-6 r_0 +6 r_{11} +2 \sqrt{3} r_{12} -6 t_0 +3 t_{11} +3 t_{22} -6 u_0 +3 u_{11} +3 u_{22} -\varepsilon_1 +\varepsilon_2 }\right.\\
    &\left.+\frac{9 ( r_1 -r_2 +t_1 +2 u_1 )^2}{-18 r_0 +18 r_{11} +4 \sqrt{3} r_{12} +3 (-6 t_0 +3 t_{11} +3 t_{22} -6 u_0 +3 u_{11} +3 u_{22} -\varepsilon_1 +\varepsilon_2 )}+3 r_0 +t_0 +4 u_0 \right)\\
    &\times \left(-\frac{3 ( r_1 -r_2 +t_1 +2 u_1 )^2}{6 r_0 -6 r_{11} -2 \sqrt{3} r_{12} +6 t_0 -3 t_{11} -3 t_{22} +6 u_0 -3 u_{11} -3 u_{22} +\varepsilon_1 -\varepsilon_2 }\right.\\
    &\left.-\frac{9 ( r_1 -r_2 +t_1 +2 u_1 )^2}{18 r_0 -18 r_{11} -4 \sqrt{3} r_{12} +18 t_0 -9 t_{11} -9 t_{22} +18 u_0 -9 u_{11} -9 u_{22} +3 \varepsilon_1 -3 \varepsilon_2 }+3 r_0 +t_0 +4 u_0 \right)^{-2}
    \end{aligned}
\end{equation}

Since this result is very unwieldy and not very insightful, we have decided to provide numeric results for specific choices of TMDs are given in the main text in Table \ref{tab:TMDmassmodParam}.

\section{Downfolding the three band model of a TMD subjected to circularly polarized light in the high frequency limit and near the \texorpdfstring{$K$}{K} point}
\label{App:Downfold3BandmodelK}
In this appendix, we study the impact of circularly polarized light on a twisted TMD hetero bilayer. We are interested in the effective mass near the $K$ point.  The starting point of our discussion, unlike near the $\Gamma$ point, is the first nearest neighbor three band tight binding model of a TMD that was derived in Ref.\cite{Liu_2013}. We chose this model because the band structure near the $K$ point is well captured by this simpler model - unlike the bands near the $\Gamma$ point. Furthermore, it allows for more analytical progress. The model is given as below
\begin{equation}
\begin{aligned}
&H^{\text{TNN}}(\vect k)=\begin{pmatrix}h_0 & h_{1} & h_{2}\\
h_{1}^{*} & h_{11} & h_{12}\\
h_{2}^{*} & h_{12}^{*} & h_{22}
\end{pmatrix}
\end{aligned}.
\end{equation}
This Hamiltonian much like the third nearest neighbor model from the previous section is written in the basis of d-orbitals $\{d_{z^2},d_{xy},d_{x^2-y^2}\}$.
The different contributions to the Hamiltonian are given as
\begin{equation}
    h_0=2t_0(\cos 2X_k + 2 \cos X_k \cos Y_k) + \epsilon_1
\end{equation}
\begin{equation}
    h_1 = -2\sqrt{3}t_2 \sin X_k \sin Y_k + 2it_1(\sin 2X_k + \sin X_k \cos Y_k);
\end{equation}
\begin{equation}
    h_2 = 2t_2(\cos 2X_k -\cos X_k \cos Y_k) + 2i\sqrt{3}t_1 \cos X_k \sin Y_k
\end{equation}
\begin{equation}
    h_{11} = 2t_{11} \cos 2X_k + (t_{11} + 3t_{22}) \cos X_k \cos Y_k + \epsilon_2
\end{equation}
\begin{equation}
    h_{22} = 2t_{22} \cos 2X_k + (3t_{11} + t_{22}) \cos X_k \cos Y_k + \epsilon_2
\end{equation}
\begin{equation}
    h_{12} =
\sqrt{3}(t_{22} -t_{11}) \sin X_k \sin Y_k
+ 4it_{12} \sin X_k(\cos X_k -\cos Y_k)
\end{equation}
 with the shorthand
\begin{equation}
	X_k=\frac{1}{2}k_xa;\quad Y_k=\frac{\sqrt{3}}{2}k_ya.
\end{equation}
The different couplings $\varepsilon_i$ and $t_i$ for  a variety of different TMDs can be found in \cite{Liu_2013}.

Circularly polarized light may be included using the same minimal coupling procedure as in the previous Appendix \ref{App:Downfold3Bandmodel}. We also make use of the same average Hamiltonian approach as in Appendix \ref{App:Downfold3Bandmodel} and find that circularly light renormalized the effective mass near the $K$ point according to 
\begin{equation}
    \frac{m^*(A)}{m^*(0)}=\frac{\Xi(0)}{\Xi(A)}
\end{equation}
with
\begin{equation}
\Xi(A)=4 J_0(a A) \left(-\frac{6 \left(t_1^2+2 \sqrt{3} t_1 t_2+3 t_2^2\right) J_0(a A)}{3 J_0(a A) \left(-2 t_0+t_{11}+2 \sqrt{3} t_{12}+t_{22}\right)+2 (\epsilon_1-\epsilon_2)}-\frac{\sqrt{3} (t_{11}-t_{22})^2}{4 t_{12}}+t_{11}+2 \sqrt{3} t_{12}+t_{22}\right),
\end{equation}
where $J_0$ is the 0th Bessel function of the first kind. An expansion to second order in field strength $A$ yields the result that is referenced in the main text.

\end{document}